\documentclass[floatfix,twocolumn,prl,amsmath]{revtex4-2}
\usepackage{tikz}
\usepackage{adjustbox}
\usepackage{graphicx}
\usepackage{bm}
\usepackage{amssymb}
\usepackage{dcolumn}
\usepackage{subfigure}
\usepackage{multirow}
\usepackage{mathrsfs}
\DeclareMathAlphabet{\mathscrbf}{OMS}{mdugm}{b}{n}
\usepackage{color}

\usetikzlibrary{shapes.geometric, arrows}

\tikzstyle{mybox} = [rectangle, rounded corners, minimum width=3cm, minimum height=1cm,text centered, draw=black, fill=red!30]

\tikzstyle{mybox2} = [rectangle, rounded corners, minimum width=3cm, minimum height=1cm,text centered, draw=black, fill=red!30, text width=6cm]

\tikzstyle{mybox3} = [rectangle, rounded corners, minimum width=3cm, minimum height=1cm,text centered, draw=black, fill=red!30, text width=7cm]
\tikzstyle{arrow} = [thick,->,>=stealth]
\tikzstyle{dashedarrow} = [dashed,->,>=stealth]

\begin{document}
\newcommand{\intqspa}{\int\!\!\frac{\rmd^d q}{(2\pi)^d}}
\newcommand{\intqspathr}{\int\!\!\frac{\rmd^3 q}{(2\pi)^3}}
\newcommand{\intqspatwo}{\int\!\!\frac{\rmd^2 q}{(2\pi)^2}}
\newcommand{\intkspatwo}{\int\!\!\frac{\rmd^2 k}{(2\pi)^2}}
\newcommand{\intkspa}{\int\!\!\frac{\rmd^d k}{(2\pi)^d}}
\newcommand{\intkspapri}{\int\!\!\frac{\rmd^d k'}{(2\pi)^d}}
\newcommand{\vn}[1]{{\boldsymbol{#1}}}
\newcommand{\vht}[1]{{\boldsymbol{#1}}}
\newcommand{\matn}[1]{{\bf{#1}}}
\newcommand{\matnht}[1]{{\boldsymbol{#1}}}
\newcommand{\bege}{\begin{equation}}
\newcommand{\gretke}{G_{\vn{k} }^{\rm R}(\mathcal{E})}
\newcommand{\gret}{G^{\rm R}}
\newcommand{\gadv}{G^{\rm A}}
\newcommand{\gmat}{G^{\rm M}}
\newcommand{\gles}{G^{<}}
\newcommand{\ghat}{\hat{G}}
\newcommand{\sigmahat}{\hat{\Sigma}}
\newcommand{\glesone}{G^{<,{\rm I}}}
\newcommand{\glestwo}{G^{<,{\rm II}}}
\newcommand{\gspec}{G^{\rm S}}
\newcommand{\glesthree}{G^{<,{\rm III}}}
\newcommand{\magdir}{\hat{\vn{n}}}
\newcommand{\sigmaret}{\Sigma^{\rm R}}
\newcommand{\sigmales}{\Sigma^{<}}
\newcommand{\sigmalesone}{\Sigma^{<,{\rm I}}}
\newcommand{\sigmalestwo}{\Sigma^{<,{\rm II}}}
\newcommand{\sigmalesthree}{\Sigma^{<,{\rm III}}}
\newcommand{\sigmaadv}{\Sigma^{A}}
\newcommand{\ee}{\end{equation}}
\newcommand{\bal}{\begin{aligned}}
\newcommand{\defbar}{\overline}
\newcommand{\SM}{\scriptstyle}
\newcommand{\rmd}{{\rm d}}
\newcommand{\rme}{{\rm e}}
\newcommand{\eal}{\end{aligned}}
\newcommand{\crea}[1]{{c_{#1}^{\dagger}}}
\newcommand{\annihi}[1]{{c_{#1}^{\phantom{\dagger}}}}
\newcommand{\udot}{\overset{.}{u}}
\newcommand{\exponential}[1]{{\exp(#1)}}
\newcommand{\phandot}[1]{\overset{\phantom{.}}{#1}}
\newcommand{\phandag}{\phantom{\dagger}}
\newcommand{\Trace}{\text{Tr}}
\setcounter{secnumdepth}{2}
\title{Moment-functional based spectral density-functional theory}
\author{Frank Freimuth$^{1,2}$}
\email[Corresp.~author:~]{f.freimuth@fz-juelich.de}
\author{Stefan Bl\"ugel$^{1}$}
\author{Yuriy Mokrousov$^{1,2}$}
\affiliation{$^1$Peter Gr\"unberg Institut and Institute for Advanced Simulation,
Forschungszentrum J\"ulich and JARA, 52425 J\"ulich, Germany}
\affiliation{$^2$ Institute of Physics, Johannes Gutenberg University Mainz, 55099 Mainz, Germany
}
\begin{abstract}
  We describe a density-functional method which aims at computing the ground
  state electron density and the spectral function at the same
  time. One basic ingredient of our method is the construction of the
  spectral function from the first four spectral moment matrices.
  The second basic ingredient is the construction of
  the spectral moment matrices from density functionals. We call our method
  moment-functional based spectral density-functional theory (MFbSDFT),
  because it is based on density-functionals for the spectral moments and
  because it allows us to compute the spectral function.
  If it is implemented in second variation our method consumes only a fraction
  more computer time than a standard DFT
  calculation with the PBE functional.
  We show that MFbSDFT captures correlation effects such as the valence-band
  satellite in Ni and the formation of lower and upper
  Hubbard bands in SrVO$_3$.
  For the purpose of constructing the spectral function from the first four
  $N\times N$ spectral moment matrices we describe an efficient algorithm
  based on the diagonalization of one hermitean $2N\times 2N$ matrix.
\end{abstract}
\maketitle
\section{Introduction}
In density-functional theory (DFT) the
ground state electron density is determined
by minimizing the total energy
functional~\cite{PhysRev.136.B864}.
While most contributions to the total energy functional,
such as the Hartree energy, the exchange energy,
and the correlation
energy, can be expressed as functionals
of the electron density, it is difficult to express
the kinetic energy directly in this way.
This is why
within the
most popular kind of DFT -- the Kohn-Sham (KS) DFT -- 
the KS-Hamiltonian~\cite{PhysRev.140.A1133}
is set up and solved with the main purpose
to provide the kinetic energy. 

However, the KS energy bands agree very
often fairly well with photoemission data~\cite{PhysRevLett.103.267203}
and the KS spectrum is therefore even used to
compute response properties such as
the anomalous Hall effect~\cite{PhysRevLett.92.037204},
the Gilbert damping~\cite{PhysRevLett.99.027204},
the direct and inverse spin-orbit torque~\cite{invsot},
and the inverse Faraday effect~\cite{ife_berritta}
in metallic systems.
These KS response functions are often in 
good agreement with the corresponding material property tensors
measured experimentally.

Well-known deficiencies of this approach are the underestimation
of the band gap, which may require the application
of band shifts when computing optical
responses such as photocurrents~\cite{PhysRevB.97.245143}
in
semiconductors such as GaAs.
Instead of shifting the bands to match the band gap known from experiments,
one may use the $GW$ approximation~\cite{PhysRev.139.A796},
which is a parameter-free technique based on many-body
perturbation theory and which often predicts gaps that are closer to experiment
than KS-DFT.
However, since one deals then directly with a many-body
Hamiltonian, one forsakes the DFT idea of obtaining all properties
as directly as possible from the ground state density in order to
avoid the complexity and factorial growth of the many-body Hilbert space.
Another short-coming of KS-spectra is the overestimation of the
magnetic moment and the resulting overestimation
of the exchange splitting of some
weak itinerant ferromagnets such as MnSi, which requires us to
reduce the exchange field by a scaling factor
in order to compute the topological Hall effect
in MnSi~\cite{PhysRevLett.112.186601}.

Moreover, a well-known deficiency of the KS spectrum
is the absence of the splitting of bands into
lower and upper Hubbard bands due to strong
electron correlations~\cite{PhysRevB.72.155106}.
Such a splitting of the single-particle bands
leads for example to the appearance of a
satellite peak roughly 6~eV below the Fermi
energy in Ni~\cite{PhysRevLett.43.1431,nickel_PhysRevB.40.5015,PhysRevB.85.235136}.
In order to compute the spectrum in such cases
of strongly interacting
electrons one often uses DFT only to obtain
the KS wavefunctions of a small manifold
corresponding to a small energy window around the
Fermi energy and constructs an interacting
Hamiltonian for this manifold, which one solves
by dynamical mean-field theory (DMFT)~\cite{rmp_dmft,RevModPhys.78.865}
in order to
obtain the spectral function. In other words,
one remains within the DFT-framework in order
to determine the ground state density, but
similarly to $GW$ one leaves this framework
and directly solves an interacting many-electron
Hamiltonian in order to obtain the spectrum
of the correlated system instead of evaluating
a density functional.
However, one may also take a different viewpoint:
The local spectral function of DMFT minimizes the
effective action. In this sense DMFT is a
spectral density-functional approach~\cite{RevModPhys.78.865}.

Nevertheless, the question still poses itself if it is
possible to obtain both ground state density
and correlated spectral function within a
density-functional approach which avoids the direct use
of many-body techniques such as $GW$ and DMFT.
A hermitean $N\times N$ matrix has
$N$ real-valued eigenvalues. This well-known
fact from linear algebra is exploited in many
electronic structure programs based on
density-functional theory, where the
KS equations are solved numerically
by diagonalizing a hermitean matrix.
The direct construction of $2N$ state vectors,
$2N$ state energies, and $2N$ spectral
weight factors from 4 hermitean $N\times N$ spectral moment
matrices~\cite{spectral}
may be considered as a generalization
of the diagonalization of hermitean matrices.
It may also be interpreted as a generalization
of the two-pole approximation~\cite{PhysRevB.38.2608,PhysRevB.50.17980}
used in the
self-consistent spectral moment method of the
single-band Hubbard model to the case of many bands~\cite{spectral}.
Since the self-consistent moment method based on the first four spectral moments
captures the Ni satellite peak~\cite{nickel_PhysRevB.40.5015},
such a generalization may be useful when the
description of the spectral properties
obtained from standard DFT needs to be improved because of
strong correlation effects, which split the electronic
bands into lower and upper Hubbard bands.

The $I$-th spectral moment matrix
is defined by~\cite{bcc_iron_Nolting1995,nickel_PhysRevB.40.5015,PhysRevB.38.2608,spectral}
\bege\label{eq_specmoms_eneint}
\vn{M}^{(I)}_{\sigma }=\frac{1}{\hbar}\int_{-\infty}^{\infty} \vn{S}_{\sigma}(E) E^{I}  dE,
\ee
where $\vn{S}_{\sigma}(E)$ is the spectral density matrix at energy $E$.
In this paper we discuss only the magnetically collinear case without spin-orbit coupling.
Therefore, there are only spectral density matrices $\vn{S}_{\uparrow}(E)$
with spin $\sigma=\uparrow$ and spectral density matrices
$\vn{S}_{\downarrow}(E)$
with spin $\sigma=\downarrow$.
So far, the direct construction of the spectral function of
interacting fermionic many-particle systems from the
first four spectral moments has not yet been investigated
intensively. Well-explored are only the single-band
case with
the first four spectral moments~\cite{PhysRevB.38.2608,PhysRevB.50.17980},
the many-band case with the first two spectral
moments~\cite{Kalashnikov},
which has been shown to provide a
Hartree-Fock type approximation, 
and the option to use 
the spectral moments as sum rules in order to 
guide the construction of the spectral function
by other means~\cite{PhysRevB.44.4670} -- a concept
which one may extend even to nonequilibrium
conditions~\cite{PhysRevB.77.205102}.

Recently, we have demonstrated how to solve the
Hubbard-Rashba model within the many-band
generalization of the two-pole approximation of
the spectral density and of the self-consistent
moment method~\cite{spectral}.
For this purpose we did not make use of the DFT
concept, but instead we computed the higher-order
correlation functions $\langle c^{\dagger}_{i\alpha}c^{\dagger}_{j\beta}c_{l\gamma} c_{m\delta}  \rangle$
self-consistently based on the
spectral theorem~\cite{book_Nolting}.
Such higher-order correlation functions are needed to
compute the spectral
moment $\vn{M}^{(3)}_{\sigma}$, for example.
While the spectral moment matrices $\vn{M}^{(I)}_{\sigma}$ 
all have two orbital indices, the higher-order correlation
functions such as $\langle c^{\dagger}_{i\alpha}c^{\dagger}_{j\beta}c_{l\gamma} c_{m\delta}  \rangle$
have at least four orbital indices.
The computational effort of standard KS DFT scales with
the third power of the number of basis functions $N_{\rm B}$.
Obviously, the many-band self-consistent moment method scales worse,
namely at least $\propto N_{\rm B}^{4}$.
In order to keep the computational effort low,
we therefore proposed in Ref.~\cite{spectral}
to map the KS electronic structure of the valence
bands and the first few conduction bands 
first onto Wannier functions. The resulting Wannier
Hamiltonian may then be supplemented by Hubbard-type interactions
and this interacting Hamiltonian may be treated with
the many-band generalization of the self-consistent
moment method. However, similarly to the $GW$ and LDA+DMFT approaches 
discussed above, one thereby leaves the DFT
framework, because one computes the spectrum using
a many-body Hamiltonian technique instead of a density functional. 

In this paper we combine basic ideas of DFT with the
many-band generalization of the self-consistent moment
method in order to develop an approach which aims
at computing both the ground state density and the
spectral function at the same time without forsaking the
DFT framework. The first Hohenberg-Kohn theorem states
that the ground state electron density determines the
Hamiltonian up to a constant. Since the spectral moments
can be computed from the Hamiltonian, the Hohenberg-Kohn
theorem implies therefore that also the spectral moments may be
expressed as density-functionals. To explore how this can be
done in practice is the central goal of this paper.
Combining this with our recipe~\cite{spectral} to
construct the spectral function from the first four
spectral moments we obtain a 
moment-functional based spectral density-functional
theory (MFbSDFT).

The rest of this paper is structured as follows.
In Sec.~\ref{sec_theory} we explain the theory of MFbSDFT.
In Sec.~\ref{sec_constru_specfun} we describe an efficient
algorithm for computing the spectral function from the first
four spectral moments.
In Sec.~\ref{sec_func} we explain how we construct the moment functionals.
In Sec.~\ref{sec_secvar}
we explain how the MFbSDFT method may be implemented
within the full-potential linearized augmented plane
wave method (FLAPW) within a second variation approach.
In Sec.~\ref{sec_results} we present applications
of our method to fcc Ni and SrVO$_3$.
In
Sec.~\ref{sec_dis_out}
we discuss some open questions of MFbSDFT and strategies of how to
develop it further.
This paper ends with a summary in Sec.~\ref{sec_summary}.

\section{Theory}
\label{sec_theory}
\subsection{The concept of moment functionals}
The ground state charge density defines the Hamiltonian
uniquely (up to a constant)~\cite{PhysRev.136.B864}.
Consequently, it determines also the spectral
function uniquely.
In order to write the spectral function in matrix form we need
a suitable set of orthonormal basis functions
$\phi_{n}(\vn{r})$. Denoting the creation and annihilation operators
corresponding to state $\phi_{n}(\vn{r})|\sigma\rangle$ -- where $|\sigma\rangle$ is a spinor --
by $c^{\dagger}_{\sigma n}$ and $c_{\sigma n}$, respectively,
the matrix elements of the spectral function matrix are
\bege\label{eq_specfun_ft}
S_{\sigma nm}(E)=\frac{1}{2\pi}\int d\,t e^{\frac{i}{\hbar}E t}
\langle[c_{\sigma n}(t), c^{\dagger}_{\sigma m}  ]_{+}
\rangle.
\ee
When periodic boundary conditions are used, the spectral function and the spectral moments
aquire an additional $\vn{k}$-index for the $k$ point $\vn{k}$, which we often
suppress in this manuscript for notational convenience.

The spectral moments may be obtained by plugging
Eq.~\eqref{eq_specfun_ft} into Eq.~\eqref{eq_specmoms_eneint}.
Since the spectral function is uniquely determined by the
ground state density, also the spectral moments are
uniquely defined by it.
The spectral moments may also be expressed in terms of
real-space coordinates:
\bege
M^{(I)}_{\sigma }(\vn{r},\vn{r}')=\frac{1}{\hbar}\int d\, E E^I \sum_{nm}S_{\sigma nm}(E)\phi_{n}(\vn{r})\phi^{*}_{m}(\vn{r}').
\ee
We may consider $M^{(I)}_{\sigma }(\vn{r},\vn{r}')$
as a non-local potential, from which we may obtain the
spectral moment matrices by computing the matrix elements:
\bege
M^{(I)}_{\sigma nm}=\int d^3 r d^3 r'
M^{(I)}_{\sigma }(\vn{r},\vn{r}')\phi^{*}_{n}(\vn{r})\phi_{m}(\vn{r}').
\ee
According to our arguments above, the non-local
potentials $M^{(I)}_{\sigma }(\vn{r},\vn{r}')$
are unique functionals of the electron density.

In KS-DFT the total energy functional
is split into the kinetic energy, the Hartree energy,
and the exchange-correlation energy~\cite{PhysRev.140.A1133}.
The kinetic energy is computed from the KS-wavefunctions,
the Hartree energy is computed from the charge density,
and for the exchange-correlation energy one often uses
analytical expressions in terms of the charge density, which
have been derived for the uniform electron gas~\cite{PhysRevB.45.13244}.
Similarly, the potentials $M^{(I)}_{\sigma }(\vn{r},\vn{r}')$
($I=1,2,3$ if we use the first four moments)
contain contributions from the kinetic energy and from the
Hartree term. In the following section we show
that these contributions may be identified and separated
from a remainder, which thus plays a similar role
in MFbSDFT like the exchange-correlation potential does in KS-DFT.
We expect that useful expressions for this remainder can be
found by evaluating it for the uniform electron gas.

\subsection{Explicit expressions for the moments}
\label{sec_explicit_expr_mom}
We consider the Hamiltonian
\bege
\begin{aligned}
&H=\sum_{\sigma nm}T_{ nm} c^{\dagger}_{\sigma n}c_{\sigma m}\\
&+\frac{1}{2}\sum_{\sigma \sigma' n m n' m'}
V_{nmn'm'} c^{\dagger}_{\sigma n}c^{\dagger}_{\sigma' m}c_{\sigma' m'}c_{\sigma n'},
\end{aligned}
\ee
where
\bege
T_{nm}=\int d^3 r \phi^{*}_{n}(\vn{r})
\left[
  -\frac{1}{2}\Delta + V(\vn{r})
\right]
\phi_{m}(\vn{r})
\ee
and
\bege
V_{nmn'm'}=\int d^3 r_{1} d^3 r_{2}
\frac{\phi_{n}^{*}(\vn{r}_{1})\phi_{m}^{*}(\vn{r}_{2})\phi_{m'}(\vn{r}_{2})\phi_{n'}(\vn{r}_{1})}
{|\vn{r}_{1}-\vn{r}_{2}|},
\ee
and $V(\vn{r})$ is the lattice potential.
Note that in the entire Sec.~\ref{sec_explicit_expr_mom}
we use Hartree atomic units for notational convenience.

Many-body approaches such as
LDA+DMFT often take into account the
Coulomb matrix element $V_{nmn'm'}$
only when all orbitals, i.e., $n$, $m$, $n'$ and
$m'$, describe the same crystal lattice site.
In the simplest approximation $V_{nmn'm'}$ is
described by a single parameter, the so-called
Hubbard-$U$. Components of $V_{nmn'm'}$ that are
neglected hereby are of course partly treated
in LDA+DMFT, because the lattice potential
$V(\vn{r})$ is replaced by the KS potential
in this case. Therefore, the Hubbard-$U$ only
describes the Coulomb interaction from strong
localization of electrons. These effects are
underestimated by KS-DFT and become important
when $U$ approaches or exceeds the bandwidth.
In contrast, we do not restrict $V_{nmn'm'}$
at this point, i.e., both local and non-local
contributions are described by it in MFbSDFT and
$V(\vn{r})$ is the pure lattice potential without
exchange-correlation terms.

The zeroth moment is given by
\bege\label{eq_muba_zero}
M^{(0)}_{\sigma nm}=\langle[c_{\sigma n},c^{\dagger}_{ \sigma m}]_{+}\rangle=\delta_{nm},
\ee
where $[\dots]_{+}$ denotes the anticommutator,
and the
first moment evaluates to
\bege\label{eq_muba_one}
\begin{aligned}
  &M^{(1)}_{\sigma nm}=\langle[[c_{\sigma n},H]_{-},c^{\dagger}_{ \sigma m}]_{+}\rangle=
  T_{nm}+\\
  &+\sum_{n'm'\sigma'}V_{nn'mm'}\langle c^{\dagger}_{ \sigma' n'} c_{\sigma' m'} \rangle\\
  &-\sum_{n'm'}V_{nn'm'm}\langle c^{\dagger}_{ \sigma n'} c_{\sigma m'} \rangle.
\end{aligned}
\ee
Defining the Hartree potential
by
\bege\label{eq_hartree_potential}
V^{\rm H}(\vn{r})=\sum_{\sigma' n' m'}\int d^3 r_{2}
\frac{\phi_{n'}^{*}(\vn{r}_{2})\phi_{m'}(\vn{r}_{2})}
     {|\vn{r}-\vn{r}_{2}|}
\langle c^{\dagger}_{ \sigma' n'} c_{\sigma' m'}\rangle
\ee
and the non-local exchange potential
by 
\bege\label{xc_nonloc}
V_{\sigma}^{\rm X}(\vn{r}_{1},\vn{r}_{2})=-\sum_{ n' m'}
\frac{\phi_{n'}^{*}(\vn{r}_{2})\phi_{m'}(\vn{r}_{1})}
     {|\vn{r}_{1}-\vn{r}_{2}|}
\langle c^{\dagger}_{ \sigma n'} c_{\sigma m'}\rangle
\ee
we may write the first moment as
\bege
\begin{aligned}
  &M^{(1)}_{\sigma nm}=T_{nm}+V^{\rm H}_{nm}+V^{\rm X}_{\sigma nm}=
  \mathcal{M}^{\rm HF}_{\sigma nm},
\end{aligned}
\ee
where $V^{\rm H}_{nm}$ and $V^{\rm X}_{\sigma nm}$
are the matrix elements of the Hartree potential
and of the non-local exchange potential, respectively.
Thus, one obtains a method of Hartree-Fock (HF) type
if one considers only the first two
moments. It differs from the exact Hartree-Fock method
by the self-interaction error (SIE)~\cite{sic_perdew_zunger} (see also
Sec.~\ref{sec_dis_out} for a brief discussion of SIE from the
perspective of MFbSDFT).
Therefore, we introduced the alternative
label $\mathcal{M}^{\rm HF}_{\sigma nm}$
for the first moment, which expresses concisely what this
first moment contains.
Instead of using the non-local exchange potential
Eq.~\eqref{xc_nonloc}
one may use the local exchange potential~\cite{PhysRev.140.A1133}
\bege\label{eq_xc_loc}
V_{\sigma}^{\rm locX}(\vn{r})=
\frac{\partial }{\partial n(\vn{r})}[n(\vn{r})\epsilon^{\rm X}(n(\vn{r}))],
\ee
where
\bege\label{eq_charge_dens_in_out}
n(\vn{r})=\sum_{\sigma n m}\phi^{*}_{n}(\vn{r})\phi_{m}(\vn{r})
\langle c^{\dagger}_{ \sigma n} c_{\sigma m}\rangle
\ee
is the electron density at position $\vn{r}$ and
$\epsilon^{\rm X}(n(\vn{r}))$ is the exchange energy density
for electron density $n(\vn{r})$.
The local potential Eq.~\eqref{eq_xc_loc}
has the advantage that it is computationally very cheap to
evaluate in contrast to the non-local version Eq.~\eqref{xc_nonloc}.
However, hybrid density functionals, which admix exact exchange,
are often more precise than density functionals that use only the
local approximation Eq.~\eqref{eq_xc_loc}.
Fortunately, one may reduce the computational burden of exact
exchange by screening the Coulomb potential~\cite{hse_functional}.
In the numerical calculations in this work we will use
only the local expression Eq.~\eqref{eq_xc_loc}, but, similar to KS-DFT, we expect
that the precision of the MFbSDFT
approach can be increased by avoiding the approximation of
non-local potentials by local potentials.
We leave it for future work to explore how the MFbSDFT approach may be
combined with non-local potentials.

In Ref.~\cite{spectral} we explain that for
independent electrons the spectral moment matrices
commute, i.e.,
\bege
[\vn{M}^{(I)}_{\sigma},\vn{M}_{\sigma}^{(J)}]_{-}=0
\ee
for all $I$ and $J$, and that the eigenvalues of the spectral moment matrix
$\vn{M}_{\sigma}^{(I)}$ are simply the eigenvalues of the single-particle
Hamiltonian raised to the $I$-th power, i.e., $(E_{\sigma n})^{I}$.
For correlated electrons this is not the case~\cite{spectral}.
However, we may expect that the moment $\vn{M}_{\sigma}^{(I)}$ contains
a term $[\vn{\mathcal{M}}_{\sigma}^{\rm HF}]^I$, because there may be cases
where Hartree-Fock provides an excellent description because correlation
effects are small, and these special cases have to be accommodated by
the general theory.
We may therefore expect that the second moment should contain a term
\bege\label{eq_secmom_hfhf}
\begin{aligned}
  &\vn{\mathcal{M}}_{\sigma}^{\rm HF}\vn{\mathcal{M}}_{\sigma}^{\rm HF}=
  \vn{T}\vn{T}+\vn{T}\vn{V}^{\rm H}+\vn{T}\vn{V}^{\rm X}_{\sigma}
  +\vn{V}^{\rm H}\vn{T}+\vn{V}^{\rm H}\vn{V}^{\rm H}+\\
&+  \vn{V}^{\rm H}\vn{V}^{\rm X}_{\sigma}+\vn{V}^{\rm X}_{\sigma} \vn{T}
  +\vn{V}^{\rm X}_{\sigma}\vn{V}^{\rm H}+\vn{V}^{\rm X}_{\sigma}\vn{V}^{\rm X}_{\sigma},
\end{aligned}  
\ee
which is indeed what we find.
We may use this observation to split the second moment into
the anticipated part $\vn{\mathcal{M}}^{\rm HF}_{\sigma}\vn{\mathcal{M}}^{\rm HF}_{\sigma}$
plus additional new terms, which we denote
by $\vn{M}^{(2+)}_{\sigma}$, i.e.,
\bege\label{eq_m2_mhfhf_m2p}
\vn{M}^{(2)}_{\sigma}=\vn{\mathcal{M}}^{\rm HF}_{\sigma}\vn{\mathcal{M}}_{\sigma}^{\rm HF}+\vn{M}_{\sigma}^{(2+)}.
\ee

In contrast to the single-band Hubbard model
with on-site Coulomb interaction, where higher-order correlation
functions appear in the third moment and in the higher moments,
already the second moment $\vn{M}^{(2)}_{\sigma}$ of the many-band case with the full
Coulomb interaction contains the higher-order correlation
function $\langle c^{\dagger}_{\sigma n} c^{\dagger}_{\sigma' m} c_{\sigma n'} c_{\sigma' m'} \rangle$.
In order to identify the terms $\vn{V}^{\rm H}\vn{V}^{\rm H}$, $\vn{V}^{\rm H}\vn{V}_{\sigma}^{\rm X}$,
$\vn{V}^{\rm X}_{\sigma}\vn{V}^{\rm H}$, and $\vn{V}^{\rm X}_{\sigma}\vn{V}^{\rm X}_{\sigma}$ predicted by
Eq.~\eqref{eq_secmom_hfhf} we need to
evaluate $\langle c^{\dagger}_{\sigma n} c^{\dagger}_{\sigma' m} c_{\sigma n'} c_{\sigma' m'} \rangle$
in perturbation theory. In contrast, the terms $\vn{T}\vn{V}^{\rm H}$,
$\vn{T}\vn{V}^{\rm X}_{\sigma}$, $\vn{V}^{\rm H}\vn{T}$, and $\vn{V}^{\rm X}_{\sigma} \vn{T}$
can be identified without using perturbation theory, because they appear with the
correlation function $ \langle c^{\dagger}_{\sigma' m} c_{\sigma' m'} \rangle$.
The term $\vn{T}\vn{T}$ appears even without any correlation function.
Therefore,
we
define
\bege\label{eq_perturbi_split}
\begin{aligned}
&\langle\langle c^{\dagger}_{\sigma n} c^{\dagger}_{\sigma' m} c_{\sigma n'} c_{\sigma' m'} \rangle\rangle=
  \langle c^{\dagger}_{\sigma n} c^{\dagger}_{\sigma' m} c_{\sigma n'} c_{\sigma' m'} \rangle\\
  &-  \langle c^{\dagger}_{\sigma n}c_{\sigma' m'} \rangle\langle c^{\dagger}_{\sigma' m}c_{\sigma n'} \rangle
  +  \langle c^{\dagger}_{\sigma n}c_{\sigma n'} \rangle\langle c^{\dagger}_{\sigma' m}c_{\sigma' m'} \rangle.
\end{aligned}
\ee
The idea behind Eq.~\eqref{eq_perturbi_split} is
that the diagrammatic expression
of $\langle c^{\dagger}_{\sigma n} c^{\dagger}_{\sigma' m} c_{\sigma n'} c_{\sigma' m'} \rangle$
as obtained within perturbation theory
contains terms that may be written
as $\langle c^{\dagger}_{\sigma n}c_{\sigma' m'} \rangle\langle c^{\dagger}_{\sigma' m}c_{\sigma n'} \rangle$
and $-\langle c^{\dagger}_{\sigma n}c_{\sigma n'} \rangle\langle c^{\dagger}_{\sigma' m}c_{\sigma' m'} \rangle$.
Since these latter two terms occur sometimes
in  $\vn{\mathcal{M}}^{\rm HF}_{\sigma}\vn{\mathcal{M}}_{\sigma}^{\rm HF}$
we introduce the notation of Eq.~\eqref{eq_perturbi_split}
in order to split
$\vn{M}^{(2)}_{\sigma}$
into
$\vn{\mathcal{M}}^{\rm HF}_{\sigma}\vn{\mathcal{M}}_{\sigma}^{\rm HF}$
and
$\vn{M}_{\sigma}^{(2+)}$.
Using this notation  we may
write $\vn{M}_{\sigma}^{(2+)}$ as a sum of 17 terms:
\bege\label{eq_m2p_first}
\begin{aligned}
  M_{nm}^{(2+,1)}=\sum_{n'm'tt'\sigma'}V_{nn'tt'}V_{t'tm'm}
  \langle c^{\dagger}_{\sigma' n'} c_{\sigma' m'} \rangle,
\end{aligned}
\ee
which is spin-independent,
\bege\label{eq_m2p_second}
\begin{aligned}
 M_{\sigma nm}^{(2+,2)}=-\sum_{n'm'tt'}V_{nn'tt'}V_{tt'm'm}
  \langle c^{\dagger}_{\sigma n'} c_{\sigma m'} \rangle,
\end{aligned}
\ee  
\bege\label{eq_m2p_third}
\begin{aligned}
  M_{ nm}^{(2+,3)}=-\!\!\!\sum_{n'm'tt'z\sigma'}\!\!\!V_{nn'zt'}V_{tzm'm}
  \langle\langle
  c^{\dagger}_{\sigma' n'} c^{\dagger}_{\sigma' t} c_{\sigma' t'} c_{\sigma' m'} 
\rangle \rangle,
\end{aligned}
\ee    
which does not depend on the spin,
\bege\label{eq_m2p_fourth}
\begin{aligned}
  M_{\sigma nm}^{(2+,4)}=-\sum_{n'm'tt'z}V_{nn't't}V_{tm'zm}
  \langle\langle
  c^{\dagger}_{\sigma n'} c^{\dagger}_{\sigma m'} c_{\sigma t'} c_{\sigma z} 
\rangle \rangle,
\end{aligned}
\ee 
\bege\label{eq_m2p_fifth}
\begin{aligned}
  M_{\sigma nm}^{(2+,5)}=\sum_{n'm'tt'z}V_{nn'tt'}V_{tm'zm}
  \langle\langle
  c^{\dagger}_{\sigma n'} c^{\dagger}_{\sigma m'} c_{\sigma t'} c_{\sigma z} 
\rangle \rangle,
\end{aligned}
\ee 
\bege\label{eq_m2p_sixth}
\begin{aligned}
  M_{\sigma nm}^{(2+,6)}=\sum_{n'm'tt'z}V_{nn'tt'}V_{zm'n'm}
  \langle
  c^{\dagger}_{\sigma z} c^{\dagger}_{\sigma m'} c_{\sigma t} c_{\sigma t'}
  \rangle, 
\end{aligned}
\ee 
\bege\label{eq_m2p_seventh}
\begin{aligned}
  M_{\sigma nm}^{(2+,7)}=\sum_{n'm'tt'z}V_{nn'tz}V_{m'zt'm}
  \langle\langle
  c^{\dagger}_{\sigma n'} c^{\dagger}_{\sigma m'} c_{\sigma t} c_{\sigma t'}
  \rangle \rangle,
\end{aligned}
\ee 
\bege\label{eq_m2p_eighth}
\begin{aligned}
  M_{\sigma nm}^{(2+,8)}=-\sum_{n'm'tt'z}V_{nn'tt'}V_{zm'n'm}
  \langle
  c^{\dagger}_{\sigma m'} c^{\dagger}_{-\sigma z} c_{\sigma t} c_{-\sigma t'}
  \rangle, 
\end{aligned}
\ee 
\bege\label{eq_m2p_nineth}
\begin{aligned}
  M_{\sigma nm}^{(2+,9)}=-\!\!\!\sum_{n'm'tt'z}\!\!\!V_{nn'tt'}V_{m'tzm}
  \langle\langle
  c^{\dagger}_{\sigma n'} c^{\dagger}_{-\sigma m'} c_{\sigma t'} c_{-\sigma z}
    \rangle \rangle,
\end{aligned}
\ee 
\bege\label{eq_m2p_tenth}
\begin{aligned}
  M_{\sigma nm}^{(2+,10)}=-\!\!\!\sum_{n'm'tt'z}\!\!\!V_{nn'tt'}V_{m'tzm}
  \langle\langle
  c^{\dagger}_{\sigma m'} c^{\dagger}_{-\sigma n'} c_{\sigma z} c_{-\sigma t'}
      \rangle \rangle,
\end{aligned}
\ee 
\bege\label{eq_m2p_eleventh}
\begin{aligned}
  M_{\sigma nm}^{(2+,11)}=\sum_{n'm'tt'z}V_{nn'tt'}V_{tm'zm}
  \langle\langle
  c^{\dagger}_{\sigma m'} c^{\dagger}_{-\sigma n'} c_{\sigma z} c_{-\sigma t'}
      \rangle \rangle,
\end{aligned}
\ee 
\bege\label{eq_m2p_twelfth}
\begin{aligned}
  M_{\sigma nm}^{(2+,12)}=\sum_{n'm'tt'z}V_{nn'tt'}V_{t'm'zm}
  \langle
  c^{\dagger}_{\sigma m'} c^{\dagger}_{-\sigma n'} c_{\sigma t} c_{-\sigma z}
      \rangle, 
\end{aligned}
\ee 
\bege\label{eq_m2p_thirteenth}
\begin{aligned}
  M_{\sigma nm}^{(2+,13)}=\sum_{n'm'tt'z}V_{nn'tt'}V_{m't'zm}
  \langle\langle
  c^{\dagger}_{\sigma n'} c^{\dagger}_{-\sigma m'} c_{\sigma t} c_{-\sigma z}
      \rangle \rangle,
\end{aligned}
\ee 
\bege\label{eq_m2p_fourteenth}
\begin{aligned}
  M_{ nm}^{(2+,14)}=-\!\!\!\sum_{\sigma' n'm'tt'z}\!\!\!V_{nn'tt'}V_{m'tzm}
  \langle c^{\dagger}_{\sigma' m'} c_{\sigma' t'}
  \rangle
  \langle c^{\dagger}_{\sigma' n'} c_{\sigma' z}
  \rangle,
\end{aligned}
\ee
which is spin-independent,
\bege\label{eq_m2p_fifteenth}
\begin{aligned}
  M_{\sigma nm}^{(2+,15)}=-\sum_{n'm'tt'z}V_{nn'tt'}V_{t'm'zm}
  \langle c^{\dagger}_{\sigma m'} c_{\sigma t}
  \rangle
  \langle c^{\dagger}_{\sigma n'} c_{\sigma z}
  \rangle,
\end{aligned}
\ee
\bege\label{eq_m2p_sixteenth}
\begin{aligned}
  M_{\sigma nm}^{(2+,16)}=\sum_{n'm'tt'z}V_{nn'tt'}V_{tm'zm}
    \langle c^{\dagger}_{\sigma m'} c_{\sigma t'}
  \rangle
  \langle c^{\dagger}_{\sigma n'} c_{\sigma z}
  \rangle,
\end{aligned}
\ee
and
\bege\label{eq_m2p_seventeenth}
\begin{aligned}
  M_{\sigma nm}^{(2+,17)}=\sum_{n'm'tt'z}V_{nn'tt'}V_{m't'zm}
      \langle c^{\dagger}_{\sigma m'} c_{\sigma t}
  \rangle
  \langle c^{\dagger}_{\sigma n'} c_{\sigma z}
  \rangle.
\end{aligned}
\ee

In order to evaluate the contributions to $\vn{M}^{(2+)}$ in a
way similar to Eq.~\eqref{eq_xc_loc}, we suggest to consider the contractions
\bege\label{eq_dft_contraction}
\mathcal{C}_{\sigma}^{(2+,j)}=\sum_{nm}M_{\sigma nm}^{(2+,j)}\langle c_{\sigma n}^{\dagger} c_{\sigma m} \rangle
\ee
and to compute them for the uniform electron gas as a function of electron density.
Similarly to Eq.~\eqref{eq_xc_loc} we
assume that we may derive local potentials
\bege\label{eq_dft_locpot}
\mathcal{V}_{\sigma}^{(2+,j)}(\vn{r})=\frac{\partial}{\partial n(\vn{r})}[\mathcal{C}_{\sigma}^{(2+,j)}]
\ee
from these contractions and compute
the moments from these local potentials:
\bege\label{eq_dft_style_mat}
M_{\sigma nm}^{(2+,j)}=\int d^3 r \mathcal{V}_{\sigma}^{(2+,j)}(\vn{r}) \phi_{n}^{*}(\vn{r})\phi_{m}(\vn{r}).
\ee
Many popular exchange-correlation potentials are constructed
with the help of Green's function Monte Carlo
simulations of the energy of the uniform
electron gas~\cite{PhysRevLett.45.566,PhysRevB.45.13244},
because the universality of the exchange correlation potential implies
that it may be constructed from a uniform system.
However, Green's function Monte Carlo data
are not yet available for our expressions Eq.~\eqref{eq_m2p_first}
through Eq.~\eqref{eq_m2p_seventeenth}.
On the other hand, diagrammatic perturbation theory
has been used to derive expressions for the energy of
the uniform electron gas in the limit of low and high
density and these results are considered in the construction
of exchange correlation potentials
as well~\cite{PhysRev.106.364,PhysRevB.45.13244}.
For the purpose of demonstrating the feasibility of the
MFbSDFT approach it is sufficient to find simple approximate
expressions for the contractions Eq.~\eqref{eq_dft_contraction}.
Therefore, we evaluate
these contractions
for the uniform electron gas using perturbation theory
in Appendix~\ref{feg_2p}.
We leave it for future work to find accurate
analytic representations of the contractions
of Eq.~\eqref{eq_m2p_first}
through Eq.~\eqref{eq_m2p_seventeenth}
based on techniques such as Green's function Monte Carlo
simulations and diagrammatic expansions for the high-density limit.

Similar to Eq.~\eqref{eq_m2_mhfhf_m2p},
one may anticipate that the third moment
may be decomposed as
\bege
\vn{M}^{(3)}_{\sigma}=\vn{\mathcal{M}}^{\rm HF}_{\sigma}\vn{\mathcal{M}}_{\sigma}^{\rm HF}\vn{\mathcal{M}}^{\rm HF}_{\sigma}+\vn{M}_{\sigma}^{(3+)},
\ee
where
\bege\label{eq_hfhfhf}
\begin{aligned}
  &\vn{\mathcal{M}}^{\rm HF}_{\sigma}\vn{\mathcal{M}}_{\sigma}^{\rm HF}\vn{\mathcal{M}}^{\rm HF}_{\sigma}=
  \vn{T}\vn{T}\vn{T}+\vn{T}\vn{T}\vn{V}^{\rm H}+\vn{T}\vn{T}\vn{V}^{\rm X}_{\sigma}\\
&+\vn{T}\vn{V}^{\rm H}\vn{T}+\vn{T}\vn{V}^{\rm H}\vn{V}^{\rm H}
+  \vn{T}\vn{V}^{\rm H}\vn{V}^{\rm X}_{\sigma}\\
  &+\vn{T}\vn{V}^{\rm X}_{\sigma} \vn{T}
+\vn{V}^{\rm H}\vn{T}\vn{V}^{\rm X}_{\sigma}\vn{V}^{\rm H}+\vn{V}^{\rm H}\vn{T}\vn{V}^{\rm X}_{\sigma}\vn{V}^{\rm X}_{\sigma}\\
&+\vn{V}^{\rm H}\vn{T}\vn{T}+\vn{V}^{\rm H}\vn{T}\vn{V}^{\rm H}+\vn{V}^{\rm H}\vn{T}\vn{V}^{\rm X}_{\sigma}\\
&+\vn{V}^{\rm H}\vn{V}^{\rm H}\vn{T}+\vn{V}^{\rm H}\vn{V}^{\rm H}\vn{V}^{\rm H}
+  \vn{V}^{\rm H}\vn{V}^{\rm H}\vn{V}^{\rm X}_{\sigma}\\
  &+\vn{V}^{\rm H}\vn{V}^{\rm X}_{\sigma} \vn{T}
+\vn{V}^{\rm H}\vn{V}^{\rm X}_{\sigma}\vn{V}^{\rm H}+\vn{V}^{\rm H}\vn{V}^{\rm X}_{\sigma}\vn{V}^{\rm X}_{\sigma}\\
  &+\vn{V}^{\rm X}_{\sigma}\vn{T}\vn{T}+\vn{V}^{\rm X}_{\sigma}\vn{T}\vn{V}^{\rm H}+\vn{V}^{\rm X}_{\sigma}\vn{T}\vn{V}^{\rm X}_{\sigma}\\
&+\vn{V}^{\rm X}_{\sigma}\vn{V}^{\rm H}\vn{T}+\vn{V}^{\rm X}_{\sigma}\vn{V}^{\rm H}\vn{V}^{\rm H}
+  \vn{V}^{\rm X}_{\sigma}\vn{V}^{\rm H}\vn{V}^{\rm X}_{\sigma}\\
  &+\vn{V}^{\rm X}_{\sigma}\vn{V}^{\rm X}_{\sigma} \vn{T}
  +\vn{V}^{\rm X}_{\sigma}\vn{V}^{\rm X}_{\sigma}\vn{V}^{\rm H}+\vn{V}^{\rm X}_{\sigma}\vn{V}^{\rm X}_{\sigma}\vn{V}^{\rm X}_{\sigma}, 
\end{aligned}  
\ee
which is indeed what we find: To identify $\vn{T}\vn{T}\vn{T}$ in $\vn{M}^{(3)}_{\sigma}$
one needs to check the terms without correlation functions.
To find the terms that contain two factors of the matrix $\vn{T}$, i.e.,
the terms $\vn{T}\vn{T}\vn{V}^{\rm H}$, $\vn{T}\vn{V}^{\rm H}\vn{T}$, $\vn{V}^{\rm H}\vn{T}\vn{T}$,
$\vn{T}\vn{T}\vn{V}^{\rm X}$, $\vn{T}\vn{V}^{\rm X}\vn{T}$, and $\vn{V}^{\rm X}\vn{T}\vn{T}$,
one needs to look out for the contributions to $\vn{M}^{(3)}_{\sigma}$ that contain the
correlation function $ \langle c^{\dagger}_{\sigma' m} c_{\sigma' m'} \rangle$.
To track down the terms that contain a single factor of the matrix $\vn{T}$, i.e., the
terms
$\vn{T}\vn{V}^{\rm H}\vn{V}^{\rm H}$,
$\vn{V}^{\rm H}\vn{T}\vn{V}^{\rm H}$,
$\vn{V}^{\rm H}\vn{V}^{\rm H}\vn{T}$,
$\vn{T}\vn{V}^{\rm H}\vn{V}^{\rm X}$,
$\vn{V}^{\rm H}\vn{T}\vn{V}^{\rm X}$,
$\vn{V}^{\rm H}\vn{V}^{\rm X}\vn{T}$,
$\vn{T}\vn{V}^{\rm X}\vn{V}^{\rm H}$,
$\vn{V}^{\rm X}\vn{T}\vn{V}^{\rm H}$,
$\vn{V}^{\rm X}\vn{V}^{\rm H}\vn{T}$,
$\vn{T}\vn{V}^{\rm X}\vn{V}^{\rm X}$,
$\vn{V}^{\rm X}\vn{T}\vn{V}^{\rm X}$, and
$\vn{V}^{\rm X}\vn{V}^{\rm X}\vn{T}$,
one needs to find the contributions to $\vn{M}^{(3)}_{\sigma}$ that contain the
correlation function $ \langle c^{\dagger}_{\sigma m} c^{\dagger}_{\sigma' m'} c_{\sigma n} c_{\sigma' n'}\rangle$ and
one has to evaluate this correlation function
in perturbation theory.
In order to identify all those terms in Eq.~\eqref{eq_hfhfhf}
that do not contain the matrix $\vn{T}$, i.e.,
the terms
$\vn{V}^{\rm H}\vn{V}^{\rm H}\vn{V}^{\rm H}$,
$\vn{V}^{\rm X}\vn{V}^{\rm H}\vn{V}^{\rm H}$,
$\vn{V}^{\rm H}\vn{V}^{\rm X}\vn{V}^{\rm H}$,
$\vn{V}^{\rm H}\vn{V}^{\rm H}\vn{V}^{\rm X}$,
$\vn{V}^{\rm H}\vn{V}^{\rm X}\vn{V}^{\rm X}$,
$\vn{V}^{\rm X}\vn{V}^{\rm H}\vn{V}^{\rm X}$,
$\vn{V}^{\rm X}\vn{V}^{\rm X}\vn{V}^{\rm H}$, and
$\vn{V}^{\rm X}\vn{V}^{\rm X}\vn{V}^{\rm X}$,
one needs to check the expressions
that contain the correlation
function $ \langle c^{\dagger}_{\sigma m} c^{\dagger}_{\sigma' m'}c^{\dagger}_{\sigma'' t} c_{\sigma n} c_{\sigma' n'}c_{\sigma'' t'}\rangle$ and
one has to evaluate this correlation function
in perturbation theory.

When one evaluates the
correlator $\langle c^{\dagger}_{\sigma n} c^{\dagger}_{\sigma' m} c_{\sigma n'} c_{\sigma' m'} \rangle$
in perturbation theory in order to extract the terms
discussed above, e.g.\ $\vn{V}^{\rm H}\vn{V}^{\rm H}\vn{T}$, one
may use Eq.~\eqref{eq_perturbi_split}
like for the second moment.
This procedure generates a group of
terms in $\vn{M}_{\sigma}^{(3+)}$
that contain $\langle\langle c^{\dagger}_{\sigma n} c^{\dagger}_{\sigma' m} c_{\sigma n'} c_{\sigma' m'} \rangle\rangle$.
Similarly, it is convenient to define
\bege\label{eq_perturbi_split_6ops}
\begin{aligned}
  &\langle\langle c^{\dagger}_{\sigma m} c^{\dagger}_{\sigma' m'}c^{\dagger}_{\sigma'' t} c_{\sigma n} c_{\sigma' n'}c_{\sigma'' t'}\rangle\rangle\\
  &=
  \langle c^{\dagger}_{\sigma m} c^{\dagger}_{\sigma' m'}c^{\dagger}_{\sigma'' t} c_{\sigma n} c_{\sigma' n'}c_{\sigma'' t'}\rangle\\
&-\langle c^{\dagger}_{\sigma m}c_{\sigma n} \rangle \langle  c^{\dagger}_{\sigma' m'}c^{\dagger}_{\sigma'' t}  c_{\sigma' n'}c_{\sigma'' t'}\rangle\\
&+\langle c^{\dagger}_{\sigma m} c_{\sigma' n'}\rangle  \langle  c^{\dagger}_{\sigma' m'}c^{\dagger}_{\sigma'' t} c_{\sigma n} c_{\sigma'' t'}\rangle\\
&-\langle c^{\dagger}_{\sigma m}c_{\sigma'' t'} \rangle   \langle  c^{\dagger}_{\sigma' m'}c^{\dagger}_{\sigma'' t} c_{\sigma n} c_{\sigma' n'}\rangle
\end{aligned}  
\ee
and to use Eq.~\eqref{eq_perturbi_split}
in order to replace the correlators of the type
$\langle  c^{\dagger}_{\sigma' m'}c^{\dagger}_{\sigma'' t}  c_{\sigma' n'}c_{\sigma'' t'}\rangle$ on the right-hand side of Eq.~\eqref{eq_perturbi_split_6ops}
by $\langle\langle  c^{\dagger}_{\sigma' m'}c^{\dagger}_{\sigma'' t}  c_{\sigma' n'}c_{\sigma'' t'}\rangle\rangle$
and the simpler correlators
$\langle  c^{\dagger}_{\sigma' m'}  c_{\sigma' n'}\rangle$.
When we use this procedure to express
$\langle c^{\dagger}_{\sigma m} c^{\dagger}_{\sigma' m'}c^{\dagger}_{\sigma'' t} c_{\sigma n} c_{\sigma' n'}c_{\sigma'' t'}\rangle$
in terms of
the correlators
$\langle  c^{\dagger}_{\sigma' m'}  c_{\sigma' n'}\rangle$
and thereby extract the terms
discussed above, e.g.\ $\vn{V}^{\rm H}\vn{V}^{\rm H}\vn{V}^{\rm H}$,
we generate additional groups of terms in $\vn{M}_{\sigma}^{(3+)}$,
which contain
$\langle\langle  c^{\dagger}_{\sigma' m'}c^{\dagger}_{\sigma'' t}  c_{\sigma' n'}c_{\sigma'' t'}\rangle\rangle$
or
$\langle\langle c^{\dagger}_{\sigma m} c^{\dagger}_{\sigma' m'}c^{\dagger}_{\sigma'' t} c_{\sigma n} c_{\sigma' n'}c_{\sigma'' t'}\rangle\rangle$.

The remaining contributions to $\vn{M}_{\sigma}^{(3+)}$
may be split into groups of formally similar expressions.
The first group of two terms in $\vn{M}_{\sigma}^{(3+)}$
contains two matrices $\vn{T}$ and the correlation
function $\langle c^{\dagger}_{\sigma' m} c_{\sigma' m'} \rangle$:
\bege\label{eq_m3p_one}
  M_{ nm}^{(3+,1)}=\sum_{n'tt'\sigma'}V_{nn'tt'} T_{tm}
    S_{\sigma' t'n'},
\ee
which is spin-independent, and
\bege\label{eq_m3p_two}
 M_{\sigma nm}^{(3+,2)}=-\sum_{n'tt'}V_{nn't't} T_{tm}
     S_{\sigma t'n'},
\ee
where
\bege
S_{\sigma' t' n'}=\sum_{m'}\Bigl[
    T_{t'm'}
\langle c^{\dagger}_{\sigma' n'} c_{\sigma' m'}\rangle
    -    
    \langle c^{\dagger}_{\sigma' m'} c_{\sigma' t'}\rangle
    T_{m'n'}
\Bigr]
\ee
is the commutator between the matrix $\vn{T}$ and the density matrix.

It is desirable to rewrite Eq.~\eqref{eq_m3p_one}
in a form that permits a numerically efficient
evaluation of this term, because the direct
computation of Eq.~\eqref{eq_m3p_one} will
often be numerically demanding due to the four indices of
the Coulomb matrix element.
We may exploit that the indices $t'$ and $n'$ couple to the
Coulomb matrix element in a way that allows us to identify the
density matrix.
Therefore we may define
\bege
\mathcal{N}(\vn{r})=\sum_{n't'\sigma'}\phi^{*}_{n'}(\vn{r})\phi_{t'}(\vn{r})
S_{\sigma' t' n'},
\ee
from which we compute
the Hartree-type integral
\bege
\mathcal{F}(\vn{r})=\int d^3 r_{2} \frac{\mathcal{N}(\vn{r}_{2})}{|\vn{r}-\vn{r}_{2}|}.
\ee
Eq.~\eqref{eq_m3p_one} may now be written as
\bege
M_{ nm}^{(3+,1)}=\sum_{t}\mathcal{F}_{nt}T_{tm},
\ee
where $\mathcal{F}_{nt}$ are the matrix elements of the
Hartree-type potential $\mathcal{F}(\vn{r})$.

Similarly, in order to evaluate Eq.~\eqref{eq_m3p_two}
we may exploit that the indices $t'$ and $n'$ couple to the
Coulomb matrix element in a way that allows us to identify
a non-local potential.
Therefore we may define
\bege
\mathcal{L}_{\sigma}(\vn{r}_{1},\vn{r}_{2})=\sum_{n't'}
\frac{\phi^{*}_{n'}(\vn{r}_{1})\phi_{t'}(\vn{r}_{2})}{|\vn{r}_{1}-\vn{r}_{2}|}S_{\sigma' t' n'}.
\ee
Eq.~\eqref{eq_m3p_two} may now be written as
\bege
M_{\sigma nm}^{(3+,2)}=-\sum_{t}\mathcal{L}_{\sigma nt}T_{tm},
\ee
where $\mathcal{L}_{\sigma nt}$ are the matrix elements of the
non-local Fock-type potential $\mathcal{L}_{\sigma}(\vn{r}_{1},\vn{r}_{2})$.

The next group of two terms in $\vn{M}_{\sigma}^{(3+)}$
does not contain the matrix $\vn{T}$, but it contains the correlation
function $\langle c^{\dagger}_{\sigma' m} c_{\sigma' m'} \rangle$:
\bege\label{eq_m3p_3}
M_{nm}^{(3+,3)}=\sum_{\sigma'n'tt'}\sum_{zz'm'} V_{nn'tt'}V_{tt'zz'}V_{z'zm'm}\langle c^{\dagger}_{\sigma' n'} c_{\sigma' m'} \rangle,
\ee
which does not depend on the spin, and
\bege\label{eq_m3p_4}
M_{\sigma nm}^{(3+,4)}=-\sum_{n'tt'}\sum_{zz'm'} V_{nn'tt'}V_{tt'z'z}V_{z'zm'm}\langle c^{\dagger}_{\sigma n'} c_{\sigma m'} \rangle.
\ee
In Appendix~\ref{sec_app_compare} we show that the sum of Eq.~\eqref{eq_m3p_3} and Eq.~\eqref{eq_m3p_4}
turns into the very simple result $U\langle n_{-\sigma} \rangle$ in the case of the single-band Hubbard model.
However, for realistic many-band systems, the direct evaluation of these expressions
may be numerically demanding due to the four indices of the Coulomb matrix element.
Therefore, we may  try to use concepts from DFT to simplify the calculations.
In order to approximate these contributions
by local functionals, we may use the recipe
described above in Eq.~\eqref{eq_dft_contraction},
Eq.~\eqref{eq_dft_locpot},
and Eq.~\eqref{eq_dft_style_mat}.
In Appendix~\ref{feg_3p} we evaluate the
corresponding contractions for the uniform electron gas.

$\vn{M}^{(3+)}$ contains several additional groups of terms that we
have not discussed yet.
One group of terms contains three factors of the
Coulomb matrix element and the correlator
$ \langle c^{\dagger}_{\sigma m} c^{\dagger}_{\sigma' m'}c^{\dagger}_{\sigma'' t} c_{\sigma n} c_{\sigma' n'}c_{\sigma'' t'}\rangle$,
but the indices are not connected in a way that terms such as $\vn{H}\vn{H}\vn{H}$ or $\vn{H}\vn{X}\vn{H}$
arise, which we have already discussed above.
An example from this group of terms is
\bege\label{eq_example5}
\begin{aligned}
M_{\sigma nm}^{(3+,5)}&=  -\sum_{n'tt'}
\sum_{uu'}
\sum_{zz'm'}
  V_{nn'tt'}
  V_{t'm'uu'}
  V_{zz'm'm}\\
  &\times\langle c^{\dagger}_{\sigma n'} c^{\dagger}_{\sigma z}c^{\dagger}_{\sigma z'} c_{\sigma t} c_{\sigma u}c_{\sigma u'}\rangle.
\end{aligned}  
\ee
The first index $t'$ of $V_{t'm'uu'}$ is shared with
$V_{nn'tt'}$, while the second index $m'$ of $V_{t'm'uu'}$
is shared with $V_{zz'm'm}$. The last two indices, $u$ and
$u'$ are contracted with the correlation
function $\langle c^{\dagger}_{\sigma n'} c^{\dagger}_{\sigma z}c^{\dagger}_{\sigma z'} c_{\sigma t} c_{\sigma u}c_{\sigma u'}\rangle$. In this term, it is therefore not possible to
express $V_{t'm'uu'}$ through the matrices $\vn{H}$ or $\vn{X}$
when perturbation theory is used.
Other terms in this group differ from Eq.~\eqref{eq_example5} for example
due to different spin quantum numbers in the correlation function, e.g.\
$\langle c^{\dagger}_{-\sigma n'} c^{\dagger}_{\sigma z}c^{\dagger}_{\sigma z'} c_{\sigma t} c_{-\sigma u}c_{\sigma u'}\rangle$, or they differ due to different indices of the Coulomb matrix elements,
e.g.\ $V_{nt'uu'}V_{m'n'tt'} V_{zz'm'm}$.

There is a second group of terms that contains
three factors of the Coulomb matrix element as well. However,
it contains the correlator $\langle c^{\dagger}_{\sigma m} c^{\dagger}_{\sigma' m'} c_{\sigma n} c_{\sigma' n'}\rangle$ instead.
An example from this group of terms is
\bege\label{eq_example6}
\begin{aligned}
M_{\sigma nm}^{(3+,6)}&=  -\sum_{n'tt'}
\sum_{zz'}\sum_{uu'}
  V_{nn'tt'}
  V_{tt'zz'}
  V_{z'uu'm}\\
  &\times\langle c^{\dagger}_{\sigma n'} c^{\dagger}_{\sigma u} c_{\sigma z} c_{\sigma u'}\rangle.
\end{aligned}  
\ee
Since $V_{tt'zz'}$ couples to the correlation function only through the
index $z$, it cannot be expressed through the matrices $\vn{H}$ or $\vn{X}$
when perturbation theory is used. Similar to the previous group,
the other members in this group differ from this example due to different spin quantum
numbers in the correlation function, or due to different indices of the Coulomb
matrix elements.

Another group of terms contains the matrix $\vn{T}$ once,
the Coulomb matrix elements twice, and the correlation
function $\langle c^{\dagger}_{\sigma m'} c_{\sigma n'}\rangle$.
An example from this group of terms is
\bege\label{eq_example7}
M_{\sigma nm}^{(3+,7)}=  -\sum_{tt'}
\sum_{n'm'z}
  V_{nn'tt'}
  V_{tt'zm'}T_{m'm}
 \langle c^{\dagger}_{\sigma n'} c_{\sigma z}\rangle.
\ee
Similar to the previous two groups, the other members in this group
differ from the example of Eq.~\eqref{eq_example7} due to different
spin quantum numbers in the correlator, i.e., $\langle c^{\dagger}_{-\sigma n'} c_{-\sigma z}\rangle$,
and due to different indices in the Coulomb matrix elements.

A similar group of terms contains the
correlator  $\langle c^{\dagger}_{\sigma' z'}c^{\dagger}_{\sigma u} c_{\sigma' z}c_{\sigma t'}\rangle$
instead of $\langle c^{\dagger}_{\sigma n'} c_{\sigma z}\rangle$.
An example is given by
\bege\label{eq_example8}
M_{\sigma nm}^{(3+,8)}=  \sum_{tt'}
\sum_{n'm'z}
  V_{ntm't'}
  V_{z'uzt}
  T_{m'm}
 \langle c^{\dagger}_{-\sigma z'}c^{\dagger}_{\sigma u} c_{-\sigma z}c_{\sigma t'}\rangle.
\ee

\section{Construction of the spectral function from the spectral moments}
\label{sec_constru_specfun}
In Ref.~\cite{spectral}
we have shown
that the
spectral function of 4 spectral moment matrices of size $N\times N$
may be obtained by solving $4 N^2$ coupled non-linear equations.
While this approach is efficient for small $N$, it may become
inefficient for large $N$. The reason may be understood from
the amount of computer memory needed to store the Jacobian of
the system of non-linear equations. The size of the Jacobian
scales like $16 N^4$. In contrast, the size of the KS Hamiltonian
matrix used in DFT codes scales like $N^2$ with the number
$N$ of basis functions.
Therefore, we describe an alternative algorithm in this section, which
is more efficient than solving systems of coupled non-linear equations
when $N$ is large. Since we discuss in Ref.~\cite{spectral} that
finding the spectral function from non-commuting spectral matrices
can be interpreted as a generalization of matrix diagonalization,
it is perhaps not surprising that the new algorithm that
we describe in this section uses such concepts.

In the following we describe the algorithm to construct the spectral
function from the spectral moment matrices $\vn{M}^{(1)}$, $\vn{M}^{(2)}$, $\vn{M}^{(3)}$,
where we assume that the zeroth spectral moment matrix is simply the unit matrix.
Only the final result is described here, while the detailed proof is given in
Appendix~\ref{sec_app_diag}.
First, construct the hermitean $N\times N$ matrix
\bege
\vn{M}^{(2+)}=\vn{M}^{(2)}-\vn{M}^{(1)}\vn{M}^{(1)}.
\ee
Next, diagonalize $\vn{M}^{(2+)}$:
\bege
\vn{M}^{(2+)}=\vn{U}\vn{D}\vn{U}^{\dagger},
\ee
where $\vn{U}$ is a unitary matrix and $\vn{D}$ is a diagonal matrix.
Using $\vn{D}$ and $\vn{U}$ construct the matrix
\bege\label{eq_b1}
\vn{B}_{1}=\vn{U}\sqrt{\vn{D}}.
\ee
Employ the inverse of its hermitean adjoint together with the moment matrices to compute the matrix
\bege\label{eq_b2}
\vn{B}_{2}=[\vn{M}^{(3)}-\vn{M}^{(2)} \vn{M}^{(1)}][\vn{B}_{1}^{\dagger}]^{-1}.
\ee
Use it to obtain the matrix
\bege\label{eq_d1}
\vn{D}_{1}=\vn{B}_{1}^{-1}[\vn{B}_2-\vn{M}^{(1)}\vn{B}_1].
\ee
Finally, take $\vn{B}_{1}$, $\vn{D}_{1}$ and the first moment matrix to
construct the $2N\times 2N$ matrix
\bege
\vn{\mathcal{B}}^{(1)}=\left(
\begin{array}{cc}
\vn{M}^{(1)}
&\vn{B}_{1} \\
\vn{B}_{1}^{\dagger} &\vn{D}_{1}
\end{array}
\right)
\ee
and 
diagonalize it:
\bege\label{eq_b1_ududag}
\vn{\mathcal{B}}^{(1)}=\vn{\mathcal{U}}\vn{\mathcal{D}}\vn{\mathcal{U}}^{\dagger}.
\ee

The unitary matrix $\vn{\mathcal{U}}$ contains the normalized eigenvectors
of $\vn{\mathcal{B}}^{(1)}$ as its columns.
Compute the spectral weight of state $j$
from
\bege\label{eq_spec_wei}
a_{j}=\sum_{i=1}^{N} \mathcal{U}_{i j} [\mathcal{U}_{i j}]^{*}.
\ee
Note that $a_{j}$ may be smaller than one,
because the summation over the index
$i$ goes only from 1 to $N$ and not from 1 to $2N$.
Therefore, spectral weights
smaller than 1 may occur when bands split into lower and
upper Hubbard bands.

Construct the $N\times 2 N$ matrix $\vn{\mathcal{V}}$
according to
\bege\label{eq_statevecs}
\mathcal{V}_{i j}=\frac{\mathcal{U}_{i j}}{\sqrt{a_{j}}}.
\ee
Note that $i=1,...,N$, i.e., only the first $N$ entries of the $j$-th column of
$\vn{\mathcal{U}}$ are used, while every column of $\vn{\mathcal{U}}$ has of course
$2N$ entries in total.

The spectral function is given by
\bege\label{eq_final_spec}
\frac{S_{ij}(E)}{\hbar}=\sum_{l=1}^{2N}a_{l}\mathcal{V}_{il}\mathcal{V}^{*}_{jl}\delta(E-E_{l}),
\ee
where $E_l$ is the $l$-th diagonal element of $\vn{\mathcal{D}}$, i.e., $E_l=\mathcal{D}_{ll}$.
Here, $1\leq i,j \leq N$, because in Eq.~\eqref{eq_statevecs}
we utilize only the first $N\times 2N$ block of the
$2N \times 2N$ matrix $\vn{\mathcal{U}}$
to construct the matrix $\vn{\mathcal{V}}$.

Note that in this paper we do not use the
grand canonical Hamiltonian $\mathcal{H}=H-\mu\hat{N}$,
where $\mu$ is the chemical potential, but instead
we use $H$, because most DFT codes do not work with $\mathcal{H}=H-\mu\hat{N}$,
as the chemical potential $\mu$ is typically redetermined
only before the end of every iteration in the selfconsistency loop
to achieve matching between electronic and nuclear charge, i.e., charge neutrality.
When comparing our result Eq.~\eqref{eq_final_spec}
to the literature, one therefore needs to be aware of this
difference by $\mu$ in the expressions for the spectral function.

\section{Choice of the moment functionals}
\label{sec_func}
In Appendix~\ref{feg_2p} we
have shown that
\bege\label{eq_poor_firstmom}
\mathcal{V}^{(2+)}_{\sigma}(\vn{r})=\frac{c^{(2+)}_{\sigma}}{[r_{s}(\vn{r})]^{2}}+\dots
\ee
and in Appendix~\ref{feg_3p} we have  found
\bege\label{eq_poor_secmom}
\mathcal{V}^{(3+)}_{\sigma}(\vn{r})=\frac{c^{(3+)}_{\sigma}}{[r_{s}(\vn{r})]^{3}}+\dots,
\ee
where
\bege\label{eq_dimlessdenparam}
r_s(\vn{r})=
\frac{1}{a_{\rm B}}
\left(
\frac{9\pi}{4 [k_{\rm F}(\vn{r})]^3}
\right)^{\frac{1}{3}}=
\left(
\frac{3}{4\pi n(\vn{r})}
\right)^{\frac{1}{3}}
\ee
is the dimensionless density parameter.
The corresponding matrix elements of the moments
are obtained from these potentials according to
\bege
M_{\sigma nm}^{(2+)}=\int d^3 r \mathcal{V}_{\sigma}^{(2+)}(\vn{r}) \phi_{n}^{*}(\vn{r})\phi_{m}(\vn{r})
\ee
and
\bege
M_{\sigma nm}^{(3+)}=\int d^3 r \mathcal{V}_{\sigma}^{(3+)}(\vn{r}) \phi_{n}^{*}(\vn{r})\phi_{m}(\vn{r}).
\ee
While it might be tempting to use these
expansions, Eq.~\eqref{eq_poor_firstmom} and Eq.~\eqref{eq_poor_secmom},
to compute the moment functionals it is instructive
to recall first the parameterization of the correlation energy
of the uniform electron gas.

\begin{figure}
\includegraphics[angle=0,width=\linewidth]{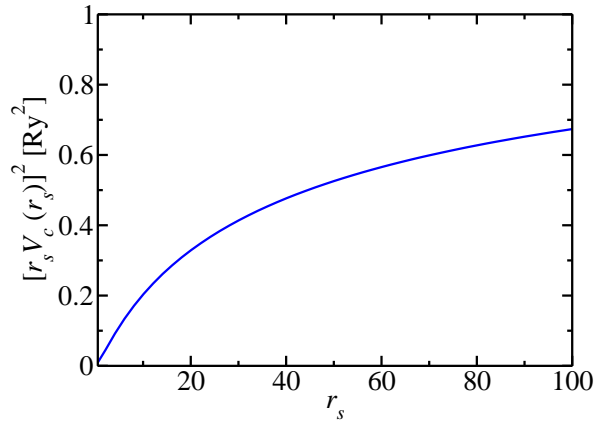}
\caption{\label{plotvwnec}
  Plot of the square of $V_{c}r_s$ vs.\ $r_s$.
  $r_s$ is the dimensionless density
  parameter defined in Eq.~\eqref{eq_dimlessdenparam}.
}
\end{figure}

In order to construct an accurate analytic representation
of the correlation energy of the uniform electron gas
one considers the high-density expansion,
the low-density expansion, and Green's-function
Monte Carlo data~\cite{PhysRevB.45.13244}.
In the low-density expansion,
the leading order is $r_{s}^{-1}$ for the exact
correlation energy.
In the high-density limit one considers instead
the parameterization $c_{0}(\zeta){\rm ln} r_s -c_1(\zeta) +c_2(\zeta)r_s {\rm ln} r_s$.
Since these two functional forms for the low and high density limits
differ considerably, we cannot expect good results,
if we construct moment functionals only based on the
parameterizations
Eq.~\eqref{eq_poor_firstmom}
and Eq.~\eqref{eq_poor_secmom}, which describe the case of low density. 
To give an impression of the deviation of the
correlation energy density $\epsilon_{c}$ from the low-density
behavior $\propto r_{s}^{-1}$, we plot in Fig.~\ref{plotvwnec}
the quantity $[V_{c}r_s]^2$, where
\bege
V_{c}=\frac{d (\epsilon_c n)}{dn}.
\ee

In order to take into account Monte Carlo simulations in the construction
of the moment functionals, we would need
such calculations for correlation functions such
as
Eq.~\eqref{eq_m2p_first} through
Eq.~\eqref{eq_m2p_seventeenth} for the uniform
electron gas.
Since these data are currently not available in the literature,
we nevertheless use the parameterization Eq.~\eqref{eq_poor_firstmom}
in our applications below.
As in this paper we present our first tests
of the MFbSDFT-method, this slightly crude approach
is justified, because the development of accurate
moment functionals will probably take similarly
long as the development of the modern functionals used
in KS-DFT calculations. Therefore, it is important to
demonstrate the feasibility of the method before developing
accurate moment functionals.

Additionally, we test the following strategy to find more elaborated
moment functionals: Eq.~\eqref{eq_poor_firstmom}
suggests that the leading order at low density
is $r_s^{-2}$. Since the leading order of the
correlation energy is $r_s^{-1}$ in this limit, we try to replace $r_s^{-2}$
in Eq.~\eqref{eq_poor_firstmom} by the square of the
correlation potential $V_{c}$, i.e.,
\bege\label{eq_firstmom_elab}
\mathcal{V}^{(2+)}_{\sigma}(\vn{r})=d^{(2+)}_{\sigma}[V_{c}(r_s)]^2
\ee
and similarly
\bege\label{eq_secmom_elab}
\mathcal{V}^{(3+)}_{\sigma}(\vn{r})=d^{(3+)}_{\sigma}[V_{c}(r_s)]^3.
\ee
This strategy should yield better
results, because the $r_s^{-2}$ and
$r_s^{-3}$ of the low-density expansion
are hereby replaced by a
more realistic functional form at high density.

In Appendix~\ref{feg_2p}   we have estimated
Eq.~\eqref{eq_m2p_first} through
Eq.~\eqref{eq_m2p_seventeenth}
based on zeroth order perturbation theory.
According to this estimate, the prefactor of
$r_s^{-2}$ is of the order of 10[Ry]$^2$.
The prefactor of $r_s^{-1}$ in the low-density expansion
of the correlation potential is of the order of 1.2[Ry].
When we use the square of the correlation energy
we have to choose the prefactor $d_{\sigma}^{(2+)}$ of the square of the correlation
energy so that $d_{\sigma}^{(2+)}(1.2 r_s^{-1})^2$ becomes comparable
to $10 r_s^{-2}$. We therefore expect $d_{\sigma}^{(2+)}$ to be of the order of 10.
At this order of magnitude of $d_{\sigma}^{(2+)}$ we indeed find a strong
satellite peak in Ni (see Sec.~\ref{sec_results}).

\section{Second variation approach}
\label{sec_secvar}
In this section we describe the implementation of our
MFbSDFT method within a second variation approach.
By second variation we mean that
first a standard KS Hamiltonian is diagonalized
at a given $k$-point and only part of its eigenvectors
are used to compute the matrix elements of the moment functionals.
The computation of the state vector matrix $\vn{\mathcal{V}}$
and of the energies $E_l=\mathcal{D}_{ll}$ may therefore be considered as a
second variation step.

The size of the KS Hamiltonian matrix
depends on the number of basis functions $N_{\rm B}$.
We do not compute all eigenvectors, but only
as many eigenvectors as we need to describe the
occupied bands and a fraction of the unoccupied bands.
We call this number $N<N_{\rm B}$.
At a given $k$ point we additionally compute the $N_{\rm B}\times N_{\rm B}$
matrices $\vn{M}^{(2+)}$ and $\vn{M}^{(3+)}$ and
project them onto the $N$ eigenstates.
By $\bar{\vn{M}}^{(2+)}$ and $\bar{\vn{M}}^{(3+)}$
we denote these projections: 
\bege
\bar{\vn{M}}^{(2+)}=\bar{\vn{U}}^{\dagger}\vn{M}^{(2+)}\bar{\vn{U}}
\ee
and
\bege
\bar{\vn{M}}^{(3+)}=\bar{\vn{U}}^{\dagger}\vn{M}^{(3+)}\bar{\vn{U}},
\ee
where $\bar{\vn{U}}$ is a $N_{\rm B}\times N$ matrix, which
holds the $N$ eigenvectors in its $N$ columns.

The implementation of the moments $\vn{M}^{(2+)}$
and $\vn{M}^{(3+)}$ is easy to do: In the subroutines computing
the standard KS-Hamiltonian one needs to switch off the kinetic
energy contribution such that only the computation of the
matrix elements of the potential remains. If one additionally replaces the
exchange-correlation potential by the moment functional potential for 
$\vn{M}^{(2+)}$
or $\vn{M}^{(3+)}$, the subroutine computes the corresponding moment matrix. 

The first and zeroth moments
in the basis of the $N$ eigenstates
are diagonal matrices:
\bege\label{eq_firstmom_secvar}
\bar{M}_{nm}^{(1)}=E^{\rm HF}_{n}\delta_{nm}
\ee
and
\bege
\bar{M}_{nm}^{(0)}=\delta_{nm}.
\ee
Note that in contrast to a standard KS-DFT calculation, the KS-Hamiltonian
used in the first variation step does not use the full exchange-correlation potential,
but only the local or non-local first-order exchange, i.e., either
Eq.~\eqref{xc_nonloc}
or Eq.~\eqref{eq_xc_loc}. Therefore, we denote the band energies from the
first variation step by $E^{\rm HF}_{n}$ in Eq.~\eqref{eq_firstmom_secvar}.
Moments and band energies depend additionally on the $k$-point
if periodic boundary conditions are used, but
we suppress again the $k$ index in the moments and also in the
band energy, i.e., instead of $E^{\rm HF}_{\vn{k}n}$ we write $E^{\rm HF}_{n}$.

The second and third moments are
given by
\bege\label{eq_barmom2}
\bar{\vn{M}}^{(2)}=\bar{\vn{M}}^{(1)}\bar{\vn{M}}^{(1)} +  \bar{\vn{M}}^{(2+)}
\ee
and
\bege\label{eq_barmom3}
\bar{\vn{M}}^{(3)}= \bar{\vn{M}}^{(1)}\bar{\vn{M}}^{(1)}\bar{\vn{M}}^{(1)}  + \bar{\vn{M}}^{(3+)}
\ee
respectively.

The size of the
matrices $\bar{\vn{M}}^{(0)}$, $\bar{\vn{M}}^{(1)}$, $\bar{\vn{M}}^{(2)}$,
and $\bar{\vn{M}}^{(3)}$ is $N\times N$ and typically $N \ll N_{\rm B}$.
Therefore, the second variation approach is fast.

Close to the end of the selfconsistency cycle the Fermi energy
is determined such that the total electronic charge compensates
the nuclear charge. Typically, the subroutine computing the
Fermi energy makes use of the eigenvalues and of weights, which
are determined by the multiplicities of the $k$ points, when symmetries
are used. In order to include the spectral weights
Eq.~\eqref{eq_spec_wei} into the calculation of the Fermi energy,
one only needs to multiply the $k$-point weights with these spectral
weights. Similarly, the spectral weights need to be considered when
computing the charge density from the matrix $\vn{\mathcal{V}}$
of state vectors, Eq.~\eqref{eq_statevecs},
according to
\bege\label{eq_chargedens_final_eq}
n(\vn{r})=\sum_{\sigma n m j}
\phi^{*}_{n}(\vn{r})
\phi_{m}(\vn{r})
a_{j\sigma}
\mathcal{V}^{*}_{nj\sigma}
\mathcal{V}_{mj\sigma}f(E_{j\sigma}),
\ee
which may be derived from Eq.~\eqref{eq_charge_dens_in_out}
by using the spectral theorem~\cite{spectral}
to express the correlator $\langle c^{\dagger}_{ \sigma n} c_{\sigma m}\rangle$
in terms of the spectral function.

\begin{figure*}
\centering
\begin{adjustbox}{width=\textwidth}
\begin{tikzpicture}[node distance=2cm]
  \node (start) [mybox] {Input Charge Density $n_{\rm inp}(\vn{r})$};
  \node (hartree) [mybox3, right of=start, xshift=6cm] {Hartree and Exchange potentials $V^{\rm H}(\vn{r})$, $V^{\rm X}(\vn{r})$ \\ (Eq.~\eqref{eq_hartree_potential}, Eq.~\eqref{xc_nonloc}, Eq.~\eqref{eq_xc_loc})};
   \node (hamil) [mybox, right of=hartree,xshift=7cm] {Diagonalize $H=\left[-\frac{\hbar^2}{2m}\Delta+V(\vn{r})+V^{\rm H}(\vn{r})+V^{\rm X}(\vn{r})\right]$};
  \node (mompot) [mybox2, below of=hartree] {Moment potentials $\mathcal{V}^{(2+)}(\vn{r})$, $\mathcal{V}^{(3+)}(\vn{r})$ \\ Eq.~\eqref{eq_poor_firstmom}, Eq.~\eqref{eq_poor_secmom}, Eq.~\eqref{eq_firstmom_elab}, Eq.~\eqref{eq_secmom_elab}};
  \draw [arrow] (start) -- (hartree);
  \draw [arrow] ([xshift=4cm]start) |- (mompot);
  \draw [arrow] (hartree) -- (hamil);
\node (moments) [mybox2,below of=hamil] {Compute moments $\bar{\vn{M}}^{(1)}$, $\bar{\vn{M}}^{(2)}$, $\bar{\vn{M}}^{(3)}$\\ Eq.~\eqref{eq_firstmom_secvar}, Eq.~\eqref{eq_barmom2}, Eq.~\eqref{eq_barmom3}};
\draw [arrow] (mompot) -- (moments);
  \draw [arrow] (hamil) -- (moments);
  \node (spectral) [mybox3,below of=moments] {
    Compute spectral poles $E_{l}=\mathcal{D}_{ll}$ (Eq.~\eqref{eq_b1_ududag})
    Compute state vectors $\vn{\mathcal{V}}$ (Eq.~\eqref{eq_statevecs})\\
    Compute spectral weights $a_{j}$ (Eq.~\eqref{eq_spec_wei})\\
  };
\draw [arrow] (moments) -- (spectral);
 \node (outputdens) [mybox,below of=mompot] {Compute output charge density $n_{\rm out}(\vn{r})$ (Eq.~\eqref{eq_chargedens_final_eq})};
 \draw [arrow] (spectral) -- (outputdens);
 \node (mixing) [mybox2,left of=outputdens,xshift=-6cm] {Mix charges to obtain new charge density \\for next iteration};
 \draw [arrow] (outputdens) -- (mixing);
 \draw [arrow] (mixing) -- (start);
  \draw [dashedarrow] ([xshift=-20cm]start) -- ([xshift=-1cm]mixing);
\end{tikzpicture}  
\end{adjustbox}
\caption{\label{mfbsdft_flowchart}
Flowchart of the MFbSDFT selfconsistency cycle.
}
\end{figure*}
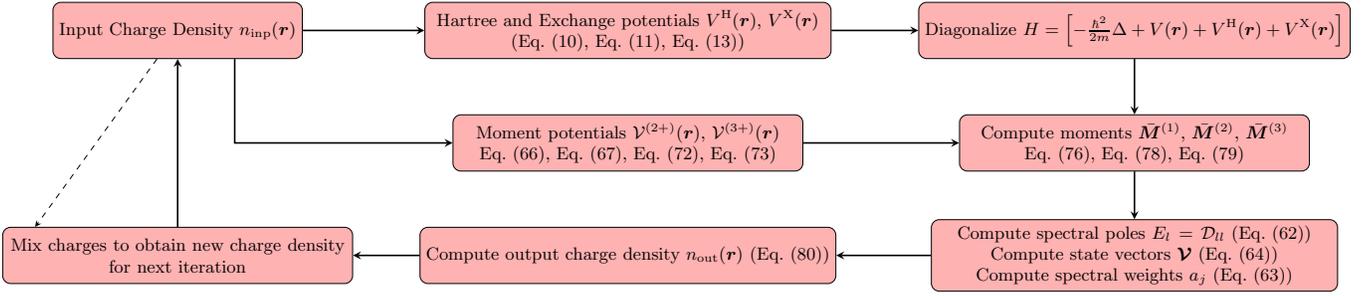

We illustrate the selfconsistency loop by the flowchart in Fig.~\ref{mfbsdft_flowchart}.
All results presented in Sec.~\ref{sec_results}
have been obtained according to the flowchart in Fig.~\ref{mfbsdft_flowchart}.

KS-DFT is so constructed that it may be used
to obtain the total energy and the charge density in principle exactly.
However, in practice the exact exchange correlation potential is not
known and therefore the charge density computed in KS-DFT is an approximation.
Since considerable progress has been made in the construction of
exchange correlation potentials, the KS charge density is a very good
approximation in many cases.
Whenever the KS charge density is sufficiently correct,
one may run MFbSDFT in a simplified mode: The converged
KS charge density is used as starting density in Fig.~\ref{mfbsdft_flowchart}
and only one iteration is performed, i.e., the output charge density is not
computed but instead the results are calculated
immediately from the state vector matrix $\vn{\mathcal{V}}$,
from the spectral poles, and from the spectral weights.

\section{Applications}
\label{sec_results}
In this section we apply the MFbSDFT method to several well-studied
materials that show genuine many-body effects such as satellite peaks.
According to the literature, the details of the spectral function of these
materials
depend strongly on the theoretical model used to study them.
According to our discussion in Sec.~\ref{sec_func}
the parameterizations that we use for the moment functionals should be
considered only as a first step towards the development of accurate
moment functionals. Consequently, if the results shown below are more
similar to one theoretical model than they are to another one this does not
imply at all that MFbSDFT confirms one particular theoretical model,
because accurate moment functionals remain to be developed.
The main purpose of this section is therefore to show that MFbSDFT is able
to reproduce spectral features qualitatively that have been identified as
genuine correlation effects before.

However, beyond validating the concept of MFbSDFT,
the results shown also hint at a practical perspective for MFbSDFT
already at this early stage of its development: Since the MFbSDFT reproduces
spectral features of correlated materials, it may be used to
compute response properties~\cite{spectral} such as the anomalous Hall effect,
which is likely to require considerably less computer time
than LDA+DMFT. For such an application one would finetune the parameters in the
parameterizations of the moment functionals to match the spectral function
known from LDA+DMFT or photoemission. While this approach is not parameter-free,
it is similar to
many applications of LDA+$U$, where the $U$ parameter is chosen to reproduce
a material property.
\subsection{fcc Ni}
\begin{figure}
\includegraphics[angle=0,width=\linewidth]{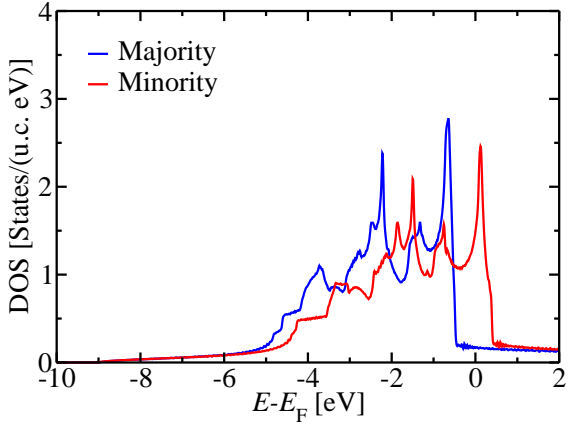}
\caption{\label{DOS_Nickel_ferromag_pbe}
  DOS of Ni vs.\ energy $E$ in the ferromagnetic state as obtained in KS-DFT.
  $E_{\rm F}$ is the Fermi energy.
}
\end{figure}

The DOS obtained in KS-DFT with the PBE functional is shown
in Fig.~\ref{DOS_Nickel_ferromag_pbe}.
The valence DOS starts to become significant
starting from $5$~eV below the Fermi energy and the exchange splitting
is around 0.75~eV. In contrast, the width of the main bands found experimentally
is significantly smaller than 5~eV, namely only 3~eV.
Additionally, a much smaller exchange splitting of around 0.3~eV is
found in photoemission experiments~\cite{PhysRevB.21.3245,PhysRevLett.40.1514}.
Moreover, the satellite peak observed in experiments at around 6~eV below
the Fermi energy is absent in the KS-DFT spectrum.

\begin{figure}
\includegraphics[angle=0,width=\linewidth]{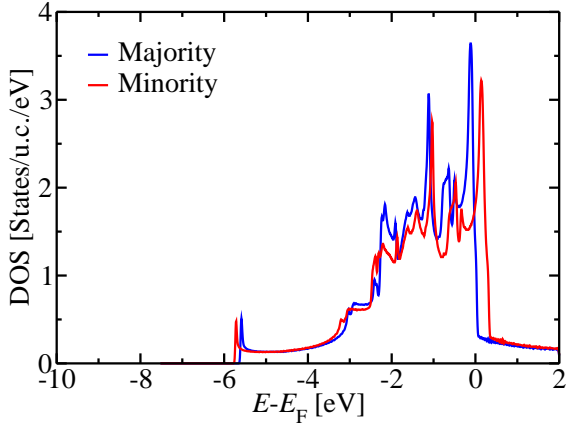}
\caption{\label{DOS_Nickel_ferromag}
  DOS of Ni vs.\ energy $E$ in the ferromagnetic state as obtained in MFbSDFT
  when the moment functional is constructed according to Eq.~\eqref{eq_firstmom_elab}.
  $E_{\rm F}$ is the Fermi energy.
}
\end{figure}

Next, we discuss the MFbSDFT-spectrum obtained with
$\mathcal{V}^{(2+)}_{\sigma}(\vn{r})=15\zeta_{\sigma}^{(5/3)}[V_{c}(r_s)]^2$
and $\mathcal{V}^{(3+)}_{\sigma}(\vn{r})=0$.
Here $\zeta_{\sigma}=(1-\sigma (n_{\uparrow}-n_{\downarrow})/n)$.
We use $N=36$.
With this choice of parameters the magnetic moment computed
self-consistently in MFbSDFT
is 0.58~$\mu_{\rm B}$.
The resulting DOS is presented in Fig.~\ref{DOS_Nickel_ferromag}.
The exchange splitting of around 0.3~eV is strongly reduced
compared to the KS-DFT calculation and close to
the experiments~\cite{PhysRevB.21.3245,PhysRevLett.40.1514}. Additionally,
the main bands
are much narrower than in KS-DFT and therefore in much better agreement
with experiments.
Moreover, satellite peaks are found close to 6~eV. However, the spectral
weight of these satellite peaks is smaller than what is found in
experiments and in LDA+DMFT
calculations (see e.g.\
Fig.~9 in Ref.~\cite{PhysRevB.51.11002},
Fig.~2 in Ref.~\cite{PhysRevB.85.235136},
and Fig.~2 in Ref.~\cite{PhysRevLett.87.067205}).

\begin{figure}
\includegraphics[angle=0,width=\linewidth]{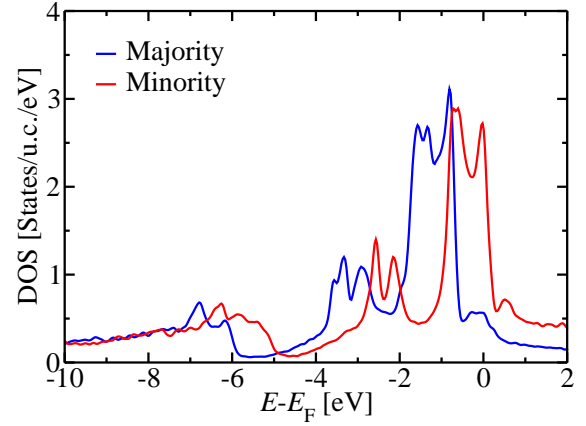}
\caption{\label{DOS_Nickel_ferromag_rs2}
  DOS of Ni vs.\ energy $E$ in the ferromagnetic state as obtained in MFbSDFT
  when the moment functional is constructed according to Eq.~\eqref{eq_poor_firstmom}
  and Eq.~\eqref{eq_poor_secmom}.
  $E_{\rm F}$ is the Fermi energy.
}
\end{figure}

Finally, we discuss the MFbSDFT-spectrum obtained with
$\mathcal{V}^{(2+)}_{\sigma}(\vn{r})=0.015 \zeta_{\sigma}^{(7/3)}r_s^{-2}$~[Ry]$^2$
and
$\mathcal{V}^{(3+)}_{\sigma}(\vn{r})=-0.00472 \zeta_{-\sigma}^{(1/3)}r_s^{-3}$~[Ry]$^3$.
We use $N=36$.
With this choice of parameters the magnetic moment
is 0.63~$\mu_{\rm B}$.
Fig.~\ref{DOS_Nickel_ferromag_rs2}
shows the density of states (DOS) of Ni in the
ferromagnetic state as computed selfconsistently in MFbSDFT.
While the exchange splitting is similar to KS-DFT, the
width of the main bands is reduced, leading to a slightly
better agreement with experiment. Around 6~eV below the Fermi energy
satellite peaks appear with a spectral weight of a similar order of
magnitude like in experiment and LDA+DMFT. However, the spin polarization
of the satellite peak structure differs from both experiment and
LDA+DMFT, which both predict the minority satellite to be strongly suppressed
(see e.g.\
Fig.~9 in Ref.~\cite{PhysRevB.51.11002},
Fig.~2 in Ref.~\cite{PhysRevB.85.235136},
and Fig.~2 in Ref.~\cite{PhysRevLett.87.067205}).
In contrast, in Fig.~\ref{DOS_Nickel_ferromag_rs2} the satellites of
the majority and minority band are comparable in magnitude and only shifted
in energy.

As discussed in Sec.~\ref{sec_func}
the parameters employed
in $\mathcal{V}^{(2+)}_{\sigma}(\vn{r})=15\zeta_{\sigma}^{(5/3)}[V_{c}(r_s)]^2$ (used
to generate the DOS shown in Fig.~\ref{DOS_Nickel_ferromag})
are of the order of magnitude expected from the estimate
given in Sec.~\ref{sec_func}.
In contrast, we determined the parameters
employed
in $\mathcal{V}^{(2+)}_{\sigma}(\vn{r})=0.015 \zeta_{\sigma}^{(7/3)}r_s^{-2}$~[Ry]$^2$
(used
to generate the DOS shown in Fig.~\ref{DOS_Nickel_ferromag_rs2})
only based on try-out, because it is unclear how to renormalize the parameters
of the low-density expansion so that it effectively describes the regimes of
intermediate and high densities as well.
However, Fig.~\ref{DOS_Nickel_ferromag_rs2} is useful nevertheless, because it shows that
satellite peaks with the correct order of magnitude of spectral weight can be produced
by MFbSDFT. Taken together with the result of Fig.~\ref{DOS_Nickel_ferromag}, which shows
that band widths, exchange splittings and location of the satellite peaks are predicted very
well if the correlation potential is used to construct the moment functionals, the overall
conclusion from this finding is that it is likely that assessing
Eq.~\eqref{eq_m2p_first}
through Eq.~\eqref{eq_m2p_seventeenth}
in the low-density and the high-density regimes and using Monte Carlo results to interpolate
between these limits will allow us to formulate a moment functional that predicts the spectral
features in Ni quite well.

\subsection{SrVO$_3$}

\begin{figure}
\includegraphics[angle=0,width=\linewidth]{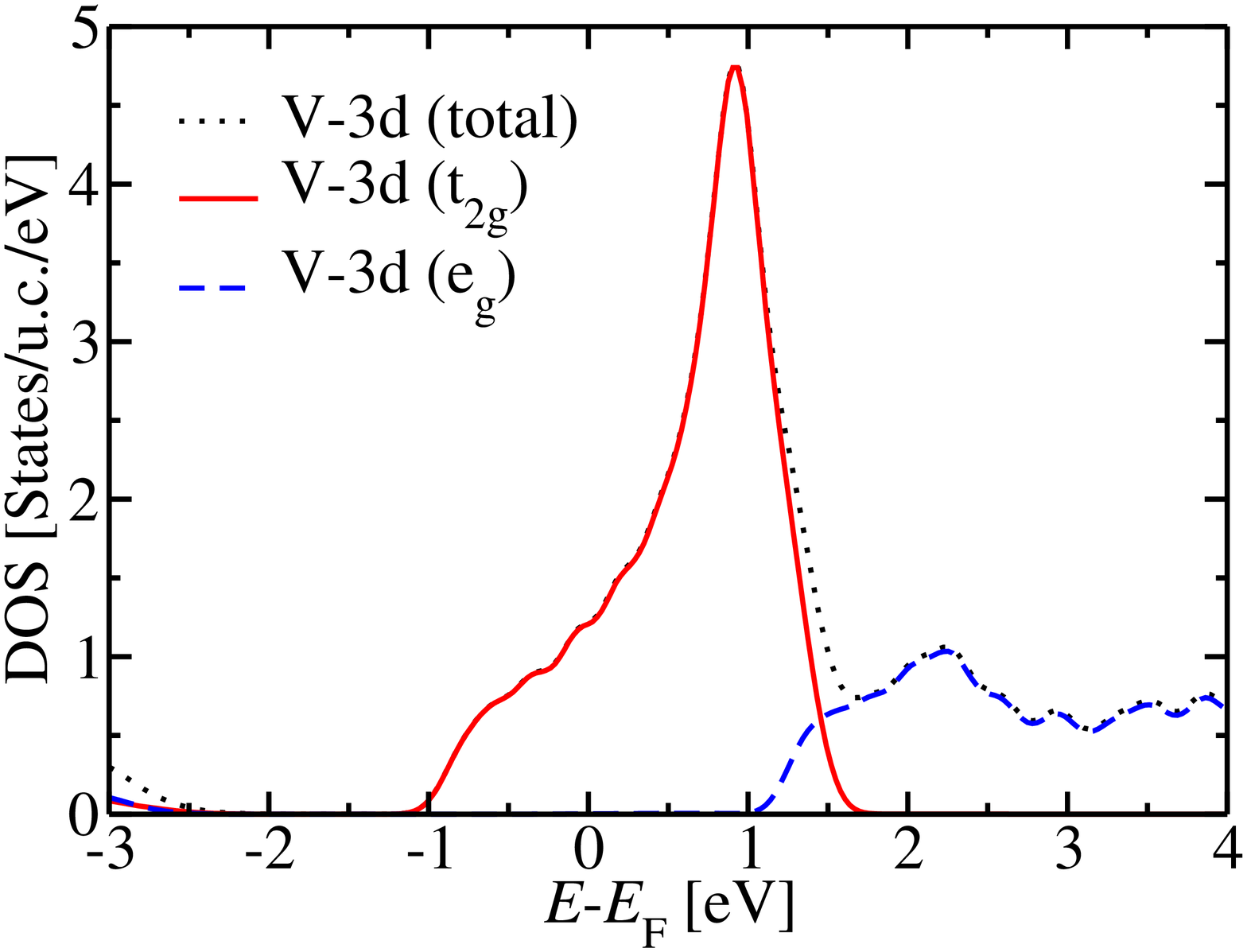}
\caption{\label{SrVO3_t2g_eg_pbe}
  Contributions of the V-3$d$ $e_g$ and $t_{2g}$ states to the DOS in SrVO$_3$.
  Results from KS-DFT using the PBE functional.
}
\end{figure}

In Fig.~\ref{SrVO3_t2g_eg_pbe}
we show the
contributions of the V-3$d$ $e_g$ and $t_{2g}$ states to the DOS
of SrVO$_3$,
as obtained with KS-DFT using the PBE functional.
The KS spectrum is not in good agreement with experiment.
It has been shown that
the agreement with
experiment is improved significantly~\cite{PhysRevB.72.155106},
when DMFT is used to supplement
these bands with correlation effects.
The detailed rearrangement of
the spectral features obtained from LDA+DMFT depends on the details
of the modelling of the correlation effects by the Hubbard model.
Ref.~\cite{PhysRevB.72.155106}
includes only the $t_{2g}$ states into the Hubbard model. In this
case the DOS of the $t_{2g}$ states obtained from LDA+DMFT is distributed into
three pronounced spectral peaks: A dominant central peak roughly 0.5~eV
above the Fermi energy, a lower Hubbard band around 2~eV below the Fermi energy and an
additional upper Hubbard band around 3~eV above the Fermi energy.
The position of these peaks is in good agreement with experiments,
which find peaks roughly
at -1.7~eV, 0.5~eV, and 2.4~eV~\cite{INOUE19941007,PhysRevLett.93.156402}.
These spectral features are also observed in Ref.~\cite{PhysRevB.77.205112},
which includes the $e_{g}$ states,
however they strongly depend on the parameters, and the intensities
of the lower and upper Hubbard bands are much smaller
for some parameters. Additionally, the intensities of the lower and upper
Hubbard bands depend strongly on the double counting correction.

\begin{figure}
\includegraphics[angle=0,width=\linewidth]{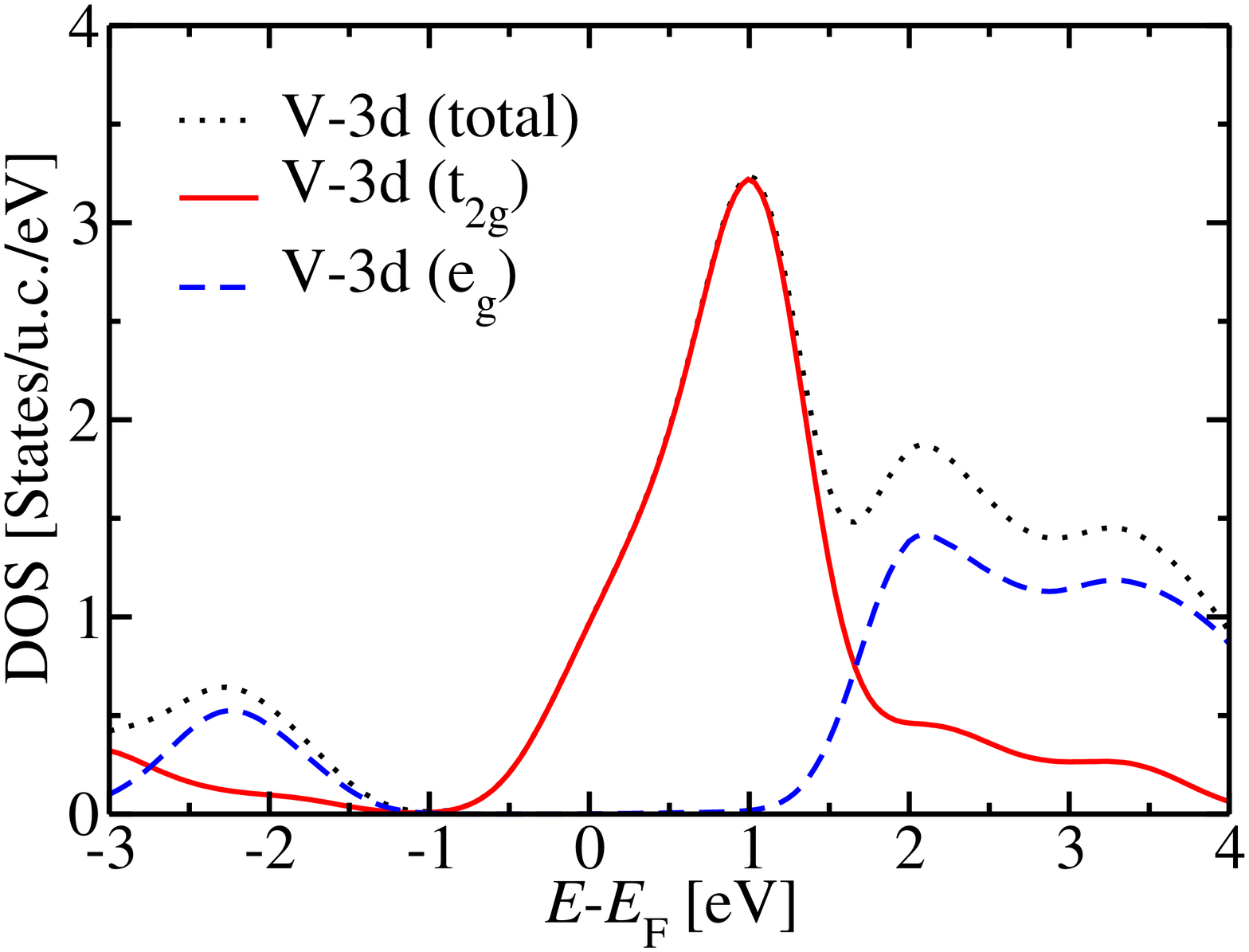}
\caption{\label{SrVO3_mfbsdft_vc}
  Contributions of the V-3$d$ $e_g$ and $t_{2g}$ states to the DOS in SrVO$_3$.
  Results obtained within MFbSDFT when the moment functionals are constructed
  according to Eq.~\eqref{eq_firstmom_elab} and Eq.~\eqref{eq_secmom_elab}.
}
\end{figure}

In Fig.~\ref{SrVO3_mfbsdft_vc}
we present the
contributions of the V-3$d$ $e_g$ and $t_{2g}$ states to the DOS,
as obtained with MFbSDFT when we use
Eq.~\eqref{eq_firstmom_elab} and Eq.~\eqref{eq_secmom_elab}, where
we
set $d^{(2+)}_{\sigma}=100$
and
$d^{(3+)}_{\sigma}=-200$.
We use $N=200$.
The total V-$d$ DOS is in good agreement with both the experimental spectrum and
the LDA+DMFT spectrum (See e.g.\ Fig.~7 in Ref.~\cite{PhysRevB.72.155106} for comparison.
Ref.~\cite{PhysRevB.72.155106} uses a broadening of 0.36~eV in order to reproduce the
experimental resolution. We use 0.36~eV in our Fig.~\ref{SrVO3_mfbsdft_vc} as well.).
However, in our case the peak between 2~eV and 2.5~eV stems mainly from the $e_{g}$ states,
which are not included into the Hubbard model in  Ref.~\cite{PhysRevB.72.155106}.
Ref.~\cite{PhysRevB.77.205112} includes the $e_{g}$ states, but still
finds a small peak from the upper Hubbard band for the $t_{2g}$ states at around 3~eV
above the Fermi energy.
Such a small peak is consistent with our Fig.~\ref{SrVO3_mfbsdft_vc},
where a shoulder in the V-3d(t$_{2g}$) is clearly visible between 2~eV and 2.5~eV.
Moreover, Ref.~\cite{PhysRevB.77.205112} finds a large contribution from the $e_g$
states to the DOS at this energy as well.
In this regard, our Fig.~\ref{SrVO3_mfbsdft_vc} resembles closely
Fig.~8 in Ref.~\cite{PhysRevB.77.205112} when the energy is above the Fermi energy.
However, Fig.~8 in Ref.~\cite{PhysRevB.77.205112} does not 
find a strong V-$d$ DOS at around 2~eV below the Fermi energy. In contrast, we
find a strong V-$d$ DOS at around 2~eV below the Fermi energy with a dominant
part from the $e_{g}$ states and a small contribution from the $t_{2g}$ states.

\begin{figure}
\includegraphics[angle=0,width=\linewidth]{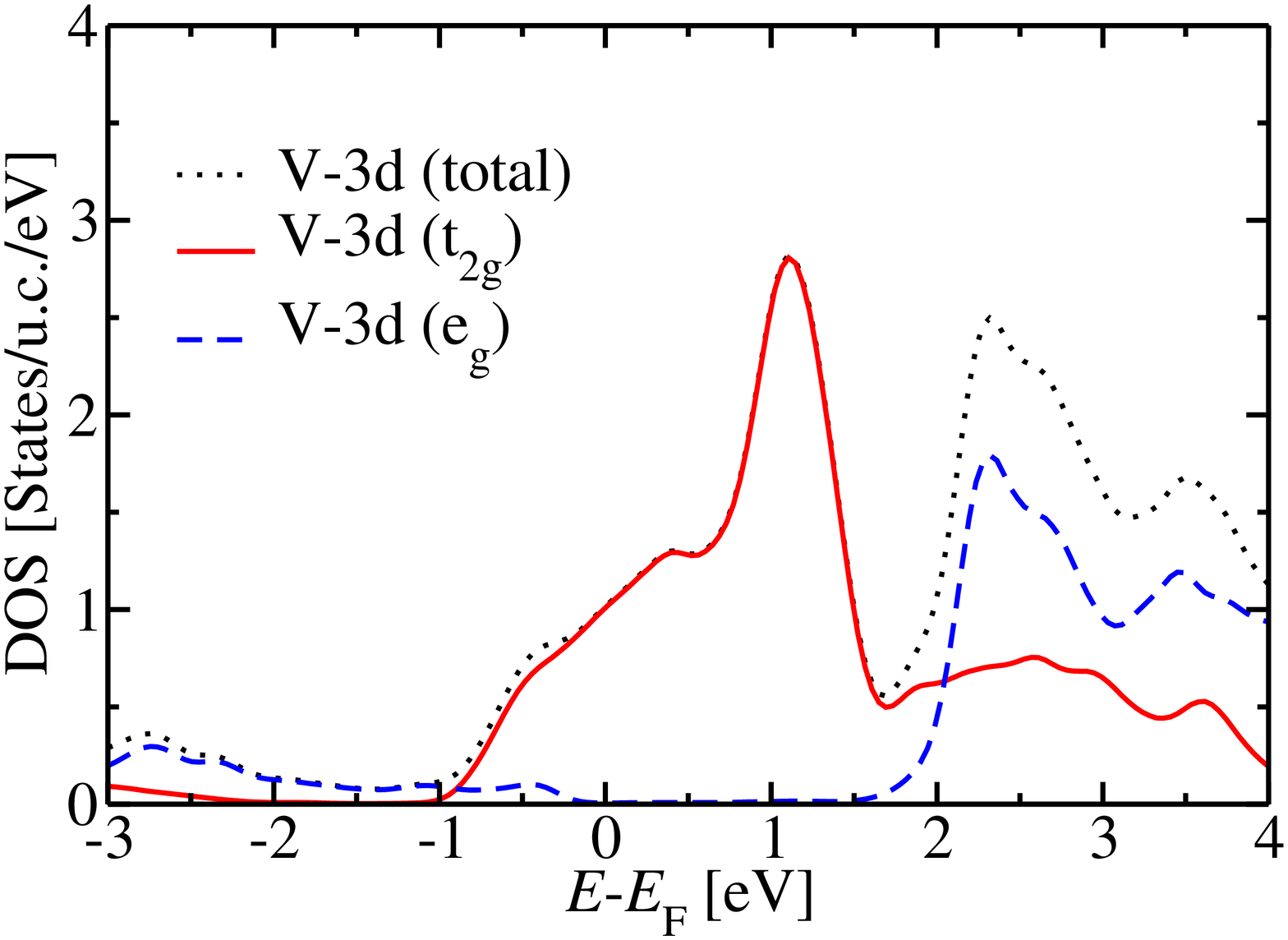}
\caption{\label{SrVO3_mfbsdft_r2}
  Contributions of the V-3$d$ $e_g$ and $t_{2g}$ states to the DOS in SrVO$_3$.
  Results obtained within MFbSDFT when the moment functionals are constructed
  according to Eq.~\eqref{eq_poor_firstmom} and Eq.~\eqref{eq_poor_secmom}.
}
\end{figure}

In Fig.~\ref{SrVO3_mfbsdft_r2}
we show the
contributions of the V-3$d$ $e_g$ and $t_{2g}$ states to the DOS,
as obtained with MFbSDFT when we
take $N=200$ and
use
Eq.~\eqref{eq_poor_firstmom} and Eq.~\eqref{eq_poor_secmom}, where
we
set $c^{(2+)}_{\sigma}=1.1$~[Ry]$^2$,
and $c^{(3+)}_{\sigma}=-1.5$~[Ry]$^3$.
The peak between 2~eV and 2.5~eV is very pronounced and
both $e_{g}$ and $t_{2g}$ bands contribute to it.
In contrast, the peak around -2~eV in Fig.~\ref{SrVO3_mfbsdft_vc}
is significantly smaller in Fig.~\ref{SrVO3_mfbsdft_r2} and
shifted to lower energy between -2.5~eV and -3~eV.
Several main features of the  $t_{2g}$ band in  Fig.~\ref{SrVO3_mfbsdft_r2}
resemble those obtained
from a LDA+DMFT calculation with a
Hubbard $U$ of 6~eV (see Fig.~8 in Ref.~\cite{PhysRevB.77.205112}).
Notably, the $t_{2g}$ band, which ends around 2~eV in Fig.~\ref{SrVO3_t2g_eg_pbe},
is expanded to higher energies like in LDA+DMFT.
Overall, the total V-$d$ DOS in Fig.~\ref{SrVO3_mfbsdft_r2}
is qualitatively similar to the one in Fig.~8 of Ref.~\cite{PhysRevB.77.205112},
which displays pronounced peaks close to 1~eV and close to 2.5~eV, while the
V-$d$ DOS close to -2~eV is small, in agreement with our result in Fig.~\ref{SrVO3_mfbsdft_r2}.
However, the peak close to 2.5~eV is much larger in our  Fig.~\ref{SrVO3_mfbsdft_r2}.

\begin{figure}
\includegraphics[angle=0,width=\linewidth]{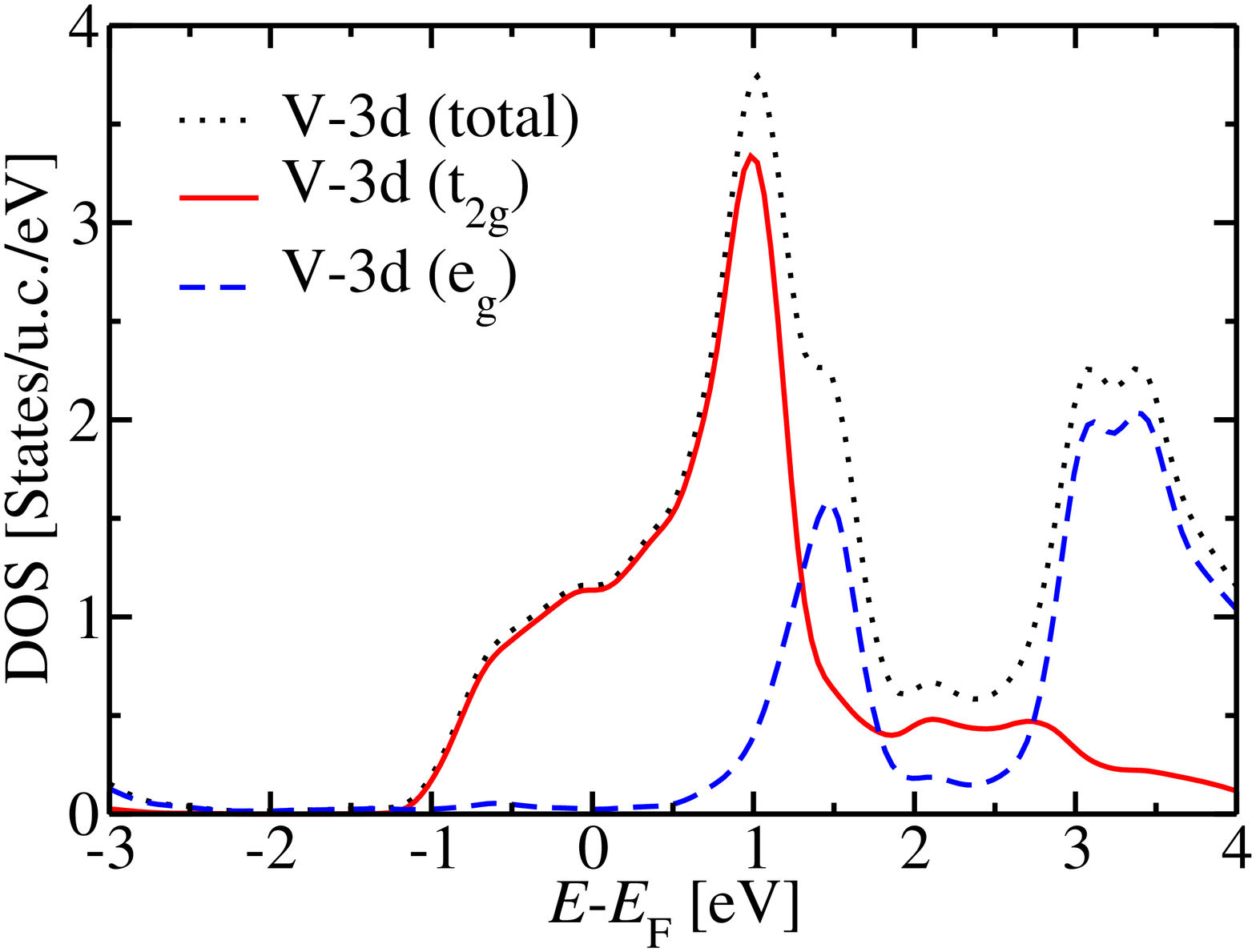}
\caption{\label{SrVO3_mfbsdft_r2_smallparam}
  Contributions of the V-3$d$ $e_g$ and $t_{2g}$ states to the DOS in SrVO$_3$.
  Results obtained within MFbSDFT when the moment functionals are constructed
  according to Eq.~\eqref{eq_poor_firstmom} and Eq.~\eqref{eq_poor_secmom}.
  In contrast to Fig.~\ref{SrVO3_mfbsdft_r2} the parameters $c^{(2+)}_{\sigma}$
  and $c^{(3+)}_{\sigma}$ are reduced by 18~\%
  and 33~\%, respectively.
}
\end{figure}

We may reduce the intensity of this peak at 2.5~eV
by reducing the parameters $c^{(2+)}_{\sigma}$ and $c^{(3+)}_{\sigma}$.
In Fig.~\ref{SrVO3_mfbsdft_r2_smallparam} we show the DOS
obtained with the parameters $c^{(2+)}_{\sigma}=0.9$[Ry]$^2$ 
and $c^{(3+)}_{\sigma}=-1$[Ry]$^3$. 
Indeed, the peak intensity is reduced and it is now significantly smaller than
the intensity of the main peak at around 1~eV, but relative to the main
peak it is still more pronounced than in Fig.~8 of Ref.~\cite{PhysRevB.77.205112}.
Moreover, the peak is shifted from 2.5~eV in our Fig.~\ref{SrVO3_mfbsdft_r2}
to higher energies and lies now between 3~eV and 3.5~eV. However, overall
the total V-$d$ DOS, the V-3$d$ (t$_{2g}$) DOS, and the V-3$d$ (e$_{g}$) DOS
in Fig.~8 of Ref.~\cite{PhysRevB.77.205112}
is in better agreement with our 
Fig.~\ref{SrVO3_mfbsdft_r2_smallparam}
than with our  Fig.~\ref{SrVO3_mfbsdft_r2}.

\section{Discussion and outlook}
\label{sec_dis_out}

In the previous section we have shown that MFbSDFT reproduces features
such as satellite peaks and spectral weight shifts which are usually
obtained by solving the correlated electron problem directly, e.g.\ by
means of LDA+DMFT. While these first MFbSDFT results look therefore very
promising there is a large number of open questions and obvious
possibilities to improve this method further.

First, accurate moment functionals are required. The construction of accurate
moment functionals should be possible based on Monte Carlo data of
correlation functions such as those given in Eq.~\eqref{eq_m2p_first}
through Eq.~\eqref{eq_m2p_seventeenth}.

Second, it is desirable to derive gradient approximations for the moment functionals.
Using local functionals that depend only on the spin densities
for MFbSDFT misses effects related to their spatial inhomogeneity.
Like GGA is an improvement over LDA, we expect that MFbSDFT will become more accurate
by adding gradient corrections to the functionals.

Third, in this work we do not explore the calculation of total energies and atomic
forces for structural relaxation.
However, since force calculations in correlated materials are possible
within LDA+DMFT
we expect that total energies and forces may also be obtained within our
MFbSDFT approach.

Fourth, one may use more than the first 4 moments.
While the first four moments are sufficient
to reproduce the
quasi-particle band structure qualitatively correctly in the strong-correlation
regime~\cite{moment_sum_rule}, the precision of MFbSDFT is expected to
increase with the number of moments used.
The moments become increasingly more complicated with increasing order.
However, one may use computer algebra systems in order to
derive the higher-order moments and to assess them for the uniform
electron gas.
This seems feasible since the complexity is probably comparable to
higher-order perturbation theory in QED, where high-order
contributions have to be tackled by computer algebra.
Assuming that computer algebra systems can manage the complexity
of the higher-order moments the question remains if the spectral function
can be found for more than 2 or 4 moments.
In Ref.~\cite{spectral} we give an argument that the spectral function
may be found from the first 4 moments, which is based on counting the number
of available equations and the number of parameters that determine the
spectral function and showing that these numbers match.
In the present paper we have explicitly constructed the spectral
function from the first 4 moments in Sec.~\ref{sec_constru_specfun}.
We may generalize the argument
given in Ref.~\cite{spectral} and show that from the first $2P$ moments ($P=1,2,\dots$)
one may construct the spectral function.
This generalization is discussed in App.~\ref{sec_app_more}.
If one uses only Delta-functions in the expression for the spectral function
(as in Eq.~\eqref{eq_final_spec})
one misses lifetime effects, which may be accommodated by employing
Gaussians instead~\cite{Nolting_1980}.
When 4 moments are required to put the spectral peaks, such as satellite peaks,
at the right energies, it is clear that more than 4 moments are generally required
to correct the spectral widths of these spectral features by lifetime effects.

Fifth, perhaps the method of spectral moments may contribute to the understanding of
SIE, because it is remarkable that setting $\mu_{c}(\vn{r})=0$ in Eq.~(2.22) of
Ref.~\cite{PhysRev.140.A1133} leads to a HF-type method that suffers from the same SIE
and is equivalent to
the method of spectral moments derived from the first two moments (see Sec.~\ref{sec_theory}).
We suspect that increasing the
number of moments used will ultimately eliminate the SIE.
However, it is an open question, how the self-interaction correction (SIC) takes
place exactly within the method of spectral moments.
Within the KS-DFT framework the explanation of SIC is that $\mu_{c}(\vn{r})$ in Eq.~(2.22)
of Ref.~\cite{PhysRev.140.A1133} has to eliminate SIE when the exact exchange correlation
functional is used.
However, within the spectral moment method a valid explanation of SIE seems to be that
using only the first two moments produces an error, which may be eliminated by using more moments.
Of course, the precise moment functionals are expected to be necessary in order to remove
SIE. However, Eq.~\eqref{eq_m2p_first} through
Eq.~\eqref{eq_m2p_seventeenth} provide explicit expressions, which may be used for the
construction of the moment functionals.

Sixth, it is an important open question how to extend the MFbSDFT approach to finite temperatures.
In Ref.~\cite{spectral} we have shown how to generalize the spectral moment method so that
it can be applied to many-band Hamiltonians.
Since the method of Ref.~\cite{spectral}
computes the correlation functions from the spectral theorem, which involves the Fermi function
and the actual excitation energies, it naturally includes finite temperature effects.
As the spectral theorem is not used for the higher-order correlation functions in MFbSDFT, which are obtained from moment functionals,
it is currently unknown how to accomodate
finite temperatures accurately in this method.

While accurate moment functionals are currently not available yet, the MFbSDFT method may
also be used in practice in a way similar to LDA+$U$: In LDA+$U$
the $U$ and $J$ parameters are usually
chosen for a given material in order to add correlation effects that are not described by LDA.
Similarly, one may use parameterizations of the moment functionals similar to the ones that
we discussed in Sec.~\ref{sec_func} and choose the coefficients in the functional in order to optimize spectral
features.

\section{Summary}
\label{sec_summary}
We describe the concept of moment functionals, which allow us to obtain
the spectral moments from functionals of the charge density. These functionals
play a similar role in MFbSDFT as the exchange correlation functional does in KS-DFT.
We derive explicit expressions for the moment functionals and use perturbation theory
to investigate their scaling with the charge density.
We describe an efficient algorithm to obtain the spectral function from the first
four spectral moments.
We demonstrate that MFbSDFT allows us to reproduce spectral features such as
satellite peaks in Ni and lower and upper Hubbard bands in SrVO$_{3}$.
At this stage of its development, MFbSDFT may be used in a way similar to LDA+$U$:
The parameters in the moment functionals are chosen such that spectral features
found in experiments
are reproduced.

\section*{Acknowledgments}We acknowledge funding
by SPP 2137 ``Skyrmionics" of
the DFG
and by Sino-German research project DISTOMAT (DFG project MO \mbox{1731/10-1}).
We gratefully acknowledge financial support from the European Research
Council (ERC) under
the European Union's Horizon 2020 research and innovation
program (Grant No. 856538, project ``3D MAGiC'').
The work was also supported by the Deutsche Forschungsgemeinschaft
(DFG, German Research Foundation) $-$ TRR 288 $-$ 422213477 (project B06).  We  also gratefully
acknowledge the J\"ulich
Supercomputing Centre and RWTH Aachen University for providing
computational
resources under project No. jiff40.

\appendix
\section{Comparison of the spectral moments of the many-band case to the spectral moments of the single-band Hubbard model}
\label{sec_app_compare}
The spectral moments of the many-band case contain many new
contributions that do not have counterparts
in the single-band Hubbard model.
In this appendix we discuss which of the many-band terms
have a correspondence in the single-band Hubbard model.
In the single band Hubbard model the Coulomb matrix element
is simply
\bege
V_{nn'tt'}=U\delta_{nn'}\delta_{tt'}\delta_{nt},
\ee
where $U$ is the Hubbard-$U$.
For the single band Hubbard model the first four moments are:
\bege\label{eq_sibahu_zero}
\tilde{M}^{(0)}_{\vn{k}\sigma}=\frac{1}{\mathcal{N}}
\sum_{lj}e^{i\vn{k}\cdot
  (\vn{R}_{l}-\vn{R}_{j})}
\langle[c_{\sigma l},c^{\dagger}_{\sigma j}]_{+} \rangle=1,
\ee

\bege\label{eq_sibahu_one}
\begin{aligned}
&\tilde{M}^{(1)}_{\vn{k}\sigma}=\frac{1}{\mathcal{N}}
\sum_{lj}e^{i\vn{k}\cdot
  (\vn{R}_{l}-\vn{R}_{j})}
\langle[[c_{\sigma l},H]_{-},c^{\dagger}_{\sigma j}]_{+} \rangle\\
&=\epsilon(\vn{k})+U\langle n_{-\sigma} \rangle,
\end{aligned}
\ee

\bege\label{eq_sibahu_two}
\begin{aligned}
&\tilde{M}^{(2)}_{\vn{k}\sigma}=\frac{1}{\mathcal{N}}
\sum_{lj}e^{i\vn{k}\cdot
  (\vn{R}_{l}-\vn{R}_{j})}
\langle[
[c_{\sigma l},H]_{-},
[H,c^{\dagger}_{\sigma j}]_{-}
]_{+} \rangle\\
&=(\epsilon(\vn{k}))^2
+2 U\langle n_{-\sigma} \rangle \epsilon(\vn{k})
+ U^2\langle n_{-\sigma}\rangle,\\
\end{aligned}
\ee
and
\bege\label{eq_singleband_moment3}
\begin{aligned}
&\tilde{M}^{(3)}_{\vn{k}\sigma}=\frac{1}{\mathcal{N}}
\sum_{lj}e^{i\vn{k}\cdot
  (\vn{R}_{l}-\vn{R}_{j})}
\langle[
[[c_{ls},H]_{-},H]_{-},
[H,c^{\dagger}_{js}]_{-}
]_{+} \rangle\\
&=[\epsilon(\vn{k})]^3
+3U\langle n_{-\sigma} \rangle [\epsilon(\vn{k})]^2\\
&+2 U^2 \epsilon(\vn{k}) n_{-\sigma}
+2 U^2 t_{00}n_{-\sigma}
+U^3 n_{-\sigma}
\\
&-U^2\frac{1}{\mathcal{N}}
\sum_{lj}e^{i\vn{k}\cdot
  (\vn{R}_{l}-\vn{R}_{j})}t_{lj}
\langle
c^{\dagger}_{-\sigma l}c^{\dagger}_{-\sigma j}c_{-\sigma l}c_{-\sigma j}
\rangle
\\
&+U^2\frac{1}{\mathcal{N}}
\sum_{lj}
t_{lj}
\langle
(2n_{\sigma l}-1)c^{\dagger}_{-\sigma l}c_{-\sigma j}
\rangle
\\
&+U^2\frac{1}{\mathcal{N}}
\sum_{lj}e^{i\vn{k}\cdot
  (\vn{R}_{l}-\vn{R}_{j})}t_{lj}
\langle
c^{\dagger}_{\sigma j}c^{\dagger}_{-\sigma l}c_{\sigma l}c_{-\sigma j}
\rangle\\
&+U^2\frac{1}{\mathcal{N}}
\sum_{lj}e^{i\vn{k}\cdot
  (\vn{R}_{l}-\vn{R}_{j})}t_{lj}
\langle
c^{\dagger}_{\sigma j}c^{\dagger}_{-\sigma j}c_{\sigma l}c_{-\sigma l}
\rangle.\\
\end{aligned}
\ee
Here, $\mathcal{N}$ is the number of $\vn{k}$ points.

Clearly, Eq.~\eqref{eq_muba_zero}
and Eq.~\eqref{eq_muba_one}
turn into Eq.~\eqref{eq_sibahu_zero}
and Eq.~\eqref{eq_sibahu_one}, respectively,
when one evaluates them for the single-band Hubbard model and
performs a Fourier transformation.

The following contributions to $M^{(2+)}$ are zero for the
single-band Hubbard model: $M^{(2+,3)}$, $M^{(2+,4)}$, $M^{(2+,5)}$, $M^{(2+,6)}$, $M^{(2+,7)}$.
The sum $M^{(2+,1)}+M^{(2+,2)}$ turns into $U^2\langle n_{-\sigma} \rangle$
in the single-band case, which is the last term in Eq.~\eqref{eq_sibahu_two}.
The
sum $\vn{T}\vn{V}^{\rm H}+\vn{T}\vn{V}^{\rm X}_{\sigma}+\vn{V}^{\rm H}\vn{T}+\vn{V}^{\rm X}_{\sigma}\vn{T}$, which contributes to Eq.~\eqref{eq_secmom_hfhf},
evaluates to $2U\langle n_{-\sigma}\rangle\epsilon(\vn{k})$ in the case
of the single-band Hubbard model. This is the middle term in
Eq.~\eqref{eq_sibahu_two}.

For the single-band Hubbard model the sum of $M_{nm}^{(3+,3)}$ (Eq.~\eqref{eq_m3p_3}) and $M_{\sigma nm}^{(3+,4)}$  (Eq.~\eqref{eq_m3p_4})
is $U^{3}\langle n_{-\sigma} \rangle$,
which is the last term in the third line of Eq.~\eqref{eq_singleband_moment3}.

\section{Evaluation of $M^{(2+,j)}_{\sigma nm }$}
\label{feg_2p}
In order to keep the notation simple, we discuss the
contractions $\mathcal{C}^{(2+,j)}_{\sigma}$, Eq.~\eqref{eq_dft_contraction},
for the uniform electron gas without spin-polarization.

We evaluate the contraction of Eq.~\eqref{eq_m2p_second} 
by
transforming it into the momentum representation, where 
we obtain at zero temperature
\bege\label{eq_c2p2}
\begin{aligned}
&\mathcal{C}^{(2+,2)}_{\sigma}\!=\!-\!\int\! \frac{d^3 q d^3 k_{4} d^3 k_{5}}{(2\pi)^9 n V^{-2}} v(\vn{q})v(\vn{k}_{4}-\vn{k}_{5}+\vn{q})n_{\vn{k}_{5}-\vn{q}}n_{\vn{k}_{4}+\vn{q}}\\
&=-A_{2}\int d^3 q d^3 k_{4} d^3 k_{5}\frac{\Theta(k_{\rm F}-|\vn{k}_{5}-\vn{q}|)\Theta(k_{\rm F}-|\vn{k}_{4}+\vn{q}|)}{q^2|\vn{k}_{4}-\vn{k}_{5}+\vn{q}|^2}.
\end{aligned}  
\ee
Here,
\bege
k_{\rm F}=(3 \pi^2 n)^{1/3}
\ee
is the Fermi wave number,
$\Theta(k)$ is the Heaviside step function,
\bege
v(\vn{q})=\frac{8\pi}{V}[{\rm Ry}][a_{\rm B}]\frac{1}{q^2}
\ee
is the Coulomb potential expressed in terms of
the Bohr radius $a_{\rm B}$, ${\rm Ry}=13.6$~eV,
and
\bege
A_{2}=\frac{(8\pi)^2}{(2\pi)^9}[{\rm Ry}]^2[a_{\rm B}]^2\frac{1}{n}.
\ee
Scaling all momenta in Eq.~\eqref{eq_c2p2} by the factor $\xi$,
we observe that this integral is proportional to $\xi^2$, i.e., it
is proportional to $k_{\rm F}^2$. In this scaling analysis
we took into account that $n$ depends on $k_{\rm F}$ as well: $n=k_{\rm F}^3/(3\pi^2)$.
It is convenient to express $\mathcal{C}^{(2+,2)}_{\sigma}$ in terms
of the dimensionless density parameter
\bege
r_s=
\frac{1}{a_{\rm B}}
\left(
\frac{9\pi}{4 k_{\rm F}^3}
\right)^{\frac{1}{3}}.
\ee
According to the scaling analysis above, it is sufficient to evaluate the
integral for a single density parameter, e.g.\ $r'_s=1$, because
\bege
\mathcal{C}^{(2+,2)}_{\sigma}(r_s)=\frac{\mathcal{C}^{(2+,2)}_{\sigma}(r'_s=1)}{r_s^2}.
\ee
The integral can be performed numerically using the
{\tt VEGAS}~\cite{PETERLEPAGE1978192,LEPAGE2021110386}
package for the Monte-Carlo integration of
high-dimensional integrals. We obtain
\bege
\mathcal{C}^{(2+,2)}(r_s)=\frac{-2.57}{r_s^2}[\rm Ry]^2.
\ee

Next, we consider $\mathcal{C}^{(2+,16)}$.
This integral is given by
\bege\label{eq_c2p16}
\begin{aligned}
  &\mathcal{C}^{(2+,16)}_{\sigma}=\int \frac{d^3 q d^3 k_{4} d^3 k_{5}}{(2\pi)^9 n V^{-2}} v(\vn{q})\\
  &\times v(\vn{k}_{4}-\vn{k}_{5}+\vn{q})n_{\vn{k}_{5}-\vn{q}}n_{\vn{k}_{4}+\vn{q}}n_{\vn{k}_{5}}\\
  &=A_{2}\int d^3 q d^3 k_{4} d^3 k_{5} \Theta(k_{\rm F}-|\vn{k}_{5}|)\\
  &\times\frac{\Theta(k_{\rm F}-|\vn{k}_{5}-\vn{q}|)\Theta(k_{\rm F}-|\vn{k}_{4}+\vn{q}|)}{q^2|\vn{k}_{4}-\vn{k}_{5}+\vn{q}|^2},
\end{aligned}
\ee
which also scales like $r_s^{-2}$, which is easy to see with a scaling analysis.
Evaluating this integral with the {\tt VEGAS}~\cite{PETERLEPAGE1978192,LEPAGE2021110386} package
gives
\bege
\mathcal{C}^{(2+,16)}(r_s)=\frac{1.73}{r_s^2}[\rm Ry]^2.
\ee

For the contraction $\mathcal{C}^{(2+,17)}$ we
need to compute the integral
\bege\label{eq_c2p17}
\begin{aligned}
  &\mathcal{C}^{(2+,17)}_{\sigma}=\int \frac{d^3 q d^3 k_{4} d^3 k_{5}}{(2\pi)^9 n V^{-2}} v(\vn{q})\\
  &\times v(\vn{k}_{4}-\vn{k}_{5}+\vn{q})n_{\vn{k}_{5}-\vn{q}}n_{\vn{k}_{4}+\vn{q}}n_{\vn{k}_{4}}\\
  &=A_{2}\int d^3 q d^3 k_{4} d^3 k_{5} \Theta(k_{\rm F}-|\vn{k}_{4}|)\\
  &\times\frac{\Theta(k_{\rm F}-|\vn{k}_{5}-\vn{q}|)\Theta(k_{\rm F}-|\vn{k}_{4}+\vn{q}|)}{q^2|\vn{k}_{4}-\vn{k}_{5}+\vn{q}|^2}.
\end{aligned}
\ee
Using a scaling analysis, we find that this integral also scales like $r_s^{-2}$.
Employing the {\tt VEGAS}~\cite{PETERLEPAGE1978192,LEPAGE2021110386} package
yields
\bege
\mathcal{C}^{(2+,17)}(r_s)=\frac{1.73}{r_s^2}[\rm Ry]^2.
\ee

While the contractions $\mathcal{C}^{(2+,2)}$,
$\mathcal{C}^{(2+,16)}$, and $\mathcal{C}^{(2+,17)}$
above are straightforward to
evaluate with ${\tt VEGAS}$~\cite{PETERLEPAGE1978192,LEPAGE2021110386}, 
the contributions
$\mathcal{C}^{(2+,1)}$,
$\mathcal{C}^{(2+,6)}$,
$\mathcal{C}^{(2+,8)}$,
$\mathcal{C}^{(2+,12)}$,
$\mathcal{C}^{(2+,14)}$, and
$\mathcal{C}^{(2+,15)}$
require more care, because their integrands contain factors
$(v(\vn{q}))^2$, which lead to a strong divergence of the
integrands in the limit $q\rightarrow 0$.
In contrast, the integrals of the
contractions  $\mathcal{C}^{(2+,2)}$,
$\mathcal{C}^{(2+,16)}$, and $\mathcal{C}^{(2+,17)}$
contain only a single factor $v(\vn{q})$, which
does not produce a divergence, because it is compensated by the
$q^2$ of $d^3 q=q^2\sin(\theta) d \theta d \phi$.
However, these contributions may be grouped 
into pairs of two, where the two partners in a pair differ in sign.
When we replace the Coulomb potential by
\bege
v_{\eta}(\vn{q})=\frac{8\pi}{V}[{\rm Ry}][a_{\rm B}]\frac{1}{q^2+\eta^2},
\ee
we observe that the limit $\eta\rightarrow 0$ is finite for the pair,
while both partners in a pair diverge in this limit.

Consider for example the pair composed of $\mathcal{C}^{(2+,1)}$ and $\mathcal{C}^{(2+,14)}$.
Evaluating the integrals
\bege\label{eq_c2p1}
\begin{aligned}
  &\mathcal{C}^{(2+,1)}_{\eta}=2\int \frac{d^3 q d^3 k_{4} d^3 k_{5}}{(2\pi)^9 n V^{-2}} v_{\eta}(\vn{q}) v_{\eta}(\vn{q})n_{\vn{k}_{5}-\vn{q}}n_{\vn{k}_{4}+\vn{q}}\\
  &=2A_{2}\int d^3 q d^3 k_{4} d^3 k_{5} \Theta(k_{\rm F}-|\vn{k}_{4}+\vn{q}|)\\
  &\times\frac{\Theta(k_{\rm F}-|\vn{k}_{5}-\vn{q}|)}{[q^2+\eta^2]^2}
\end{aligned}
\ee
and
\bege\label{eq_c2p14}
\begin{aligned}
  &\mathcal{C}^{(2+,14)}_{\eta}=-2\int \frac{d^3 q d^3 k_{4} d^3 k_{5}}{(2\pi)^9 n V^{-2}} v_{\eta}(\vn{q}) v_{\eta}(\vn{q})n_{\vn{k}_{5}-\vn{q}}n_{\vn{k}_{4}+\vn{q}}n_{\vn{k}_{5}}\\
  &=2A_{2}\int d^3 q d^3 k_{4} d^3 k_{5} \Theta(k_{\rm F}-|\vn{k}_{5}|)\\
  &\times\frac{\Theta(k_{\rm F}-|\vn{k}_{5}-\vn{q}|)\Theta(k_{\rm F}-|\vn{k}_{4}+\vn{q}|)}{[q^2+\eta^2]^2}
\end{aligned}
\ee
with the {\tt VEGAS}~\cite{PETERLEPAGE1978192,LEPAGE2021110386} package
we obtain
\bege
\lim_{\eta\rightarrow 0}\left[
  \mathcal{C}_{\eta}^{(2+,1)}(r_s)+\mathcal{C}_{\eta}^{(2+,14)}(r_s)
  \right ]=2\frac{10.51}{r_s^2}[\rm Ry]^2.
\ee
We explicitly left the spin-degeneracy factor 2 in this equation.

\section{Evaluation of $M^{(3+,j)}_{\sigma nm }$}
\label{feg_3p}
In order to keep the notation simple, we discuss the
contractions $\mathcal{C}^{(3+,j)}_{\sigma}$ for the uniform electron
gas without spin-polarization.

Transforming Eq.~\eqref{eq_m3p_4} into the momentum representation, we obtain
the following expression for $\mathcal{C}^{(3+,4)}_{\sigma}$ in terms
of an integral:
\bege\label{eq_c3p4}
\begin{aligned}
  &\mathcal{C}^{(3+,4)}_{\sigma }=\!\!\int\!\! \frac{d^3 q d^3 q'  d^3 k_{4} d^3 k_{5}}{(2\pi)^{12} n V^{-3}} v(\vn{q})v(\vn{q}')v(\vn{k}_{4}-\vn{k}_{5}+\vn{q}-\vn{q}')\\
  &\times n_{\vn{k}_{5}-\vn{q}}n_{\vn{k}_{4}+\vn{q}}\\
  &=A_3 \int d^3 q d^3 q'  d^3 k_{4} d^3 k_{5} \frac{1}{q^2} \frac{1}{[q']^2}  \\
  &\times \frac{\Theta(k_{\rm F}-|\vn{k}_{5}-\vn{q}|)\Theta(k_{\rm F}-|\vn{k}_{4}+\vn{q}|)}{|\vn{k}_{4}-\vn{k}_{5}+\vn{q}-\vn{q}'|^2},
\end{aligned}  
\ee
where
\bege
A_3=\frac{(8\pi)^3}{(2\pi)^{12}}[{\rm Ry}]^3 [a_{\rm B}]^3\frac{1}{n}.
\ee
Scaling all momenta in Eq.~\eqref{eq_c3p4} by the factor $\xi$ one may
easily find that $\mathcal{C}^{(3+,4)}_{\sigma }\propto k_{\rm F}^3\propto r_s^{-3}$.
Using {\tt VEGAS}~\cite{PETERLEPAGE1978192,LEPAGE2021110386}
we obtain
\bege
\mathcal{C}^{(3+,4)}_{\sigma }(r_{s})=-\frac{9.81}{r_s^3}[\rm Ry]^3.
\ee

\section{Algorithm to construct the spectral function from non-commuting spectral moment matrices}
\label{sec_app_diag}
In this section we provide the derivation of the algorithm described in Sec.~\ref{sec_constru_specfun}
for the construction of the spectral function from the first four
$N\times N$ spectral moment matrices,
$\vn{M}^{(0)}$, $\vn{M}^{(1)}$, $\vn{M}^{(2)}$, and $\vn{M}^{(3)}$,
where $\vn{M}^{(0)}$
is the unit matrix.

Assume that we manage to find hermitean $2N\times 2N$ matrices
\bege
\vn{\mathcal{B}}^{(1)}=\left(
\begin{array}{cc}
\vn{M}^{(1)}
&\vn{B}_{1} \\
\vn{B}_{1}^{\dagger} &\vn{D}_{1}
\end{array}
\right),
\ee
\bege
\vn{\mathcal{B}}^{(2)}=\left(
\begin{array}{cc}
\vn{M}^{(2)}
&\vn{B}_{2} \\
\vn{B}_{2}^{\dagger} &\vn{D}_{2}
\end{array}
\right),
\ee
and
\bege
\vn{\mathcal{B}}^{(3)}=\left(
\begin{array}{cc}
\vn{M}^{(3)}
&\vn{B}_{3} \\
\vn{B}_{3}^{\dagger} &\vn{D}_{3}
\end{array}
\right),
\ee
which mutually commute, i.e.,
\bege\label{app_eq_b1b2b3_commut}
\begin{aligned}
  &[\vn{\mathcal{B}}^{(1)},\vn{\mathcal{B}}^{(2)}]_{-}=0,\\
  &[\vn{\mathcal{B}}^{(1)},\vn{\mathcal{B}}^{(3)}]_{-}=0,\\
 &[\vn{\mathcal{B}}^{(2)},\vn{\mathcal{B}}^{(3)}]_{-}=0,\\ 
\end{aligned}  
\ee
and which satisfy
\bege\label{app_eq_b2_fact}
\vn{\mathcal{B}}^{(2)}=\vn{\mathcal{B}}^{(1)}\vn{\mathcal{B}}^{(1)},
\ee
and
\bege\label{app_eq_b3_fact}
\vn{\mathcal{B}}^{(3)}=\vn{\mathcal{B}}^{(1)}\vn{\mathcal{B}}^{(2)}=\vn{\mathcal{B}}^{(1)}\vn{\mathcal{B}}^{(1)}\vn{\mathcal{B}}^{(1)}.
\ee
Note that Eq.~\eqref{app_eq_b1b2b3_commut}
is satisfied if Eq.~\eqref{app_eq_b2_fact}
and Eq.~\eqref{app_eq_b3_fact}
are satisfied. We will therefore solve only
Eq.~\eqref{app_eq_b2_fact}
and Eq.~\eqref{app_eq_b3_fact} below.
The matrices $\vn{M}^{(I)}$, $\vn{B}_{i}$ and $\vn{D}_{i}$ have the size
$N\times N$. The matrices $\vn{M}^{(I)}$ are the given hermitean spectral
moment matrices, while $\vn{B}_{i}$ and $\vn{D}_{i}$ are
matrices that need to be determined
such that Eq.~\eqref{app_eq_b2_fact}
and Eq.~\eqref{app_eq_b3_fact}  are satisfied.
While $\vn{D}_{i}$ is required to be hermitean,
$\vn{B}_{i}$ is not.

If we manage to find these matrices $\vn{\mathcal{B}}^{(1)}$,
$\vn{\mathcal{B}}^{(2)}$,
and $\vn{\mathcal{B}}^{(3)}$, we know
that they possess a common system of eigenvectors, i.e.,
they may be diagonalized by the same unitary transformation,
because
they are hermitean
and they commute mutually.
Consequently, we may find a unitary transformation $\vn{\mathcal{U}}$
so that
\bege\label{app_eq_b1_diag}
\vn{\mathcal{B}}^{(1)}=\vn{\mathcal{U}}\vn{\mathcal{D}}\vn{\mathcal{U}}^{\dagger},
\ee
where $\vn{\mathcal{D}}$ is a diagonal matrix.
Using $\vn{\mathcal{U}}$ and $\vn{\mathcal{D}}$ we may
write
\bege\label{app_eq_b2_diag}
\vn{\mathcal{B}}^{(2)}=\vn{\mathcal{U}}\vn{\mathcal{D}}^2\vn{\mathcal{U}}^{\dagger}
\ee
and
\bege\label{app_eq_b3_diag}
\vn{\mathcal{B}}^{(3)}=\vn{\mathcal{U}}\vn{\mathcal{D}}^3\vn{\mathcal{U}}^{\dagger}.
\ee

In Ref.~\cite{spectral} we have shown that the
eigenvalue problems Eq.~\eqref{app_eq_b1_diag},
Eq.~\eqref{app_eq_b2_diag}, and
Eq.~\eqref{app_eq_b3_diag}
may be rewritten in the form
\bege
\bar{\vn{\mathcal{\vn{W}}}}\vn{\mathcal{A}}^{(I)}=\bar{\vn{\mathcal{\vn{B}}}}^{(I)},
\ee
where $I=1,2,3$
(see Eq.~(13) in Ref.~\cite{spectral}).
When we denote the representation of the
unit matrix as a column vector by $\bar{\vn{\mathcal{\vn{B}}}}^{(0)}$,
we may combine Eq.~\eqref{app_eq_b1_diag},
Eq.~\eqref{app_eq_b2_diag}, and
Eq.~\eqref{app_eq_b3_diag}
into the compact expression
\bege\label{app_eq_compact_wa=b}
\bar{\vn{\mathcal{\vn{W}}}}\vn{\mathcal{A}}=\bar{\vn{\mathcal{\vn{B}}}},
\ee
where
\bege
\vn{\mathcal{A}}=[\vn{\mathcal{A}}^{(0)},\vn{\mathcal{A}}^{(1)},\vn{\mathcal{A}}^{(2)},\vn{\mathcal{A}}^{(3)}]
\ee
and
\bege
\bar{\vn{\mathcal{\vn{B}}}}=[\bar{\vn{\mathcal{\vn{B}}}}^{(0)},\bar{\vn{\mathcal{\vn{B}}}}^{(1)},\bar{\vn{\mathcal{\vn{B}}}}^{(2)},\bar{\vn{\mathcal{\vn{B}}}}^{(3)}].
\ee

Next, we rewrite $\bar{\vn{\mathcal{\vn{B}}}}$ as
\bege
\bar{\vn{\mathcal{\vn{B}}}}=\left(
\begin{array}{cc}
\vn{\mathcal{\vn{M}}}  \\
\bar{\vn{\mathcal{\vn{B}}}}_{\rm Low}
\end{array}
\right)
\ee
and
\bege
\bar{\vn{\mathcal{\vn{W}}}}=\left(
\begin{array}{cc}
\vn{\mathcal{\vn{W}}}\\
\bar{\vn{\mathcal{\vn{W}}}}_{\rm Low}
\end{array}
\right),
\ee
where $\vn{\mathcal{\vn{M}}}$
and $\vn{\mathcal{\vn{W}}}$
are the matrices defined in
Ref.~\cite{spectral} (see Eq.~(7) and Eq.~(8) in Ref.~\cite{spectral}).
$\vn{\mathcal{\vn{M}}}$ is a $N^2\times 4$
matrix, $\bar{\vn{\mathcal{\vn{B}}}}_{\rm Low}$ is
a $3N^2\times 4$ matrix, 
$\vn{\mathcal{\vn{W}}}$ is a $N^2\times 2N$ matrix,
and $\bar{\vn{\mathcal{\vn{W}}}}_{\rm Low}$
is a $3N^2\times 2N$ matrix.
Thus, we may rewrite Eq.~\eqref{app_eq_compact_wa=b}
as two equations:
\bege\label{app_eq_wa=m_upper}
\vn{\mathcal{W}}\vn{\mathcal{A}}=\vn{\mathcal{M}}
\ee
and
\bege
\bar{\vn{\mathcal{W}}}_{\rm Low}\vn{\mathcal{A}}=\bar{\vn{\mathcal{B}}}_{\rm Low}.
\ee
Eq.~\eqref{app_eq_wa=m_upper} is identical to the Eq.~(9) in
Ref.~\cite{spectral}, which needs to be solved to obtain the
spectral function. Thus, we may solve Eq.~\eqref{app_eq_wa=m_upper}
by determining the matrices $\vn{B}_{1}$ and $\vn{D}_{1}$,
and by diagonalizing the matrix $\vn{\mathcal{B}}^{(1)}$.

Therefore, in order to prove the algorithm in Sec.~\ref{sec_constru_specfun}
it remains to show that the matrices $\vn{B}_{1}$ and $\vn{D}_{1}$
may be found by solving 
Eq.~\eqref{app_eq_b2_fact} and Eq.~\eqref{app_eq_b3_fact}.
From Eq.~\eqref{app_eq_b2_fact} we obtain
the following equation for $\vn{B}_{1}$:
\bege\label{app_eq_bbdag}
\vn{B}_{1}\vn{B}_{1}^{\dagger}=\vn{M}^{(2)}-\vn{M}^{(1)}\vn{M}^{(1)}.
\ee
Since $\vn{B}_{1}\vn{B}_{1}^{\dagger}$ is a hermitean matrix,
it may be diagonalized:
\bege
\vn{B}_{1}\vn{B}_{1}^{\dagger}=\vn{U}\vn{D}\vn{U}^{\dagger},
\ee
where $\vn{U}$ is a unitary matrix and $\vn{D}$ is a diagonal matrix.
If $\vn{B}_{1}\vn{B}_{1}^{\dagger}$ is positive definite,
we obtain
\bege
\vn{B}_{1}=\vn{U}\sqrt{\vn{D}},
\ee
which is Eq.~\eqref{eq_b1} in the main text.
If $\vn{B}_{1}\vn{B}_{1}^{\dagger}$ is not positive definite,
the algorithm described in this section cannot be
used. However, in all applications discussed in this
paper,  $\vn{B}_{1}\vn{B}_{1}^{\dagger}$ is positive definite.
We suspect that the reason for this is that
$\vn{M}^{(2+)}$ is generally positive definite.

From Eq.~\eqref{app_eq_b3_fact} we obtain
the following equation for $\vn{B}_{2}$:
\bege\label{app_eq_b2}
\vn{B}_2=\left[
  \vn{M}^{(3)}-\vn{M}^{(2)}\vn{M}^{(1)}
  \right]
\left[
  \vn{B}_{1}^{\dagger}
\right]^{-1}.
\ee
This is Eq.~\eqref{eq_b2} in the main text.

From Eq.~\eqref{app_eq_b2_fact} we obtain the
following equation for $\vn{D}_{1}$:
\bege\label{app_eq_d1}
\vn{D}_{1}=\vn{B}_{1}^{-1}
\left[
\vn{B}_{2}-\vn{M}^{(1)}\vn{B}_{1}
\right].
\ee
This is Eq.~\eqref{eq_d1} in the main text.
$\vn{D}_{1}$ is required to be hermitean, which is not
directly obvious from Eq.~\eqref{app_eq_d1}.
However, making use of
Eq.~\eqref{app_eq_b2}
and Eq.~\eqref{app_eq_bbdag}
it is straightforward to show that
\bege\label{eq_d1_hermitean}
\vn{D}_{1}-\vn{D}_{1}^{\dagger}=0.
\ee

At this point we have completely determined the matrix
$\vn{\mathcal{B}}^{(1)}$, from which the spectral function
may be constructed using its eigenvalues, which are contained in
the diagonal matrix $\vn{\mathcal{D}}$, and the unitary transformation
$\vn{\mathcal{U}}$ defined in Eq.~\eqref{app_eq_b1_diag}.
However, it remains to show that all those additional
equations that follow from  
Eq.~\eqref{app_eq_b2_fact}
and
Eq.~\eqref{app_eq_b3_fact}
but that we did not use to derive the expressions for $\vn{B}_{1}$
and $\vn{D}_{1}$ can be satisfied as well.
From Eq.~\eqref{app_eq_b2_fact}
we obtain
\bege\label{app_eq_remain1}
\vn{D}_{2}=\vn{B}_{1}^{\dagger}\vn{B}_{1}+\vn{D}_{1}\vn{D}_{1},
\ee
and from
Eq.~\eqref{app_eq_b3_fact}
we obtain
\bege\label{app_eq_remain2}
\vn{B}_{3}=\vn{M}^{(1)}\vn{B}_{2}+\vn{B}_{1}\vn{D}_{2}
\ee
and
\bege\label{app_eq_remain3}
\vn{D}_{3}=\vn{B}_{1}^{\dagger}\vn{B}_{2}+\vn{D}_{1}\vn{D}_{2}.
\ee
$\vn{D}_2$ as given by Eq.~\eqref{app_eq_remain1}
is hermitean, because
$\vn{D}_1$ is hermitean according to Eq.~\eqref{eq_d1_hermitean}.
Thus, it does not violate any of the
equations above.
$\vn{B}_3$ as given by Eq.~\eqref{app_eq_remain2}
does not violate any of the equations above either.
$\vn{D}_3$ as given by Eq.~\eqref{app_eq_remain3}
should be hermitean, which is not directly obvious.
However, using Eq.~\eqref{app_eq_remain1}, Eq.~\eqref{app_eq_d1},
and Eq.~\eqref{eq_d1_hermitean},
it is straightforward to show that
\bege
\vn{D}_{3}-\vn{D}_{3}^{\dagger}=0.
\ee

Eq.~\eqref{eq_final_spec}
in the main text 
follows from Eq.~\eqref{app_eq_b1_diag},
Eq.~\eqref{app_eq_wa=m_upper},
and Ref.~\cite{spectral}.

\section{Generalization to more moments}
\label{sec_app_more}
We may generalize the argument
given in Ref.~\cite{spectral} and show that from the first $2P$ moments ($P=1,2,\dots$)
one may construct the spectral function:
We may map each moment $\tilde{\vn{M}}^{(I)}$
(where $\tilde{\vn{M}}^{(I)}$ denotes the moment computed from
the nested commutator expression -- as opposed to the moment
obtained from the explicit energy integration)
onto an $N^{2}$-dimensional real-valued vector
$\vn{\mathcal{M}}^{(I)}$, because
$N^{2}$ real-valued parameters fully define
a hermitean $N\times N$ matrix.
We introduce the $N^{2}\times 2P$ matrix
$\vn{\mathcal{M}}$ by
\bege
\vn{\mathcal{M}}=[\vn{\mathcal{M}}^{(0)},\dots,\vn{\mathcal{M}}^{(2P-1)}].
\ee
We try to approximate the spectral function by
\bege\label{eq_spectral_matrix_general}
\frac{S_{\alpha\beta}(E)}{\hbar}=\sum_{p=1}^{P}\sum_{\gamma=1}^{N}
a_{\gamma p}
\mathcal{V}_{\alpha\gamma p}\mathcal{V}^{*}_{\beta\gamma p} \delta(E-E_{\gamma p}),
\ee
because we expect that $PN$ bands can be computed from
the first $2P$ spectral moment matrices.
Inserting this approximation 
into Eq.~\eqref{eq_specmoms_eneint} yields
\bege\label{eq_spectral_matrix_general_eneint}
M^{(I)}_{\alpha\beta}=
\sum_{p=1}^{P}\sum_{\gamma=1}^{N}
a_{\gamma p}
\mathcal{W}_{\alpha\beta\gamma p}
[E_{\gamma p}]^I,
\ee
where we
defined $\mathcal{W}_{\alpha\beta\gamma p}=\mathcal{V}_{\alpha\gamma p}\mathcal{V}^{*}_{\beta\gamma p}$.
We may consider $\mathcal{W}_{\alpha\beta\gamma p}$ as the row-$\alpha$ column-$\beta$ element of
a hermitean matrix $\vn{\mathcal{W}}_{\gamma p}$.
Since $\gamma=1,\dots,N$ and $p=1,...,P$, there are $PN$ such matrices.
As the hermitean $N\times N$ matrix $\vn{\mathcal{W}}_{\gamma p}$  is
equivalent to a $N^2$-dimensional real-valued vector $\tilde{\vn{\mathcal{W}}}_{\gamma p}$,
we define the $N^2 \times PN$
matrix $\vn{\mathcal{W}}=[\tilde{\vn{\mathcal{W}}}_{11}\dots \tilde{\vn{\mathcal{W}}}_{NP}]$.
Additionally, we construct the $PN\times 2P$ matrix $\vn{\mathcal{A}}$ by setting the element
$\mathcal{A}_{\gamma p m}$ in row $(\gamma, p)$ and
column $m$ to $a_{\gamma p}(E_{\gamma p})^{m-1}$.
The requirements $\vn{M}^{(I)}=\tilde{\vn{M}}^{(I)}$ with $I=0, 1,\dots 2P-1$
(where $\tilde{\vn{M}}^{(I)}$ are the moments computed from the nested commutator expressions)
can now be formulated in compact form by
\bege\label{eq_compact_m_eq_m}
\vn{\mathcal{W}}\vn{\mathcal{A}}=\vn{\mathcal{M}}.
\ee
This is the generalization of Eq.~(9) in Ref.~\cite{spectral} for
the first $2P$ moments. The form of the equation is the same, only the
sizes of the matrices are different.
Since the matrix $\vn{\mathcal{M}}$
contains $2P N^2$ elements,
Eq.~\eqref{eq_compact_m_eq_m}
defines  $2P N^2$ nonlinear equations.
Each vector $\vn{\mathcal{V}}_{\gamma p}$ has $N$ components and there are
$PN$ such vectors. $\vn{\mathcal{V}}_{\gamma p}$ is required to be
normalized and the
gauge-transformation $\vn{\mathcal{V}}_{\gamma p}\rightarrow e^{i\Phi}\vn{\mathcal{V}}_{\gamma p} $
does not affect $\mathcal{W}_{\alpha\beta\gamma p}=\mathcal{V}_{\alpha\gamma p}\mathcal{V}^{*}_{\beta\gamma p}$. Thus,
every $\vn{\mathcal{V}}_{\gamma p}$ is determined by $2(N-1)$ real-valued unknowns, i.e.,
$2P(N^2-N)$ unknown coefficients need to be found to determine all vectors $\vn{\mathcal{V}}_{\gamma p}$.
Additionally, we need to find the $PN$ energies $E_{\gamma p}$
as well as the $PN$ spectral weights $a_{\gamma p}$. Consequently, Eq.~\eqref{eq_compact_m_eq_m}
is a system of $2PN^2$ nonlinear equations for $2PN^2$ unknowns.
Thus, one may expect that it should be possible to compute $PN$ bands
from the first $2P$ spectral moment matrices of size $N\times N$,
because the number of unknowns matches the number of available nonlinear equations.

\bibliography{momentis}

\begin{thebibliography}{40}%
\makeatletter
\providecommand \@ifxundefined [1]{%
 \@ifx{#1\undefined}
}%
\providecommand \@ifnum [1]{%
 \ifnum #1\expandafter \@firstoftwo
 \else \expandafter \@secondoftwo
 \fi
}%
\providecommand \@ifx [1]{%
 \ifx #1\expandafter \@firstoftwo
 \else \expandafter \@secondoftwo
 \fi
}%
\providecommand \natexlab [1]{#1}%
\providecommand \enquote  [1]{``#1''}%
\providecommand \bibnamefont  [1]{#1}%
\providecommand \bibfnamefont [1]{#1}%
\providecommand \citenamefont [1]{#1}%
\providecommand \href@noop [0]{\@secondoftwo}%
\providecommand \href [0]{\begingroup \@sanitize@url \@href}%
\providecommand \@href[1]{\@@startlink{#1}\@@href}%
\providecommand \@@href[1]{\endgroup#1\@@endlink}%
\providecommand \@sanitize@url [0]{\catcode `\\12\catcode `\$12\catcode
  `\&12\catcode `\#12\catcode `\^12\catcode `\_12\catcode `\%12\relax}%
\providecommand \@@startlink[1]{}%
\providecommand \@@endlink[0]{}%
\providecommand \url  [0]{\begingroup\@sanitize@url \@url }%
\providecommand \@url [1]{\endgroup\@href {#1}{\urlprefix }}%
\providecommand \urlprefix  [0]{URL }%
\providecommand \Eprint [0]{\href }%
\providecommand \doibase [0]{https://doi.org/}%
\providecommand \selectlanguage [0]{\@gobble}%
\providecommand \bibinfo  [0]{\@secondoftwo}%
\providecommand \bibfield  [0]{\@secondoftwo}%
\providecommand \translation [1]{[#1]}%
\providecommand \BibitemOpen [0]{}%
\providecommand \bibitemStop [0]{}%
\providecommand \bibitemNoStop [0]{.\EOS\space}%
\providecommand \EOS [0]{\spacefactor3000\relax}%
\providecommand \BibitemShut  [1]{\csname bibitem#1\endcsname}%
\let\auto@bib@innerbib\@empty
\bibitem [{\citenamefont {Hohenberg}\ and\ \citenamefont
  {Kohn}(1964)}]{PhysRev.136.B864}%
  \BibitemOpen
  \bibfield  {author} {\bibinfo {author} {\bibfnamefont {P.}~\bibnamefont
  {Hohenberg}}\ and\ \bibinfo {author} {\bibfnamefont {W.}~\bibnamefont
  {Kohn}},\ }\bibfield  {title} {\bibinfo {title} {Inhomogeneous electron
  gas},\ }\href {https://doi.org/10.1103/PhysRev.136.B864} {\bibfield
  {journal} {\bibinfo  {journal} {Phys. Rev.}\ }\textbf {\bibinfo {volume}
  {136}},\ \bibinfo {pages} {B864} (\bibinfo {year} {1964})}\BibitemShut
  {NoStop}%
\bibitem [{\citenamefont {Kohn}\ and\ \citenamefont
  {Sham}(1965)}]{PhysRev.140.A1133}%
  \BibitemOpen
  \bibfield  {author} {\bibinfo {author} {\bibfnamefont {W.}~\bibnamefont
  {Kohn}}\ and\ \bibinfo {author} {\bibfnamefont {L.~J.}\ \bibnamefont
  {Sham}},\ }\bibfield  {title} {\bibinfo {title} {Self-consistent equations
  including exchange and correlation effects},\ }\href
  {https://doi.org/10.1103/PhysRev.140.A1133} {\bibfield  {journal} {\bibinfo
  {journal} {Phys. Rev.}\ }\textbf {\bibinfo {volume} {140}},\ \bibinfo {pages}
  {A1133} (\bibinfo {year} {1965})}\BibitemShut {NoStop}%
\bibitem [{\citenamefont {S\'anchez-Barriga}\ \emph {et~al.}(2009)\citenamefont
  {S\'anchez-Barriga}, \citenamefont {Fink}, \citenamefont {Boni},
  \citenamefont {Di~Marco}, \citenamefont {Braun}, \citenamefont {Min\'ar},
  \citenamefont {Varykhalov}, \citenamefont {Rader}, \citenamefont {Bellini},
  \citenamefont {Manghi}, \citenamefont {Ebert}, \citenamefont {Katsnelson},
  \citenamefont {Lichtenstein}, \citenamefont {Eriksson}, \citenamefont
  {Eberhardt},\ and\ \citenamefont {D\"urr}}]{PhysRevLett.103.267203}%
  \BibitemOpen
  \bibfield  {author} {\bibinfo {author} {\bibfnamefont {J.}~\bibnamefont
  {S\'anchez-Barriga}}, \bibinfo {author} {\bibfnamefont {J.}~\bibnamefont
  {Fink}}, \bibinfo {author} {\bibfnamefont {V.}~\bibnamefont {Boni}}, \bibinfo
  {author} {\bibfnamefont {I.}~\bibnamefont {Di~Marco}}, \bibinfo {author}
  {\bibfnamefont {J.}~\bibnamefont {Braun}}, \bibinfo {author} {\bibfnamefont
  {J.}~\bibnamefont {Min\'ar}}, \bibinfo {author} {\bibfnamefont
  {A.}~\bibnamefont {Varykhalov}}, \bibinfo {author} {\bibfnamefont
  {O.}~\bibnamefont {Rader}}, \bibinfo {author} {\bibfnamefont
  {V.}~\bibnamefont {Bellini}}, \bibinfo {author} {\bibfnamefont
  {F.}~\bibnamefont {Manghi}}, \bibinfo {author} {\bibfnamefont
  {H.}~\bibnamefont {Ebert}}, \bibinfo {author} {\bibfnamefont {M.~I.}\
  \bibnamefont {Katsnelson}}, \bibinfo {author} {\bibfnamefont {A.~I.}\
  \bibnamefont {Lichtenstein}}, \bibinfo {author} {\bibfnamefont
  {O.}~\bibnamefont {Eriksson}}, \bibinfo {author} {\bibfnamefont
  {W.}~\bibnamefont {Eberhardt}},\ and\ \bibinfo {author} {\bibfnamefont
  {H.~A.}\ \bibnamefont {D\"urr}},\ }\bibfield  {title} {\bibinfo {title}
  {Strength of correlation effects in the electronic structure of iron},\
  }\href {https://doi.org/10.1103/PhysRevLett.103.267203} {\bibfield  {journal}
  {\bibinfo  {journal} {Phys. Rev. Lett.}\ }\textbf {\bibinfo {volume} {103}},\
  \bibinfo {pages} {267203} (\bibinfo {year} {2009})}\BibitemShut {NoStop}%
\bibitem [{\citenamefont {Yao}\ \emph {et~al.}(2004)\citenamefont {Yao},
  \citenamefont {Kleinman}, \citenamefont {MacDonald}, \citenamefont {Sinova},
  \citenamefont {Jungwirth}, \citenamefont {Wang}, \citenamefont {Wang},\ and\
  \citenamefont {Niu}}]{PhysRevLett.92.037204}%
  \BibitemOpen
  \bibfield  {author} {\bibinfo {author} {\bibfnamefont {Y.}~\bibnamefont
  {Yao}}, \bibinfo {author} {\bibfnamefont {L.}~\bibnamefont {Kleinman}},
  \bibinfo {author} {\bibfnamefont {A.~H.}\ \bibnamefont {MacDonald}}, \bibinfo
  {author} {\bibfnamefont {J.}~\bibnamefont {Sinova}}, \bibinfo {author}
  {\bibfnamefont {T.}~\bibnamefont {Jungwirth}}, \bibinfo {author}
  {\bibfnamefont {D.-s.}\ \bibnamefont {Wang}}, \bibinfo {author}
  {\bibfnamefont {E.}~\bibnamefont {Wang}},\ and\ \bibinfo {author}
  {\bibfnamefont {Q.}~\bibnamefont {Niu}},\ }\bibfield  {title} {\bibinfo
  {title} {First principles calculation of anomalous hall conductivity in
  ferromagnetic bcc fe},\ }\href
  {https://doi.org/10.1103/PhysRevLett.92.037204} {\bibfield  {journal}
  {\bibinfo  {journal} {Phys. Rev. Lett.}\ }\textbf {\bibinfo {volume} {92}},\
  \bibinfo {pages} {037204} (\bibinfo {year} {2004})}\BibitemShut {NoStop}%
\bibitem [{\citenamefont {Gilmore}\ \emph {et~al.}(2007)\citenamefont
  {Gilmore}, \citenamefont {Idzerda},\ and\ \citenamefont
  {Stiles}}]{PhysRevLett.99.027204}%
  \BibitemOpen
  \bibfield  {author} {\bibinfo {author} {\bibfnamefont {K.}~\bibnamefont
  {Gilmore}}, \bibinfo {author} {\bibfnamefont {Y.~U.}\ \bibnamefont
  {Idzerda}},\ and\ \bibinfo {author} {\bibfnamefont {M.~D.}\ \bibnamefont
  {Stiles}},\ }\bibfield  {title} {\bibinfo {title} {Identification of the
  dominant precession-damping mechanism in fe, co, and ni by first-principles
  calculations},\ }\href {https://doi.org/10.1103/PhysRevLett.99.027204}
  {\bibfield  {journal} {\bibinfo  {journal} {Phys. Rev. Lett.}\ }\textbf
  {\bibinfo {volume} {99}},\ \bibinfo {pages} {027204} (\bibinfo {year}
  {2007})}\BibitemShut {NoStop}%
\bibitem [{\citenamefont {Freimuth}\ \emph {et~al.}(2015)\citenamefont
  {Freimuth}, \citenamefont {Bl\"ugel},\ and\ \citenamefont
  {Mokrousov}}]{invsot}%
  \BibitemOpen
  \bibfield  {author} {\bibinfo {author} {\bibfnamefont {F.}~\bibnamefont
  {Freimuth}}, \bibinfo {author} {\bibfnamefont {S.}~\bibnamefont {Bl\"ugel}},\
  and\ \bibinfo {author} {\bibfnamefont {Y.}~\bibnamefont {Mokrousov}},\
  }\bibfield  {title} {\bibinfo {title} {Direct and inverse spin-orbit
  torques},\ }\href@noop {} {\bibfield  {journal} {\bibinfo  {journal} {Phys.
  Rev. B}\ }\textbf {\bibinfo {volume} {92}},\ \bibinfo {pages} {064415}
  (\bibinfo {year} {2015})}\BibitemShut {NoStop}%
\bibitem [{\citenamefont {Berritta}\ \emph {et~al.}(2016)\citenamefont
  {Berritta}, \citenamefont {Mondal}, \citenamefont {Carva},\ and\
  \citenamefont {Oppeneer}}]{ife_berritta}%
  \BibitemOpen
  \bibfield  {author} {\bibinfo {author} {\bibfnamefont {M.}~\bibnamefont
  {Berritta}}, \bibinfo {author} {\bibfnamefont {R.}~\bibnamefont {Mondal}},
  \bibinfo {author} {\bibfnamefont {K.}~\bibnamefont {Carva}},\ and\ \bibinfo
  {author} {\bibfnamefont {P.~M.}\ \bibnamefont {Oppeneer}},\ }\bibfield
  {title} {\bibinfo {title} {\textit{Ab Initio} theory of coherent
  laser-induced magnetization in metals},\ }\href@noop {} {\bibfield  {journal}
  {\bibinfo  {journal} {Phys. Rev. Lett.}\ }\textbf {\bibinfo {volume} {117}},\
  \bibinfo {pages} {137203} (\bibinfo {year} {2016})}\BibitemShut {NoStop}%
\bibitem [{\citenamefont {Iba\~nez Azpiroz}\ \emph {et~al.}(2018)\citenamefont
  {Iba\~nez Azpiroz}, \citenamefont {Tsirkin},\ and\ \citenamefont
  {Souza}}]{PhysRevB.97.245143}%
  \BibitemOpen
  \bibfield  {author} {\bibinfo {author} {\bibfnamefont {J.}~\bibnamefont
  {Iba\~nez Azpiroz}}, \bibinfo {author} {\bibfnamefont {S.~S.}\ \bibnamefont
  {Tsirkin}},\ and\ \bibinfo {author} {\bibfnamefont {I.}~\bibnamefont
  {Souza}},\ }\bibfield  {title} {\bibinfo {title} {Ab initio calculation of
  the shift photocurrent by wannier interpolation},\ }\href
  {https://doi.org/10.1103/PhysRevB.97.245143} {\bibfield  {journal} {\bibinfo
  {journal} {Phys. Rev. B}\ }\textbf {\bibinfo {volume} {97}},\ \bibinfo
  {pages} {245143} (\bibinfo {year} {2018})}\BibitemShut {NoStop}%
\bibitem [{\citenamefont {Hedin}(1965)}]{PhysRev.139.A796}%
  \BibitemOpen
  \bibfield  {author} {\bibinfo {author} {\bibfnamefont {L.}~\bibnamefont
  {Hedin}},\ }\bibfield  {title} {\bibinfo {title} {New method for calculating
  the one-particle green's function with application to the electron-gas
  problem},\ }\href {https://doi.org/10.1103/PhysRev.139.A796} {\bibfield
  {journal} {\bibinfo  {journal} {Phys. Rev.}\ }\textbf {\bibinfo {volume}
  {139}},\ \bibinfo {pages} {A796} (\bibinfo {year} {1965})}\BibitemShut
  {NoStop}%
\bibitem [{\citenamefont {Franz}\ \emph {et~al.}(2014)\citenamefont {Franz},
  \citenamefont {Freimuth}, \citenamefont {Bauer}, \citenamefont {Ritz},
  \citenamefont {Schnarr}, \citenamefont {Duvinage}, \citenamefont {Adams},
  \citenamefont {Bl\"ugel}, \citenamefont {Rosch}, \citenamefont {Mokrousov},\
  and\ \citenamefont {Pfleiderer}}]{PhysRevLett.112.186601}%
  \BibitemOpen
  \bibfield  {author} {\bibinfo {author} {\bibfnamefont {C.}~\bibnamefont
  {Franz}}, \bibinfo {author} {\bibfnamefont {F.}~\bibnamefont {Freimuth}},
  \bibinfo {author} {\bibfnamefont {A.}~\bibnamefont {Bauer}}, \bibinfo
  {author} {\bibfnamefont {R.}~\bibnamefont {Ritz}}, \bibinfo {author}
  {\bibfnamefont {C.}~\bibnamefont {Schnarr}}, \bibinfo {author} {\bibfnamefont
  {C.}~\bibnamefont {Duvinage}}, \bibinfo {author} {\bibfnamefont
  {T.}~\bibnamefont {Adams}}, \bibinfo {author} {\bibfnamefont
  {S.}~\bibnamefont {Bl\"ugel}}, \bibinfo {author} {\bibfnamefont
  {A.}~\bibnamefont {Rosch}}, \bibinfo {author} {\bibfnamefont
  {Y.}~\bibnamefont {Mokrousov}},\ and\ \bibinfo {author} {\bibfnamefont
  {C.}~\bibnamefont {Pfleiderer}},\ }\bibfield  {title} {\bibinfo {title}
  {Real-space and reciprocal-space berry phases in the hall effect of
  ${\mathrm{mn}}_{1\ensuremath{-}x}{\mathrm{fe}}_{x}\mathrm{Si}$},\ }\href
  {https://doi.org/10.1103/PhysRevLett.112.186601} {\bibfield  {journal}
  {\bibinfo  {journal} {Phys. Rev. Lett.}\ }\textbf {\bibinfo {volume} {112}},\
  \bibinfo {pages} {186601} (\bibinfo {year} {2014})}\BibitemShut {NoStop}%
\bibitem [{\citenamefont {Nekrasov}\ \emph {et~al.}(2005)\citenamefont
  {Nekrasov}, \citenamefont {Keller}, \citenamefont {Kondakov}, \citenamefont
  {Kozhevnikov}, \citenamefont {Pruschke}, \citenamefont {Held}, \citenamefont
  {Vollhardt},\ and\ \citenamefont {Anisimov}}]{PhysRevB.72.155106}%
  \BibitemOpen
  \bibfield  {author} {\bibinfo {author} {\bibfnamefont {I.~A.}\ \bibnamefont
  {Nekrasov}}, \bibinfo {author} {\bibfnamefont {G.}~\bibnamefont {Keller}},
  \bibinfo {author} {\bibfnamefont {D.~E.}\ \bibnamefont {Kondakov}}, \bibinfo
  {author} {\bibfnamefont {A.~V.}\ \bibnamefont {Kozhevnikov}}, \bibinfo
  {author} {\bibfnamefont {T.}~\bibnamefont {Pruschke}}, \bibinfo {author}
  {\bibfnamefont {K.}~\bibnamefont {Held}}, \bibinfo {author} {\bibfnamefont
  {D.}~\bibnamefont {Vollhardt}},\ and\ \bibinfo {author} {\bibfnamefont
  {V.~I.}\ \bibnamefont {Anisimov}},\ }\bibfield  {title} {\bibinfo {title}
  {Comparative study of correlation effects in
  $\mathrm{Ca}\mathrm{V}{\mathrm{o}}_{3}$ and
  $\mathrm{Sr}\mathrm{V}{\mathrm{o}}_{3}$},\ }\href
  {https://doi.org/10.1103/PhysRevB.72.155106} {\bibfield  {journal} {\bibinfo
  {journal} {Phys. Rev. B}\ }\textbf {\bibinfo {volume} {72}},\ \bibinfo
  {pages} {155106} (\bibinfo {year} {2005})}\BibitemShut {NoStop}%
\bibitem [{\citenamefont {Liebsch}(1979)}]{PhysRevLett.43.1431}%
  \BibitemOpen
  \bibfield  {author} {\bibinfo {author} {\bibfnamefont {A.}~\bibnamefont
  {Liebsch}},\ }\bibfield  {title} {\bibinfo {title} {Effect of self-energy
  corrections on the valence-band photoemission spectra of ni},\ }\href
  {https://doi.org/10.1103/PhysRevLett.43.1431} {\bibfield  {journal} {\bibinfo
   {journal} {Phys. Rev. Lett.}\ }\textbf {\bibinfo {volume} {43}},\ \bibinfo
  {pages} {1431} (\bibinfo {year} {1979})}\BibitemShut {NoStop}%
\bibitem [{\citenamefont {Nolting}\ \emph {et~al.}(1989)\citenamefont
  {Nolting}, \citenamefont {Borgiel/}, \citenamefont {Dose},\ and\
  \citenamefont {Fauster}}]{nickel_PhysRevB.40.5015}%
  \BibitemOpen
  \bibfield  {author} {\bibinfo {author} {\bibfnamefont {W.}~\bibnamefont
  {Nolting}}, \bibinfo {author} {\bibfnamefont {W.}~\bibnamefont {Borgiel/}},
  \bibinfo {author} {\bibfnamefont {V.}~\bibnamefont {Dose}},\ and\ \bibinfo
  {author} {\bibfnamefont {T.}~\bibnamefont {Fauster}},\ }\bibfield  {title}
  {\bibinfo {title} {Finite-temperature ferromagnetism of nickel},\ }\href
  {https://doi.org/10.1103/PhysRevB.40.5015} {\bibfield  {journal} {\bibinfo
  {journal} {Phys. Rev. B}\ }\textbf {\bibinfo {volume} {40}},\ \bibinfo
  {pages} {5015} (\bibinfo {year} {1989})}\BibitemShut {NoStop}%
\bibitem [{\citenamefont {Koloren\ifmmode~\check{c}\else \v{c}\fi{}}\ \emph
  {et~al.}(2012)\citenamefont {Koloren\ifmmode~\check{c}\else \v{c}\fi{}},
  \citenamefont {Poteryaev},\ and\ \citenamefont
  {Lichtenstein}}]{PhysRevB.85.235136}%
  \BibitemOpen
  \bibfield  {author} {\bibinfo {author} {\bibfnamefont {J.~c.~v.}\
  \bibnamefont {Koloren\ifmmode~\check{c}\else \v{c}\fi{}}}, \bibinfo {author}
  {\bibfnamefont {A.~I.}\ \bibnamefont {Poteryaev}},\ and\ \bibinfo {author}
  {\bibfnamefont {A.~I.}\ \bibnamefont {Lichtenstein}},\ }\bibfield  {title}
  {\bibinfo {title} {Valence-band satellite in ferromagnetic nickel: Lda+dmft
  study with exact diagonalization},\ }\href
  {https://doi.org/10.1103/PhysRevB.85.235136} {\bibfield  {journal} {\bibinfo
  {journal} {Phys. Rev. B}\ }\textbf {\bibinfo {volume} {85}},\ \bibinfo
  {pages} {235136} (\bibinfo {year} {2012})}\BibitemShut {NoStop}%
\bibitem [{\citenamefont {Georges}\ \emph {et~al.}(1996)\citenamefont
  {Georges}, \citenamefont {Kotliar}, \citenamefont {Krauth},\ and\
  \citenamefont {Rozenberg}}]{rmp_dmft}%
  \BibitemOpen
  \bibfield  {author} {\bibinfo {author} {\bibfnamefont {A.}~\bibnamefont
  {Georges}}, \bibinfo {author} {\bibfnamefont {G.}~\bibnamefont {Kotliar}},
  \bibinfo {author} {\bibfnamefont {W.}~\bibnamefont {Krauth}},\ and\ \bibinfo
  {author} {\bibfnamefont {M.~J.}\ \bibnamefont {Rozenberg}},\ }\bibfield
  {title} {\bibinfo {title} {Dynamical mean-field theory of strongly correlated
  fermion systems and the limit of infinite dimensions},\ }\href
  {https://doi.org/10.1103/RevModPhys.68.13} {\bibfield  {journal} {\bibinfo
  {journal} {Rev. Mod. Phys.}\ }\textbf {\bibinfo {volume} {68}},\ \bibinfo
  {pages} {13} (\bibinfo {year} {1996})}\BibitemShut {NoStop}%
\bibitem [{\citenamefont {Kotliar}\ \emph {et~al.}(2006)\citenamefont
  {Kotliar}, \citenamefont {Savrasov}, \citenamefont {Haule}, \citenamefont
  {Oudovenko}, \citenamefont {Parcollet},\ and\ \citenamefont
  {Marianetti}}]{RevModPhys.78.865}%
  \BibitemOpen
  \bibfield  {author} {\bibinfo {author} {\bibfnamefont {G.}~\bibnamefont
  {Kotliar}}, \bibinfo {author} {\bibfnamefont {S.~Y.}\ \bibnamefont
  {Savrasov}}, \bibinfo {author} {\bibfnamefont {K.}~\bibnamefont {Haule}},
  \bibinfo {author} {\bibfnamefont {V.~S.}\ \bibnamefont {Oudovenko}}, \bibinfo
  {author} {\bibfnamefont {O.}~\bibnamefont {Parcollet}},\ and\ \bibinfo
  {author} {\bibfnamefont {C.~A.}\ \bibnamefont {Marianetti}},\ }\bibfield
  {title} {\bibinfo {title} {Electronic structure calculations with dynamical
  mean-field theory},\ }\href {https://doi.org/10.1103/RevModPhys.78.865}
  {\bibfield  {journal} {\bibinfo  {journal} {Rev. Mod. Phys.}\ }\textbf
  {\bibinfo {volume} {78}},\ \bibinfo {pages} {865} (\bibinfo {year}
  {2006})}\BibitemShut {NoStop}%
\bibitem [{\citenamefont {Freimuth}\ \emph {et~al.}(2022)\citenamefont
  {Freimuth}, \citenamefont {Bl\"ugel},\ and\ \citenamefont
  {Mokrousov}}]{spectral}%
  \BibitemOpen
  \bibfield  {author} {\bibinfo {author} {\bibfnamefont {F.}~\bibnamefont
  {Freimuth}}, \bibinfo {author} {\bibfnamefont {S.}~\bibnamefont {Bl\"ugel}},\
  and\ \bibinfo {author} {\bibfnamefont {Y.}~\bibnamefont {Mokrousov}},\
  }\bibfield  {title} {\bibinfo {title} {Construction of the spectral function
  from noncommuting spectral moment matrices},\ }\href
  {https://doi.org/10.1103/PhysRevB.106.045135} {\bibfield  {journal} {\bibinfo
   {journal} {Phys. Rev. B}\ }\textbf {\bibinfo {volume} {106}},\ \bibinfo
  {pages} {045135} (\bibinfo {year} {2022})}\BibitemShut {NoStop}%
\bibitem [{\citenamefont {Geipel}\ and\ \citenamefont
  {Nolting}(1988)}]{PhysRevB.38.2608}%
  \BibitemOpen
  \bibfield  {author} {\bibinfo {author} {\bibfnamefont {G.}~\bibnamefont
  {Geipel}}\ and\ \bibinfo {author} {\bibfnamefont {W.}~\bibnamefont
  {Nolting}},\ }\bibfield  {title} {\bibinfo {title} {Ferromagnetism in the
  strongly correlated hubbard model},\ }\href
  {https://doi.org/10.1103/PhysRevB.38.2608} {\bibfield  {journal} {\bibinfo
  {journal} {Phys. Rev. B}\ }\textbf {\bibinfo {volume} {38}},\ \bibinfo
  {pages} {2608} (\bibinfo {year} {1988})}\BibitemShut {NoStop}%
\bibitem [{\citenamefont {Eskes}\ \emph {et~al.}(1994)\citenamefont {Eskes},
  \citenamefont {Ole\ifmmode~\acute{s}\else \'{s}\fi{}}, \citenamefont
  {Meinders},\ and\ \citenamefont {Stephan}}]{PhysRevB.50.17980}%
  \BibitemOpen
  \bibfield  {author} {\bibinfo {author} {\bibfnamefont {H.}~\bibnamefont
  {Eskes}}, \bibinfo {author} {\bibfnamefont {A.~M.}\ \bibnamefont
  {Ole\ifmmode~\acute{s}\else \'{s}\fi{}}}, \bibinfo {author} {\bibfnamefont
  {M.~B.~J.}\ \bibnamefont {Meinders}},\ and\ \bibinfo {author} {\bibfnamefont
  {W.}~\bibnamefont {Stephan}},\ }\bibfield  {title} {\bibinfo {title}
  {Spectral properties of the hubbard bands},\ }\href
  {https://doi.org/10.1103/PhysRevB.50.17980} {\bibfield  {journal} {\bibinfo
  {journal} {Phys. Rev. B}\ }\textbf {\bibinfo {volume} {50}},\ \bibinfo
  {pages} {17980} (\bibinfo {year} {1994})}\BibitemShut {NoStop}%
\bibitem [{\citenamefont {Nolting}\ \emph {et~al.}(1995)\citenamefont
  {Nolting}, \citenamefont {Vega},\ and\ \citenamefont
  {Fauster}}]{bcc_iron_Nolting1995}%
  \BibitemOpen
  \bibfield  {author} {\bibinfo {author} {\bibfnamefont {W.}~\bibnamefont
  {Nolting}}, \bibinfo {author} {\bibfnamefont {A.}~\bibnamefont {Vega}},\ and\
  \bibinfo {author} {\bibfnamefont {T.}~\bibnamefont {Fauster}},\ }\bibfield
  {title} {\bibinfo {title} {Electronic quasiparticle structure of
  ferromagnetic bcc iron},\ }\href {https://doi.org/10.1007/BF01313058}
  {\bibfield  {journal} {\bibinfo  {journal} {Zeitschrift f{\"u}r Physik B
  Condensed Matter}\ }\textbf {\bibinfo {volume} {96}},\ \bibinfo {pages} {357}
  (\bibinfo {year} {1995})}\BibitemShut {NoStop}%
\bibitem [{\citenamefont {Kalashnikov}\ and\ \citenamefont
  {Fradkin}(1973)}]{Kalashnikov}%
  \BibitemOpen
  \bibfield  {author} {\bibinfo {author} {\bibfnamefont {O.~K.}\ \bibnamefont
  {Kalashnikov}}\ and\ \bibinfo {author} {\bibfnamefont {E.~S.}\ \bibnamefont
  {Fradkin}},\ }\bibfield  {title} {\bibinfo {title} {The spectral density
  method applied to systems showing phase transitions},\ }\href
  {https://doi.org/https://doi.org/10.1002/pssb.2220590102} {\bibfield
  {journal} {\bibinfo  {journal} {physica status solidi (b)}\ }\textbf
  {\bibinfo {volume} {59}},\ \bibinfo {pages} {9} (\bibinfo {year}
  {1973})}\BibitemShut {NoStop}%
\bibitem [{\citenamefont {White}(1991)}]{PhysRevB.44.4670}%
  \BibitemOpen
  \bibfield  {author} {\bibinfo {author} {\bibfnamefont {S.~R.}\ \bibnamefont
  {White}},\ }\bibfield  {title} {\bibinfo {title} {Spectral weight function
  for the two-dimensional hubbard model},\ }\href
  {https://doi.org/10.1103/PhysRevB.44.4670} {\bibfield  {journal} {\bibinfo
  {journal} {Phys. Rev. B}\ }\textbf {\bibinfo {volume} {44}},\ \bibinfo
  {pages} {4670} (\bibinfo {year} {1991})}\BibitemShut {NoStop}%
\bibitem [{\citenamefont {Turkowski}\ and\ \citenamefont
  {Freericks}(2008)}]{PhysRevB.77.205102}%
  \BibitemOpen
  \bibfield  {author} {\bibinfo {author} {\bibfnamefont {V.}~\bibnamefont
  {Turkowski}}\ and\ \bibinfo {author} {\bibfnamefont {J.~K.}\ \bibnamefont
  {Freericks}},\ }\bibfield  {title} {\bibinfo {title} {Nonequilibrium sum
  rules for the retarded self-energy of strongly correlated electrons},\ }\href
  {https://doi.org/10.1103/PhysRevB.77.205102} {\bibfield  {journal} {\bibinfo
  {journal} {Phys. Rev. B}\ }\textbf {\bibinfo {volume} {77}},\ \bibinfo
  {pages} {205102} (\bibinfo {year} {2008})}\BibitemShut {NoStop}%
\bibitem [{\citenamefont {Nolting}\ and\ \citenamefont
  {Brewer}(2009)}]{book_Nolting}%
  \BibitemOpen
  \bibfield  {author} {\bibinfo {author} {\bibfnamefont {W.}~\bibnamefont
  {Nolting}}\ and\ \bibinfo {author} {\bibfnamefont {W.}~\bibnamefont
  {Brewer}},\ }\href@noop {} {\emph {\bibinfo {title} {Fundamentals of
  Many-body Physics: Principles and Methods}}}\ (\bibinfo  {publisher}
  {Springer Berlin Heidelberg},\ \bibinfo {year} {2009})\BibitemShut {NoStop}%
\bibitem [{\citenamefont {Perdew}\ and\ \citenamefont
  {Wang}(1992)}]{PhysRevB.45.13244}%
  \BibitemOpen
  \bibfield  {author} {\bibinfo {author} {\bibfnamefont {J.~P.}\ \bibnamefont
  {Perdew}}\ and\ \bibinfo {author} {\bibfnamefont {Y.}~\bibnamefont {Wang}},\
  }\bibfield  {title} {\bibinfo {title} {Accurate and simple analytic
  representation of the electron-gas correlation energy},\ }\href
  {https://doi.org/10.1103/PhysRevB.45.13244} {\bibfield  {journal} {\bibinfo
  {journal} {Phys. Rev. B}\ }\textbf {\bibinfo {volume} {45}},\ \bibinfo
  {pages} {13244} (\bibinfo {year} {1992})}\BibitemShut {NoStop}%
\bibitem [{\citenamefont {Perdew}\ and\ \citenamefont
  {Zunger}(1981)}]{sic_perdew_zunger}%
  \BibitemOpen
  \bibfield  {author} {\bibinfo {author} {\bibfnamefont {J.~P.}\ \bibnamefont
  {Perdew}}\ and\ \bibinfo {author} {\bibfnamefont {A.}~\bibnamefont
  {Zunger}},\ }\bibfield  {title} {\bibinfo {title} {Self-interaction
  correction to density-functional approximations for many-electron systems},\
  }\href {https://doi.org/10.1103/PhysRevB.23.5048} {\bibfield  {journal}
  {\bibinfo  {journal} {Phys. Rev. B}\ }\textbf {\bibinfo {volume} {23}},\
  \bibinfo {pages} {5048} (\bibinfo {year} {1981})}\BibitemShut {NoStop}%
\bibitem [{\citenamefont {Heyd}\ \emph {et~al.}(2003)\citenamefont {Heyd},
  \citenamefont {Scuseria},\ and\ \citenamefont {Ernzerhof}}]{hse_functional}%
  \BibitemOpen
  \bibfield  {author} {\bibinfo {author} {\bibfnamefont {J.}~\bibnamefont
  {Heyd}}, \bibinfo {author} {\bibfnamefont {G.~E.}\ \bibnamefont {Scuseria}},\
  and\ \bibinfo {author} {\bibfnamefont {M.}~\bibnamefont {Ernzerhof}},\
  }\bibfield  {title} {\bibinfo {title} {Hybrid functionals based on a screened
  coulomb potential},\ }\href {https://doi.org/10.1063/1.1564060} {\bibfield
  {journal} {\bibinfo  {journal} {The Journal of Chemical Physics}\ }\textbf
  {\bibinfo {volume} {118}},\ \bibinfo {pages} {8207} (\bibinfo {year}
  {2003})}\BibitemShut {NoStop}%
\bibitem [{\citenamefont {Ceperley}\ and\ \citenamefont
  {Alder}(1980)}]{PhysRevLett.45.566}%
  \BibitemOpen
  \bibfield  {author} {\bibinfo {author} {\bibfnamefont {D.~M.}\ \bibnamefont
  {Ceperley}}\ and\ \bibinfo {author} {\bibfnamefont {B.~J.}\ \bibnamefont
  {Alder}},\ }\bibfield  {title} {\bibinfo {title} {Ground state of the
  electron gas by a stochastic method},\ }\href
  {https://doi.org/10.1103/PhysRevLett.45.566} {\bibfield  {journal} {\bibinfo
  {journal} {Phys. Rev. Lett.}\ }\textbf {\bibinfo {volume} {45}},\ \bibinfo
  {pages} {566} (\bibinfo {year} {1980})}\BibitemShut {NoStop}%
\bibitem [{\citenamefont {Gell-Mann}\ and\ \citenamefont
  {Brueckner}(1957)}]{PhysRev.106.364}%
  \BibitemOpen
  \bibfield  {author} {\bibinfo {author} {\bibfnamefont {M.}~\bibnamefont
  {Gell-Mann}}\ and\ \bibinfo {author} {\bibfnamefont {K.~A.}\ \bibnamefont
  {Brueckner}},\ }\bibfield  {title} {\bibinfo {title} {Correlation energy of
  an electron gas at high density},\ }\href
  {https://doi.org/10.1103/PhysRev.106.364} {\bibfield  {journal} {\bibinfo
  {journal} {Phys. Rev.}\ }\textbf {\bibinfo {volume} {106}},\ \bibinfo {pages}
  {364} (\bibinfo {year} {1957})}\BibitemShut {NoStop}%
\bibitem [{\citenamefont {Eberhardt}\ and\ \citenamefont
  {Plummer}(1980)}]{PhysRevB.21.3245}%
  \BibitemOpen
  \bibfield  {author} {\bibinfo {author} {\bibfnamefont {W.}~\bibnamefont
  {Eberhardt}}\ and\ \bibinfo {author} {\bibfnamefont {E.~W.}\ \bibnamefont
  {Plummer}},\ }\bibfield  {title} {\bibinfo {title} {Angle-resolved
  photoemission determination of the band structure and multielectron
  excitations in ni},\ }\href {https://doi.org/10.1103/PhysRevB.21.3245}
  {\bibfield  {journal} {\bibinfo  {journal} {Phys. Rev. B}\ }\textbf {\bibinfo
  {volume} {21}},\ \bibinfo {pages} {3245} (\bibinfo {year}
  {1980})}\BibitemShut {NoStop}%
\bibitem [{\citenamefont {Eastman}\ \emph {et~al.}(1978)\citenamefont
  {Eastman}, \citenamefont {Himpsel},\ and\ \citenamefont
  {Knapp}}]{PhysRevLett.40.1514}%
  \BibitemOpen
  \bibfield  {author} {\bibinfo {author} {\bibfnamefont {D.~E.}\ \bibnamefont
  {Eastman}}, \bibinfo {author} {\bibfnamefont {F.~J.}\ \bibnamefont
  {Himpsel}},\ and\ \bibinfo {author} {\bibfnamefont {J.~A.}\ \bibnamefont
  {Knapp}},\ }\bibfield  {title} {\bibinfo {title} {Experimental band structure
  and temperature-dependent magnetic exchange splitting of nickel using
  angle-resolved photoemission},\ }\href
  {https://doi.org/10.1103/PhysRevLett.40.1514} {\bibfield  {journal} {\bibinfo
   {journal} {Phys. Rev. Lett.}\ }\textbf {\bibinfo {volume} {40}},\ \bibinfo
  {pages} {1514} (\bibinfo {year} {1978})}\BibitemShut {NoStop}%
\bibitem [{\citenamefont {See}\ and\ \citenamefont
  {Klebanoff}(1995)}]{PhysRevB.51.11002}%
  \BibitemOpen
  \bibfield  {author} {\bibinfo {author} {\bibfnamefont {A.~K.}\ \bibnamefont
  {See}}\ and\ \bibinfo {author} {\bibfnamefont {L.~E.}\ \bibnamefont
  {Klebanoff}},\ }\bibfield  {title} {\bibinfo {title} {Spin-resolved x-ray
  photoemission from ferromagnetic nickel},\ }\href
  {https://doi.org/10.1103/PhysRevB.51.11002} {\bibfield  {journal} {\bibinfo
  {journal} {Phys. Rev. B}\ }\textbf {\bibinfo {volume} {51}},\ \bibinfo
  {pages} {11002} (\bibinfo {year} {1995})}\BibitemShut {NoStop}%
\bibitem [{\citenamefont {Lichtenstein}\ \emph {et~al.}(2001)\citenamefont
  {Lichtenstein}, \citenamefont {Katsnelson},\ and\ \citenamefont
  {Kotliar}}]{PhysRevLett.87.067205}%
  \BibitemOpen
  \bibfield  {author} {\bibinfo {author} {\bibfnamefont {A.~I.}\ \bibnamefont
  {Lichtenstein}}, \bibinfo {author} {\bibfnamefont {M.~I.}\ \bibnamefont
  {Katsnelson}},\ and\ \bibinfo {author} {\bibfnamefont {G.}~\bibnamefont
  {Kotliar}},\ }\bibfield  {title} {\bibinfo {title} {Finite-temperature
  magnetism of transition metals: An ab initio dynamical mean-field theory},\
  }\href {https://doi.org/10.1103/PhysRevLett.87.067205} {\bibfield  {journal}
  {\bibinfo  {journal} {Phys. Rev. Lett.}\ }\textbf {\bibinfo {volume} {87}},\
  \bibinfo {pages} {067205} (\bibinfo {year} {2001})}\BibitemShut {NoStop}%
\bibitem [{\citenamefont {Inoue}\ \emph {et~al.}(1994)\citenamefont {Inoue},
  \citenamefont {Hase}, \citenamefont {Aiura}, \citenamefont {Fujimori},
  \citenamefont {Morikawa}, \citenamefont {Mizokawa}, \citenamefont {Haruyama},
  \citenamefont {Maruyama},\ and\ \citenamefont {Nishihara}}]{INOUE19941007}%
  \BibitemOpen
  \bibfield  {author} {\bibinfo {author} {\bibfnamefont {I.}~\bibnamefont
  {Inoue}}, \bibinfo {author} {\bibfnamefont {I.}~\bibnamefont {Hase}},
  \bibinfo {author} {\bibfnamefont {Y.}~\bibnamefont {Aiura}}, \bibinfo
  {author} {\bibfnamefont {A.}~\bibnamefont {Fujimori}}, \bibinfo {author}
  {\bibfnamefont {K.}~\bibnamefont {Morikawa}}, \bibinfo {author}
  {\bibfnamefont {T.}~\bibnamefont {Mizokawa}}, \bibinfo {author}
  {\bibfnamefont {Y.}~\bibnamefont {Haruyama}}, \bibinfo {author}
  {\bibfnamefont {T.}~\bibnamefont {Maruyama}},\ and\ \bibinfo {author}
  {\bibfnamefont {Y.}~\bibnamefont {Nishihara}},\ }\bibfield  {title} {\bibinfo
  {title} {Systematic change of spectral function observed by controlling
  electron correlation in casrvo3 with fixed 3d1 configuration.},\ }\href
  {https://doi.org/https://doi.org/10.1016/0921-4534(94)91728-0} {\bibfield
  {journal} {\bibinfo  {journal} {Physica C: Superconductivity}\ }\textbf
  {\bibinfo {volume} {235}},\ \bibinfo {pages} {1007} (\bibinfo {year}
  {1994})}\BibitemShut {NoStop}%
\bibitem [{\citenamefont {Sekiyama}\ \emph {et~al.}(2004)\citenamefont
  {Sekiyama}, \citenamefont {Fujiwara}, \citenamefont {Imada}, \citenamefont
  {Suga}, \citenamefont {Eisaki}, \citenamefont {Uchida}, \citenamefont
  {Takegahara}, \citenamefont {Harima}, \citenamefont {Saitoh}, \citenamefont
  {Nekrasov}, \citenamefont {Keller}, \citenamefont {Kondakov}, \citenamefont
  {Kozhevnikov}, \citenamefont {Pruschke}, \citenamefont {Held}, \citenamefont
  {Vollhardt},\ and\ \citenamefont {Anisimov}}]{PhysRevLett.93.156402}%
  \BibitemOpen
  \bibfield  {author} {\bibinfo {author} {\bibfnamefont {A.}~\bibnamefont
  {Sekiyama}}, \bibinfo {author} {\bibfnamefont {H.}~\bibnamefont {Fujiwara}},
  \bibinfo {author} {\bibfnamefont {S.}~\bibnamefont {Imada}}, \bibinfo
  {author} {\bibfnamefont {S.}~\bibnamefont {Suga}}, \bibinfo {author}
  {\bibfnamefont {H.}~\bibnamefont {Eisaki}}, \bibinfo {author} {\bibfnamefont
  {S.~I.}\ \bibnamefont {Uchida}}, \bibinfo {author} {\bibfnamefont
  {K.}~\bibnamefont {Takegahara}}, \bibinfo {author} {\bibfnamefont
  {H.}~\bibnamefont {Harima}}, \bibinfo {author} {\bibfnamefont
  {Y.}~\bibnamefont {Saitoh}}, \bibinfo {author} {\bibfnamefont {I.~A.}\
  \bibnamefont {Nekrasov}}, \bibinfo {author} {\bibfnamefont {G.}~\bibnamefont
  {Keller}}, \bibinfo {author} {\bibfnamefont {D.~E.}\ \bibnamefont
  {Kondakov}}, \bibinfo {author} {\bibfnamefont {A.~V.}\ \bibnamefont
  {Kozhevnikov}}, \bibinfo {author} {\bibfnamefont {T.}~\bibnamefont
  {Pruschke}}, \bibinfo {author} {\bibfnamefont {K.}~\bibnamefont {Held}},
  \bibinfo {author} {\bibfnamefont {D.}~\bibnamefont {Vollhardt}},\ and\
  \bibinfo {author} {\bibfnamefont {V.~I.}\ \bibnamefont {Anisimov}},\
  }\bibfield  {title} {\bibinfo {title} {Mutual experimental and theoretical
  validation of bulk photoemission spectra of
  ${\mathrm{s}\mathrm{r}}_{1\ensuremath{-}x}{\mathrm{c}\mathrm{a}}_{x}{\mathrm{v}\mathrm{o}}_{3}$},\
  }\href {https://doi.org/10.1103/PhysRevLett.93.156402} {\bibfield  {journal}
  {\bibinfo  {journal} {Phys. Rev. Lett.}\ }\textbf {\bibinfo {volume} {93}},\
  \bibinfo {pages} {156402} (\bibinfo {year} {2004})}\BibitemShut {NoStop}%
\bibitem [{\citenamefont {Amadon}\ \emph {et~al.}(2008)\citenamefont {Amadon},
  \citenamefont {Lechermann}, \citenamefont {Georges}, \citenamefont {Jollet},
  \citenamefont {Wehling},\ and\ \citenamefont
  {Lichtenstein}}]{PhysRevB.77.205112}%
  \BibitemOpen
  \bibfield  {author} {\bibinfo {author} {\bibfnamefont {B.}~\bibnamefont
  {Amadon}}, \bibinfo {author} {\bibfnamefont {F.}~\bibnamefont {Lechermann}},
  \bibinfo {author} {\bibfnamefont {A.}~\bibnamefont {Georges}}, \bibinfo
  {author} {\bibfnamefont {F.}~\bibnamefont {Jollet}}, \bibinfo {author}
  {\bibfnamefont {T.~O.}\ \bibnamefont {Wehling}},\ and\ \bibinfo {author}
  {\bibfnamefont {A.~I.}\ \bibnamefont {Lichtenstein}},\ }\bibfield  {title}
  {\bibinfo {title} {Plane-wave based electronic structure calculations for
  correlated materials using dynamical mean-field theory and projected local
  orbitals},\ }\href {https://doi.org/10.1103/PhysRevB.77.205112} {\bibfield
  {journal} {\bibinfo  {journal} {Phys. Rev. B}\ }\textbf {\bibinfo {volume}
  {77}},\ \bibinfo {pages} {205112} (\bibinfo {year} {2008})}\BibitemShut
  {NoStop}%
\bibitem [{\citenamefont {Potthoff}\ \emph {et~al.}(1998)\citenamefont
  {Potthoff}, \citenamefont {Herrmann}, \citenamefont {Wegner},\ and\
  \citenamefont {Nolting}}]{moment_sum_rule}%
  \BibitemOpen
  \bibfield  {author} {\bibinfo {author} {\bibfnamefont {M.}~\bibnamefont
  {Potthoff}}, \bibinfo {author} {\bibfnamefont {T.}~\bibnamefont {Herrmann}},
  \bibinfo {author} {\bibfnamefont {T.}~\bibnamefont {Wegner}},\ and\ \bibinfo
  {author} {\bibfnamefont {W.}~\bibnamefont {Nolting}},\ }\bibfield  {title}
  {\bibinfo {title} {The moment sum rule and its consequences for
  ferromagnetism in the hubbard model},\ }\href
  {https://doi.org/https://doi.org/10.1002/(SICI)1521-3951(199811)210:1<199::AID-PSSB199>3.0.CO;2-3}
  {\bibfield  {journal} {\bibinfo  {journal} {physica status solidi (b)}\
  }\textbf {\bibinfo {volume} {210}},\ \bibinfo {pages} {199} (\bibinfo {year}
  {1998})}\BibitemShut {NoStop}%
\bibitem [{\citenamefont {Nolting}\ and\ \citenamefont
  {Oles}(1980)}]{Nolting_1980}%
  \BibitemOpen
  \bibfield  {author} {\bibinfo {author} {\bibfnamefont {W.}~\bibnamefont
  {Nolting}}\ and\ \bibinfo {author} {\bibfnamefont {A.~M.}\ \bibnamefont
  {Oles}},\ }\bibfield  {title} {\bibinfo {title} {Spectral density approach
  for the quasiparticle concept in the s-f model (ferromagnetic
  semiconductors)},\ }\href {https://doi.org/10.1088/0022-3719/13/12/012}
  {\bibfield  {journal} {\bibinfo  {journal} {Journal of Physics C: Solid State
  Physics}\ }\textbf {\bibinfo {volume} {13}},\ \bibinfo {pages} {2295}
  (\bibinfo {year} {1980})}\BibitemShut {NoStop}%
\bibitem [{\citenamefont {{Peter Lepage}}(1978)}]{PETERLEPAGE1978192}%
  \BibitemOpen
  \bibfield  {author} {\bibinfo {author} {\bibfnamefont {G.}~\bibnamefont
  {{Peter Lepage}}},\ }\bibfield  {title} {\bibinfo {title} {A new algorithm
  for adaptive multidimensional integration},\ }\href
  {https://doi.org/https://doi.org/10.1016/0021-9991(78)90004-9} {\bibfield
  {journal} {\bibinfo  {journal} {Journal of Computational Physics}\ }\textbf
  {\bibinfo {volume} {27}},\ \bibinfo {pages} {192} (\bibinfo {year}
  {1978})}\BibitemShut {NoStop}%
\bibitem [{\citenamefont {{Peter Lepage}}(2021)}]{LEPAGE2021110386}%
  \BibitemOpen
  \bibfield  {author} {\bibinfo {author} {\bibfnamefont {G.}~\bibnamefont
  {{Peter Lepage}}},\ }\bibfield  {title} {\bibinfo {title} {Adaptive
  multidimensional integration: vegas enhanced},\ }\href
  {https://doi.org/https://doi.org/10.1016/j.jcp.2021.110386} {\bibfield
  {journal} {\bibinfo  {journal} {Journal of Computational Physics}\ }\textbf
  {\bibinfo {volume} {439}},\ \bibinfo {pages} {110386} (\bibinfo {year}
  {2021})}\BibitemShut {NoStop}%
\end{thebibliography}%

\end{document}